\documentclass[
nonacm,  acmsmall
]{acmart}
\setcopyright{none}

 \usepackage{amsmath}
 \usepackage{proof}
\usepackage{wrapfig}
 \usepackage{amsthm}
\usepackage[T1]{fontenc}
\usepackage[utf8]{inputenc}
 \usepackage{array,etoolbox, tikz}
\usepackage{algorithm}
\usepackage{algorithmic}
\usetikzlibrary{shapes,fit,calc,positioning,decorations.pathreplacing}
 \usepackage{blindtext}
 \usepackage{bussproofs}
 \usepackage{makecell}
 \usepackage{multicol}
 \usepackage{multirow}
 \usepackage{pdfpages}
\usepackage{float} 
 \usepackage{color,colortbl}
  \tikzstyle{bl}=[ rectangle, fill =red, minimum width=.4cm, minimum height=.4cm]
  \tikzstyle{un}=[ rectangle, fill =red, minimum width=.4cm, minimum height=.4cm]
  \tikzstyle{mun}=[ rectangle, fill =blue, minimum width=.4cm, minimum height=.4cm]
  \tikzstyle{ghost}=[ rectangle, minimum width=.4cm, minimum height=.4cm]

  \usepackage{enumitem}
 \usepackage{algorithm}
 \usepackage{pifont}
\usepackage[all]{xy}

\newcommand{\polyshor}{\ensuremath{\texttt{Shor-poly}}}
\newcommand{\hrange}{\ensuremath{\texttt{hops\_range}}}
\newcommand{\hwidth}{\ensuremath{\texttt{hops\_width}}}
\newcommand{\hangle}{\ensuremath{\texttt{hops\_angle}}}
\newcommand{\hket}{\ensuremath{\texttt{hops\_ket}}}
\newcommand{\flathangle}{\ensuremath{\texttt{flat\_h\_angle}}}
\newcommand{\flathket}{\ensuremath{\texttt{flat\_h\_ket}}}
\newcommand{\flatangle}{\ensuremath{\texttt{flat\_angle}}}
\newcommand{\flatket}{\ensuremath{\texttt{flat\_ket}}}
\newcommand{\iogate}{\ensuremath{\texttt{circuit\_io}}}
\newcommand{\ketlength}{\ensuremath{\texttt{ket\_length}}}
\newcommand{\bvtoket}{\ensuremath{\texttt{bv\_to\_ket}}}

\newcommand{\coprime}{\ensuremath{\texttt{co\_prime}}}
\newcommand{\bvlength}{\ensuremath{\texttt{bv\_length}}}
\newcommand{\applyps}{\ensuremath{\texttt{hops\_apply}}}

\newcommand{\noeg}{\ensuremath{\texttt{size}}}

\newcommand{\getket}{\ensuremath{\texttt{ket\_get}}}
\newcommand{\getbv}{\ensuremath{\texttt{bv\_get}}}
\newcommand{\makebv}{\ensuremath{\texttt{make\_bv}}}

\newcommand{\qsize}{\ensuremath{\texttt{qsize}}}

\newcommand{\hopsequiv}{\ensuremath{\texttt{hopsequiv}}}

\newcommand{\hopssequence}{\ensuremath{\texttt{hops\_seq}}}
\newcommand{\hopsctl}{\ensuremath{\texttt{hops\_ctl}}}
\newcommand{\hopsanc}{\ensuremath{\texttt{hops\_ancilla}}}
\newcommand{\hopspar}{\ensuremath{\texttt{hops\_par}}}
\newcommand{\hopshad}{\ensuremath{\texttt{hops\_had}}}
\newcommand{\hopsrz}{\ensuremath{\texttt{hops\_rz}}}
\newcommand{\hopsphase}{\ensuremath{\texttt{hops\_phase}}}
\newcommand{\hopsswap}{\ensuremath{\texttt{hops\_swap}}}
\newcommand{\hopsid}{\ensuremath{\texttt{hops\_id}}}
\newcommand{\hopscnot}{\ensuremath{\texttt{hops\_cnot}}}

\newcommand{\ancillas}{\ensuremath{\texttt{ancillas}}\xspace}

\newcommand{\nogates}{\ensuremath{\texttt{size}}\xspace}

\newcommand{\probmeaspart}{\ensuremath{\texttt{proba\_partial\_measure}}\xspace}
\newcommand{\probmeaspartp}{\ensuremath{\texttt{proba\_partial\_measure\_p}}\xspace}
\newcommand{\implements}{\ensuremath{\texttt{implements}}\xspace}
\newcommand{\balanced}{\ensuremath{\texttt{balanced}}\xspace}
\newcommand{\constant}{\ensuremath{\texttt{constant}}\xspace}

\newcommand{\gc}{\ensuremath{\texttt{GC}}\xspace}

\newcommand{\bvtoint}{\ensuremath{\texttt{bv\_to\_int}}\xspace}

\newcommand{\rtoc}{\ensuremath{\text{r\_to\_c}}}
\newcommand{\ctoa}{\ensuremath{\text{c\_to\_a}}}

\newcommand{\atoc}{\ensuremath{\text{a\_to\_c}}}

\newcommand{\rcorrect}[2]{\ensuremath{(#2 \triangleright  #1)}}

\newcommand{\defin}{\ensuremath{\mathit{def}}}
\newcommand{\tofset}[1]{\ensuremath{\llbracket0, #1 \llbracket}\xspace }
\newcommand{\tofsett}[2]{\ensuremath{\llbracket #1, #2 \llbracket}\xspace }

\newcommand{\bitrev}[1]{\ensuremath{\overleftrightarrow{#1}}\xspace }

\newcommand{\bin}{\ensuremath{\texttt{binary}}\xspace }

\newcommand{\intounitm}[2]{\ensuremath{\omega_n^{-k}}}

\newcommand{\checkm}{\ensuremath{\includegraphics[width = 0.02\textwidth]{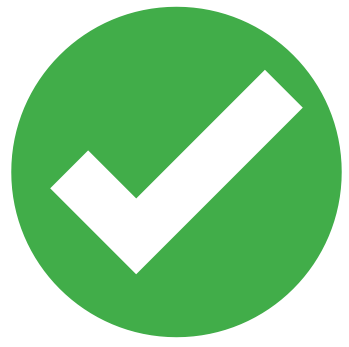}}}
\newcommand{\xmark}{\ensuremath{\includegraphics[width = 0.015\textwidth]{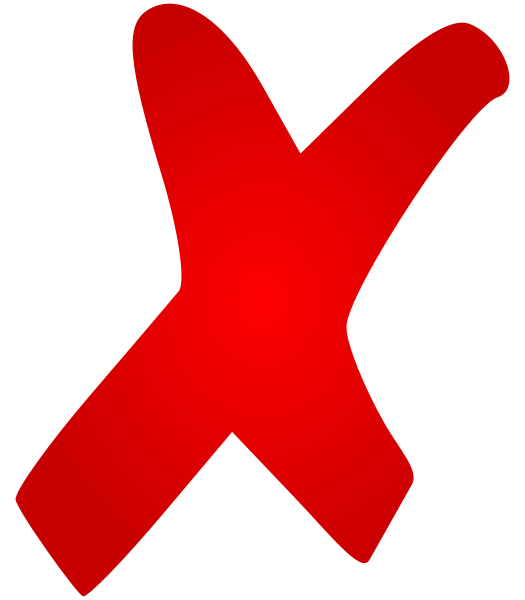}}}
\newcommand{\size}[1]{\ensuremath{\texttt{width}(#1)}\xspace }
\newcommand{\width}{\ensuremath{\texttt{width}}\xspace }
\newcommand{\ctlconst}{\ensuremath{\texttt{ctl-size-const}}\xspace }

\newcommand{\kettype}{\ensuremath{\texttt{ket}} }

\renewcommand{\mod}{\ensuremath{\texttt{mod}}}
\renewcommand{\div}{\ensuremath{\texttt{div}}\xspace}

\newcommand{\qbrickCORE}{\ensuremath{\textsc{Qbricks}}}
\newcommand{\qbrick}{\qbrickCORE\xspace}

\newcommand{\qbricks}{\qbrickCORE\xspace}
\newcommand{\qbrickDSL}{\qbrickCORE-\textsc{DSL}\xspace}

\newcommand{\qbrickSPEC}{\qbrickCORE-\textsc{Spec}\xspace}
\newcommand{\hqhl}{HQHL}
\newcommand{\hqhllong}{Hybrid Quantum Hoare Logic}

\newcommand{\circuit}{\ensuremath{\text{circuit}}\ }

\renewcommand{\circuit}{\ensuremath{\text{circ}}\ }

\newcommand{\Int}{\ensuremath{\text{int}}\ }
\newcommand{\complex}{\ensuremath{\text{complex}}\ }

\newcommand{\qft}{\ensuremath{\text{QFT}}\ }

\newcommand{\ghost}{\texttt{ghost}\xspace }

\newcounter{magicrownumbers}

\newcommand{\ket}[2]{\ensuremath{\vert #1 \rangle}_{#2}}

 \newcommand{\real}{\ensuremath{\texttt{Re}}\xspace }

 \newcommand{\imag}{\ensuremath{\texttt{Im}}\xspace }

 \newcommand{\eigen}{\ensuremath{\texttt{Eigen}}\xspace }

 \renewcommand{\flat}{\ensuremath{\texttt{flat}}\xspace }
 \newcommand{\result}{\ensuremath{\texttt{result}}\xspace }

\newcommand{\reverse}[1]{\ensuremath{\texttt{inverse}(#1)}\ }

\newcommand{\mycomment}[1]{}

\newcommand{\bor}{\mid}
\newcommand{\tuple}[1]{\langle#1\rangle}

\newcommand{\ttrue}{\texttt{t\!t}}
\newcommand{\ffalse}{\texttt{f\!f}}
\newcommand{\iftermx}[3]{\texttt{if}\,{#1}\,\texttt{then}\,{#2}\,\texttt{else}\,{#3}}
\newcommand{\letinterm}[3]{\texttt{let}\,{#1}\,{=}\,{#2}\,\texttt{in}\,{#3}}
\newcommand{\decl}[3]{\texttt{let}~{#1}(#2)~{=}~{#3}}
\newcommand{\bool}{\texttt{bool}}
\newcommand{\inttype}{\texttt{int}}

\newcommand{\realtype}{\texttt{real}}
\newcommand{\complextype}{\texttt{complex}}
\newcommand{\circtype}{\texttt{circ}}
\newcommand{\unittype}{\top}
\newcommand{\hops}{\texttt{hops}}
\newcommand{\hop}{\texttt{hop}}

\newcommand{\concat}{\texttt{concat}}

\newcommand{\bitvector}{\texttt{bitvector}}
\newcommand{\inttobv}{\texttt{int\_to\_bv}}

\newcommand{\cond}[1]{\{{#1}\}}
\newcommand{\ahoare}[3]{\begin{array}{rcl}\cond{#1}&{#2}&\cond{#3}\end{array}}
\newcommand{\ahoarelr}[3]{\begin{array}{rcl}\left\{#1\right\}&#2&\left\{#3\right\}\end{array}}
\newcommand{\ahoarelrc}[4]{\begin{array}{rcl}
\multicolumn{1}{l}{#1}&&\\
\begin{array}{r}
    \left\{#2\right\}\end{array}&#3&\left\{#4\right\}\end{array}}

\newcommand{\ahoareBigr}[3]{\begin{array}{rcl}\cond{#1}&{#2}&\Big\{#3\Big\}\end{array}}

\newcommand{\hoare}[3]{\cond{#1}{#2}\cond{#3}}

\newcommand{\denot}[1]{{[\!|{#1}|\!]}}

 \definecolor{mygrey}{rgb}{0.9, 0.95, 0.95}
 \definecolor{yell}{RGB}{246,187,0}
\definecolor{yell}{RGB}{246,187,0}
\definecolor{blu}{RGB}{0,20,100}
\definecolor{re}{RGB}{148,6,39}
\definecolor{gree}{RGB}{0,100,0}
\usepackage{listings}
\usepackage{hyperref}
\newtheorem{property}{Property}[subsection]
\makeatletter

\let\c@table\c@figure
\makeatother 
\usepackage{etoolbox}

\theoremstyle{definition}

\usepackage{xkeyval}
\usepackage{wasysym }
\usepackage{textcomp}
\usepackage[disable]{todonotes}
\usepackage[framemethod=TikZ]{mdframed}
\usepackage{stmaryrd}

\newcommand{\benvoid}[1]{\color{gray!20}}

 \tikzstyle{un}=[ rectangle, fill =red, minimum width=.4cm, minimum height=.4cm]
  \tikzstyle{mun}=[ rectangle, fill =blue, minimum width=.4cm, minimum height=.4cm]
  \tikzstyle{ghost}=[ rectangle, minimum width=.4cm, minimum height=.4cm]
  \date{}
\mdfdefinestyle{listingstyle}{
  backgroundcolor=gray!20,outerlinecolor=black,
  innerleftmargin=4pt,innerrightmargin=4pt,innertopmargin=0pt,innerbottommargin=-5pt%
}

  \BeforeBeginEnvironment{lstlisting}{\begin{mdframed}[style=listingstyle]}
\AfterEndEnvironment{lstlisting}{\end{mdframed}}

\usepackage{booktabs}   
\usepackage{subcaption} 

\renewcommand{\bigcirc}{\mathcal{O}}
\def\paragraph#1{\medskip\noindent\textit{\textbf{#1.~~}}}

\makeatletter
\renewcommand\subsection{%
  \@startsection{subsection}{1}{0em}%
                            {1ex \@plus1ex \@minus.2ex}%
                            {0em}%
                            {\normalfont\normalsize\bfseries}}
\makeatother
\let\oldsubsection\subsection
\def\subsection#1{\oldsubsection{\textbf{#1.\rule{1.2ex}{0ex}}}}

\begin{document}


\title{A Deductive Verification Framework for Circuit-building Quantum Programs}

\author{Christophe Chareton}
\affiliation{
  \institution{LRI, Centrale Supélec, Université Paris-Saclay, France}            
}
\email{christophe.chareton@lri.fr}          

\author{Sébastien Bardin}
\affiliation{
  \institution{CEA, LIST, Université Paris-Saclay, France}            
}
\email{sebastien.bardin@cea.fr}          

\author{François Bobot}
\affiliation{
  \institution{CEA, LIST, Université Paris-Saclay, France}            
}
\email{francois.bobot@cea.fr}          

\author{Valentin Perrelle}
\affiliation{
  \institution{CEA, LIST, Université Paris-Saclay, France}            
}
\email{valentin.perrelle@cea.fr}          

\author{Benoit Valiron}
\affiliation{
  \institution{LRI, Centrale Supélec, Université Paris-Saclay, France}            
}
\email{benoit.valiron@lri.fr}          

\begin{abstract}
    While recent progress in quantum hardware open the door for
  significant speedup in certain key areas, quantum algorithms are
  still hard to implement right, and the validation of such quantum
  programs is a challenge.  Early attempts either suffer from the lack
  of automation 
  or parametrized reasoning, 
  or target high-level abstract algorithm description languages 
   far from the current {\it de facto} consensus of {\it circuit-building quantum programming languages}. 
   %
  As a consequence, no significant
  quantum algorithm implementation has been currently verified in a
  scale-invariant manner. 
  We propose \qbrick, the first formal verification  environment for 
  circuit-building quantum programs,  featuring clear separation between code and proof,
  parametric specifications and proofs, high degree of proof
  automation and allowing to encode quantum programs in a natural way,
  i.e.~close to textbook style. 
  \qbrick builds on best practice of formal verification for the
  classical case and tailor them to the quantum case:  
we bring a new  domain-specific
  circuit-building language for quantum programs, namely \qbrickDSL,  together with a new logical specification language
  \qbrickSPEC\ and a dedicated Hoare-style  deductive verification rule named \hqhllong.   Especially,  {\it we introduce  and intensively build upon  HOPS,  a higher-order extension of the
  recent path-sum symbolic representation}, used for  both {\it specification} (parametrized) and {\it automation} (closure properties).     
  %

 To illustrate the
  opportunity of \qbrick, we implement the first 
  verified parametric implementations of several famous and non-trivial
  quantum algorithms, including the quantum part of Shor integer factoring
  (Order Finding -- Shor-OF), 
  quantum phase estimation (QPE) -- a basic building block of many quantum algorithms,  and Grover search. 
These breakthroughs were amply facilitated by the specification and automated deduction principles introduced within \qbrick.  

\end{abstract}

\ccsdesc[500]{Software and its engineering~Formal language definitions}
\ccsdesc[500]{Theory of computation~Semantics and reasoning}
\ccsdesc[500]{Hardware~Quantum computation}

\keywords{deductive verification, quantum programming, quantum circuits}  


\maketitle

\section{Introduction}
\label{intro}

\subsection{Quantum computing} 
Quantum computing is a young research field:  its birth act is
usually dated in 1982, when \citet{Feynman:1982:SPC}
raised the idea of simulating the quantum mechanics phenomena by
storing information in particles and controlling them according to
quantum mechanics laws. This initial idea has become one over many
application fields for quantum computing, 
along with cryptography~\cite{shor1994algorithms}, deep
learning~\cite{biamonte2017quantum},
optimization~\cite{farhi2001quantum,farhi2014quantum}, solving linear
systems~\cite{harrow2009quantum}, etc. 
In all these domains there are now quantum algorithms beating the best
known classical algorithms by either quadratic or even
exponential
factors. 
In parallel to the rise of quantum algorithms, the design of quantum
hardware has moved from lab-benches~\cite{chuang1998experimental} to
programmable, 50-qubits machines designed by industrial
actors~\cite{arute2019quantum,ibmblog} reaching the point where
quantum computers
would beat classical computers for specific tasks
\cite{arute2019quantum}. \begin{wrapfigure}[7]{r}{0.4\textwidth}
      \begin{center}
        \scalebox{.6}{
        \begin{tikzpicture}
          
\begin{scope}

  \node(cc)[draw,rectangle,minimum width = 1.6cm,minimum height =
        .9cm,ultra thick,rounded corners=3pt]at(0,0){{\tiny\begin{tabular}{l}Classical\\controller\end{tabular}}};
\draw[fill= black](-.6,-.7) to[bend right](cc.230) to (cc.310) to [bend right](.6,-.7); 
\draw[very thick,rounded corners=2pt](-.2,-.8)to(.7,-.8)to[very thick](.8,-1)to[very thick](-.8,-1)to(-.7,-.8)to(.2,-.8);
\draw[fill= black,rounded corners=2pt](.85,-.95) to (.8,-.8) to (1,-.8) to (1.1,-1) to (.9,-1) to (.85,-.95); 

\begin{scope}[xshift =-.2cm]
  \node(comp)[draw,fill=black,rectangle,minimum width = .6cm,minimum height = 1.15cm,
  ultra thick,rounded corners=3pt]at(-1.25,-.15){};
  \node(hole)[fill=white,rectangle,minimum width = .55cm,minimum height = .52cm,
  ultra thick,rounded corners=3pt]at(-1.25,.1){};
  
  \draw[thick](hole.0)to (hole.180);
  \draw[thick](hole.25)to (hole.155);
  \draw[thick](hole.-25)to (hole.205);
  \draw[thick](hole.-25)to (hole.205);
  \node[very thick,rectangle,draw, minimum width = .5cm, rounded corners=1pt]at(-1.25,-.7){};
  
\end{scope}
\begin{scope}[xshift =0cm]
  \node(s1) at(1.2,.5){};
  \node(s2) at(3,.5){};
  \node(s3) at(1.2,-1){};
  \node(s4) at(3,-1){};

\begin{scope}[yshift =2.7cm,xshift =3.5cm]
\node(qram)[draw,ultra thick,rounded corners=2pt,rectangle,minimum width=1cm,minimum height=1cm] at(0,-3){{\tiny\begin{tabular}{l}Quantum\\memory\end{tabular}}};
\draw[ultra thick](.4,-2.5)to  (.4,-2.3);
\draw[ultra thick](.2,-2.5)to  (.2,-2.3);
\draw[ultra thick](0,-2.5)to  (0,-2.3);
\draw[ultra thick](-.2,-2.5)to  (-.2,-2.3);
\draw[ultra thick](-.4,-2.5)to  (-.4,-2.3);
\draw[ultra thick](.4,-3.5)to  (.4,-3.7);
\draw[ultra thick](.2,-3.5)to  (.2,-3.7);
\draw[ultra thick](0,-3.5)to  (0,-3.7);
\draw[ultra thick](-.2,-3.5)to  (-.2,-3.7);
\draw[ultra thick](-.4,-3.5)to  (-.4,-3.7);
\end{scope}
\end{scope}
\end{scope}

\draw[<-,very
          thick, bend right,color=blu] (s2) to
        node[above,color=black]{Instructions}(s1);

        \draw[->,very thick, bend left,color=re](s4) to
        node[below]{\begin{tabular}{l}Feedback\end{tabular}}(s3);

        \end{tikzpicture}
      }
    \end{center}
\vspace{-0.3cm}
\caption{Scheme of the hybrid model}
\label{qram}
\end{wrapfigure}
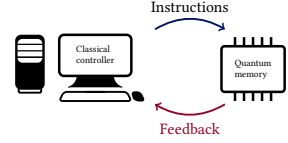
This has stirred a shift from a theoretical
standpoint on quantum algorithms to a more programming-oriented view
with the
question of their concrete coding and implementation
\cite{valiron2015programming,svore2016quantum,qecosystem}.

\textit{In this context, a particularly important problem is the adequacy
between the mathematical description of an algorithm and its concrete
implementation as a program. 
}

\subsection{The hybrid model}
The vast majority of 
quantum algorithms are 
described within the \emph{quantum
  co-processor model}~\cite{knill1996conventions}, i.e. an hybrid
model where a {\it classical} computer controls a {\it quantum}
co-processor holding a quantum memory (cf.~Figure~\ref{qram}).  The
co-processor is able to apply a fixed set of elementary operations
(buffered as {\it quantum circuits}) to
update and query ({\it measure}) the quantum memory. Importantly,
while { measurement} allows to retrieve classical (probabilistic)
information from the quantum memory, it also modifies it ({\it
  destructive effect}).  The state of the quantum memory is
represented by a linear combination of possible concrete values --- generalizing the classical notion of probabilities to the complex case,     
and  the core of a quantum
algorithm consists in successfully setting the memory in a specific
\emph{quantum state}.

\begin{wrapfigure}{l}{0.64\textwidth}
  \begin{center}
    \framebox{\begin{tikzpicture}
        \node{
          \scalebox{0.7}{\begin{minipage}{0.71\textwidth}
              
              \begin{tabular}{l}
                \textbf{Inputs:} (1) A black-box $U_{x,n}$ which performs the transformation\\
                $\ket{j}{}\ket{k}{}\rightarrow\ket{j}{}\ket{x^j k\ \text{mod } N}{}$, for $x$ co-prime to the $L-$bit number $N$,\\ (2)
                $t=2L+1 + \big\lceil \text{log}(2 + \frac{1}{2\epsilon})\big\rceil$ qubits initialized to $\ket{0}{}$, and (3) $L$ qubits\\ initialized to the state $\ket{1}{}$.
                \\[1ex]
                \textbf{Outputs:} The least integer $r>0$ such that $x^r = 1$ (mod $N$).
                \\[1ex]
                \textbf{Runtime:} $O(L^3)$ operations. Succeeds with probability $O(1)$.
                \\[1ex]
                \textbf{Procedure:}
              \end{tabular}
              \begin{alignat*}{100}
                &1. &\quad& \ket{0}{}\ket{u}{}
                && \text{\hspace{-5ex}initial state}
                \\
                &2.&& \to{} \frac1{\sqrt{2^t}}\sum_{j=0}^{2^t-1}\ket{j}{}\ket{1}{}
                && \text{\hspace{-5ex}create superposition}
                \\
                &3. && \to{}
                \frac1{\sqrt{2^t}}\sum_{j=0}^{2^t-1}\ket{j}{}\ket{x^j \text{mod } N}{}
                && \text{apply $U_{x, N}$}
                \\
                & && \approx{}
                \frac1{\sqrt{r2^t}}\sum_{s=0}^{r-1}\sum_{j=0}^{2^t-1}e^{2\pi i s j
                  /r}\ket{j}{}\ket{u_s}{}
                \\
                &4. && \to \frac{1}{\sqrt{r}}\sum_{s=0}^{r-1}\widetilde{\ket{s/r}{}}\ket{u_s}{}
                && \begin{tabular}{@{}l}\text{\hspace{-5ex}apply inverse Fourier transform}\\\text{\hspace{-5ex}to the first register}\end{tabular}
                \\
                &5. && \to{} \widetilde{\ket{s/r}{}}&& \text{\hspace{-5ex}measure first register}\\
                &6. &&\to{}  r&&\hspace{-5ex} \begin{tabular}{@{}l}apply continued fractions \\algorithm\end{tabular}
              \end{alignat*}
          \end{minipage}}
        };      \node[draw, rectangle, color=blu,thick]at
        (-1.9,1.15){\phantom{\tiny le petit chat est vivant le petit}};
        \node[draw, rectangle, color=re,thick,minimum height=.9cm]at
        (-1.9,0.45){\phantom{\tiny le petit chat est vivant le petit}};
    \end{tikzpicture}}
  \end{center}
  \caption{Bird eye view of \citet{shor1994algorithms}'s factoring algorithm \\
    (as presented in \cite[p.\,232]{nielsen2002quantum})}
  \label{ncpe}
\end{wrapfigure}

Major {\it quantum programming languages} such as
Quipper \cite{green2013quipper},
Liqui$|\rangle$ \cite{wecker2014liqui},
Q\# \cite{svore2018q},
ProjectQ \cite{projectq},
Silq \cite{silq},
and the rich
ecosystem of existing quantum programming frameworks \cite{qecosystem} 
follow  this hybrid model and 
provide dedicated features for interacting with the quantum memory and
well-suited for implementing quantum algorithms. They usually embed
these features within a standard classical programming language, with
forth (send quantum instructions) and backs (get measurement results)
between the classical control loop (classical computer) and the quantum
part (co-processor). 
Such {\it circuit-building quantum languages}  
are the current consensus for high-level executable quantum programming languages.

\subsection{The problem with quantum algorithms}
The core of a quantum algorithm ---the interaction with the quantum
co-processor--- is
usually provided in the form of Figure~\ref{ncpe}. Starting from an
initial state, the algorithm describes a series of high-level
operations which, once composed, realize the desired state. This
state is then usually measured to retrieve a classical information. Each
high-level operation may itself be described in a similar way, until
one reaches elementary operations. The benefit for using the quantum
algorithm instead of a classical counterpart lies in the fact that the
number of elementary operations is small.
The description of the algorithm is
therefore both the specification --- the global memory-state
transformation --- and the way to realize it --- the list of
elementary operations, or \emph{quantum circuit}.

{\it A major issue is then to verify that the circuit generated by the
  code written as an implementation of a given algorithm is indeed a
  run of this algorithm, and that the circuit has indeed the specified
  size.} 


%




\subsection{The case for quantum formal verification} 
While testing and debugging are the common verification practice in
classical programming, they become extremely complicated in the quantum
case.
Indeed, debugging and assertion checking are essentially impossible
due to the destructive aspect of quantum measurement. Moreover, the probabilistic
nature of quantum algorithms seriously impedes system-level quantum
testing. Finally, classical emulation of quantum algorithms is
(strongly believed to be) intractable.

On the other hand,   
nothing prevents {\it a priori} the formal verification of quantum programs. 
Formal methods and formal verification \cite{clarkew96formal} design a
wide range of techniques aiming at proving the correctness of a system
with absolute, mathematical guarantee --- reasoning over all possible
inputs and paths of the system, with methods drawn from logic,
automated reasoning and program analysis. The last two decades have
seen an extraordinary blooming of the field, with significant
case-studies ranging from pure mathematics~\cite{gonthier2008formal}
to complete software architectures~\cite{leroy2012compcert,sel4} and
industrial systems~\cite{cuoq2012frama,behm1999meteor}.
In addition to
offering an alternative to testing, formal verification has in principle the
decisive additional advantages to both enable  
parametric proof certificates and offer once-for-all absolute
guarantees for the correctness of programs.

%
%

We will focus on {\it deductive
  verification} \cite{hoare1969axiomatic,
  DBLP:journals/sttt/Filliatre11, DBLP:journals/cacm/BarnettFLMSV11,
  Filliatre:2007:WPD:1770351.1770379}, that, we argue, is well suited
for quantum formal verification (cf.~Section~\ref{overview}).

\subsection{Goal and challenges}  {\it Our goal is to provide a formal development   
  framework for circuit-building  quantum programs, including specification, programming and verification.}  Such a framework
should satisfy the following principles.

\smallskip
\noindent
\textit{Close to algorithmic description.} It should enable to
  specify and code algorithms in a way that directly matches their
  usual description from the literature, in order both to lower
  implementation \& certification time and to increase confidence in
  the specification.

\smallskip
\noindent
\textit{Separation of concerns.} It should enable a clear
  distinction between  code and  specification, in order to
  decouple implementation from certification --- in particular,
  specification should be optional and it should be possible to add it
  to a program at a later stage.


\smallskip
\noindent
\textit{Parametricity.}  It should allow parametric (i.e.~scale-invariant)
  specifications and proofs, so as to enable the generic specification and verification 
  of parametrized algorithms. This is crucial as quantum algorithms
  always describe {\it parametrized families} of circuits.

\smallskip
\noindent
\textit{Proof automation.} It should, as far as possible, provide
  automatic proof means.  Indeed, program verification 
  should be as painless as possible to the programmer in order to be
  adopted.


  



  

  \smallskip
  These requirements raise several challenges from the formal
verification point of view. Indeed, while questions about semantics,
properties, specification and efficient verification algorithms have
been largely investigated in the standard case, everything remains to
be done in the quantum case.
For example:
\begin{description}
\item[{\bf Non-standard data}:] Quantum algorithms rely heavily on {\it
  amplitudes} (generalization of probabilities to arbitrary {\it
  complex numbers}), not studied at all in standard verification;

\item[{\bf Second-order reasoning}:]
  We are interested here in {\it
  parametrized circuit-building programs}, i.e. programs that 
  describe {\it families} of circuits meeting some postcondition -- this is somehow akin to 
  addressing  dynamic code (i.e., jit)  in the classical case.

  %

\end{description}

\noindent The major scientific questions at stake here are: 
(1) How to specify quantum programs in a { natural way}? 
(2) How to support efficient proof automation for quantum programs?

{\it As a matter of fact, prior works on quantum  formal
  verification do not fully reach these goals.}

\begin{figure}
  {\scalebox{.7}{
    \begin{minipage}{0.65\linewidth}
    \centering
  {\scalebox{.9}{\begin{tabular}{@{}|l|ccccc|c|c|c|@{}}
                  \cline{2-7}
    \multicolumn{1}{c|}{}
    &
    1.&2.&3.&4.&5.&\qbrick
    \\
\cline{1-7}
    $\bullet$ Circuit-building language& \xmark &\xmark&\checkm &\checkm &\checkm &\checkm\\
    $\bullet$ Scale invariance (parametric)&\checkm &\xmark&\checkm&\checkm {\mycomment{\tiny\begin{tabular}{c}coin\\tossing \end{tabular}}}  &\xmark&\checkm\\
    $\bullet$ Proof automation& \begin{tikzpicture}\draw[->,ultra thick, color = orange](0,0)to(.3,.3);\end{tikzpicture}  &\checkm&\xmark&\xmark &\checkm& \begin{tikzpicture}\draw[->,ultra thick, color = orange](0,0)to(.3,.3);\end{tikzpicture}\\
\cline{1-7}
    $\bullet$ Separate specification from code&\checkm &\checkm&\xmark&\checkm &\xmark &\checkm\\
    $\bullet$ Specifications fitting algorithm&\xmark &\xmark&\xmark&\xmark&\checkm&\checkm\\
\cline{1-7}
  \end{tabular}
}}
  \captionof{table}{Formal verification of quantum programs }
  \label{soa}
\end{minipage}}}%
{\scalebox{0.7}{\begin{minipage}{0.6\textwidth}
  \centering
  \begin{tikzpicture}[yscale =.2,xscale =.7]
\draw[->,very thick](0,0)to (10,0);
\draw[very thick](0,0)to (0,2.9);
\draw[very thick](0,3.1)to (0,5.9);
\draw[very thick](0,6.1)to (0,8.9);
\draw[->,dashed,very thick](0,8.9)to (0,12);

\draw[very thick](0,14)to (10,14);
\node at (1,16){\textit{Size (number of qbits)}};
\node at (9.3,0.9){\textit{Difficulty}};
\node at (-.6,3){$10$};
\node at (-.6,6){$100$};
\node at (-.6,9){$1 000$};
\node at (-.6,14){$\infty$};

\draw[very thick](.5,.3)to(.5,-.3);
\draw[very thick](4.5,.3)to(4.5,-.3);
\draw[very thick](2.75,.3)to(2.75,-.3);
\draw[very thick](5.7,.3)to(5.7,-.3);
\draw[very thick](7,.3)to(7,-.3);
\draw[very thick](9.5,.3)to(9.5,-.3);
\node at (.5,-2.5){\footnotesize{\begin{tabular}{c}Superposition\\ coin flip\\teleportation\end{tabular}}};
\node at (2.75,-2.5){\footnotesize{\begin{tabular}{c}Deutsch\\ Jozsa\end{tabular}}};
\node at (4.5,-1.5){\color{re}\footnotesize{QFT}};
\node at (5.7,-1.5){\color{re}\footnotesize{Grover}};
\node at (7,-1.5){\color{re}\footnotesize{QPE}};
\node(cp) at (9.6,-1.5){\color{re}\footnotesize{Shor-OF}$\qquad\qquad$};


\node at (1,1){$\times$};
\node[] at (1,2){3.};

\node at (2.8,1){$\times$};
\node[] at (2.8,2){2.};

\node at (1,14){$\times$};
\node[] at (1,12.5){4.};

\node at (2.75,14){$\times$};
\node[] at (2.75,12.5){4.};

\node at (3.2,3.5){$\times$};
\node[] at (3.2,4.5){4.};

\node at (5,6){$\times$};
\node[] at (5,7){5.};

\node at (6.75,16){\color{re}{\textsc{This article}}};
\node at (4.5,14){\color{re}{\Huge{$\otimes$}}};
\node at (5.7,14){\color{re}{\Huge{$\otimes$}}};
\node at (7,14){\color{re}{\Huge{$\otimes$}}};
\node at (9,14){\color{re}{\Huge{$\otimes$}}};
  \end{tikzpicture}\vspace{-2ex}
  \caption{Certified quantum circuits from the literature }
  \label{stateofart}
\end{minipage}}}
{\footnotesize  \begin{tabular}{lll}
1.~ QHL~\cite{10.1007/978-3-030-25543-5_12} & 2.~
                                              QMC~\cite{10.1007/978-3-540-70545-1_51,Ying:2014:MLP:2648783.2629680}
                  \\ 3. ~\cite{DBLP:journals/corr/BoenderKN15}&
                                                                4.~Qwire~\cite{paykin2017qwire,DBLP:journals/corr/abs-1803-00699,dblp:journals/corr/abs-1904-06319}\\5.~Path-sums~\cite{amy2018towards,dblp:phd/basesearch/amy19}
                                            & We omit \cite{10.1007/978-3-030-25543-5_12} as their high-level formalism is not circuit-based

\end{tabular}
}
\end{figure}

\subsection{Prior attempts}
We summarize in Table~\ref{soa} the state of the art against the
requirements laid above for proving properties of quantum programs,
while Figure~\ref{stateofart} plots the success of the existing
methods against scale-sensitivity and algorithm difficulty.
%
%
%
Model-checking
approaches~\cite{10.1007/978-3-540-70545-1_51,Ying:2014:MLP:2648783.2629680}
are fully automatic but highly scale-sensitive. 
%
%
%
Other methods are based on interactive proofs through proof
  assistants, for instance Coq. One can cite the approach of
\citet{DBLP:journals/corr/BoenderKN15} or
Qwire~\cite{paykin2017qwire,DBLP:journals/corr/abs-1803-00699}. These however lack 
automation and deeply mix code and
specification.  Moreover, the underlying matrix semantics impedes 
scalability of the approach ---  so far it   
came only with small case-studies: coin\_flip~\cite{randthesis},
teleportation~\cite{DBLP:journals/corr/BoenderKN15}, and (recently) Deutsch-Josza\footnote{
   Deutsch-Josza consists in deciding, given an integer function that is known to be either constant or balanced, which case it is.  It has no 
practical application 
and only serves as an easy-to-understand illustration for quantum speedup in introductory courses. We see it as a toy example --- especially  compared to Grover, QPE or Shor-OF.
} 
algorithm~\cite{dblp:journals/corr/abs-1904-06319}.
%
%
%
  Recently, \citet{dblp:phd/basesearch/amy19,amy2018towards} developed a powerful framework for reasoning over quantum circuits, 
the path-sums symbolic representation.  
Thanks to their good closure properties, reasoning with path-sums is well automated and can scale up 
to large problem instances (up to 100 qubits). Yet, the method is not parametric and only addresses fixed-size circuits. 

Another explored direction tackles the formalization of {\it quantum
  programs with classical control}, where the input/outputs from the
quantum co-processor are taken as oracles.
Initiated by \cite{DBLP:journals/toplas/Ying11}, this line of work does not consider
circuit-building languages but rather high-level algorithmic models
where quantum circuits are directly given as partial density operators
of arbitrary size rather than built from elementary components ---
somehow, this is akin to verifying an algorithm expressed over sets
versus verifying an implementation of the said algorithm working over
red-black trees.
It uses an extension of Hoare Logic, called Quantum Hoare
Logic and designed for the
specification of quantum programs as mathematical objects.
Recently, the authors focused on the
definition, automatic generation~\cite{DBLP:conf/popl/YingYW17} and
proof
support~\cite{DBLP:journals/corr/LiuLWYZ16,DBLP:journals/fac/Ying19}
for loop invariants in quantum programs
invariants. In~\cite{10.1007/978-3-030-25543-5_12} the authors present
a formalization of QHL in Isabelle/HOL and illustrate it on  a
restricted case of Grover algorithm.

\subsection{Proposal and contributions}
We propose \qbrick, the first formal verification environment for  
circuit-building quantum programs,  featuring clear separation between code and proof,
parametric specification and proof, high degree of proof
automation and allowing to encode quantum programs in a natural way,
i.e.~close to textbook style. 
\qbrick builds on best practice from  formal verification for the classical
case (separation of concerns, flexible logical specification language, proof
automation, domain-based specialization) and tailors them to the
quantum case.

More precisely, while relying on the general framework of deductive verification, 
we tailor it to the specific needs of quantum, bringing several key innovations along the road. 
 \qbrick builds upon the circuit-building, first-order functional language \qbrickDSL   together with the
logical specification language \qbrickSPEC. 
While minimal, \qbrickDSL can express implementations of existing non-trivial quantum algorithms
%
and 
 \qbrickSPEC is expressive
 enough to offer \emph{higher-order}
 specification (parametrized circuit production).  
The two {\it key cornerstones} behind \qbrick   
are:  (1) a new Hoare-style proof system, called \emph{Hybrid Quantum
  Hoare Logic (\hqhl)} dedicated to circuit-building quantum languages, 
and (2)  the new  \emph{higher-order
  path-sums (HOPS)} symbolic representation of quantum states, extending   
 path-sums \citet{dblp:phd/basesearch/amy19} to the parametric case  
while keeping good closure properties --- HOPS prove extremely useful both as a specification mechanism 
and as an automation mechanism.    

In the end, we bring the following contributions.

\smallskip
\noindent
\textbf{\textit{Framework.}} A programming and verification framework, that is:
  on one hand, a core domain-specific language (\qbrickDSL, Section \ref{sec:qdsl}) for
  describing families of quantum circuits,
  with enough expressive power to describe parametric circuits from
  non-trivial quantum algorithm;
  %
  on
  the other hand, a logical, domain-specific,
  specification language (\qbrickSPEC, Section \ref{sec:qspec}), tightly integrated with
  \qbrickDSL to specify properties of 
  parametrized programs representing families of quantum circuits.

\smallskip
\noindent
\textbf{\textit{Higher-Order Path-Sums.}} A flexible symbolic representation for reasoning about quantum states, integrated
  withing \qbrickSPEC and building upon the recent path-sum 
  symbolic representation~\cite{dblp:phd/basesearch/amy19,amy2018towards}.  Our
  representation, called \emph{higher-order path-sums (HOPS)}, retains the
  compositional and closure properties of regular path-sums while
  allowing {\it genericity} and {\it parametricity} of both
  specifications and proofs: HOPS expressions not only contain regular
  path-sum constructs but also terms from \qbrickDSL. Especially, 
  HOPS provides a unified and powerful way to reason about many 
  essential quantum concepts (Section \ref{sec:qspec}).

\smallskip
\noindent
\textbf{\textit{Dedicated Proof Engine.}} We introduce  the Hybrid Quantum Hoare Logic (\hqhl) deduction system for 
deductive verification over circuit-building quantum programs -- tightly coupled with HOPS and producing proof obligations 
in the \qbrickSPEC\ logic, together with 
deduction abilities dedicated to standard quantum structures  (Section \ref{sec:deduction}).

\smallskip
\noindent
\textbf{\textit{Implementation.}} This framework is embedded in the Why3
  deductive verification tool~\cite{bobot:hal-00790310,
    Filliatre:2007:WPD:1770351.1770379} as a DSL, and provides  proof automation
  mechanisms dedicated to the quantum case --- this material is
  grounded in standard mathematics theories ---linear algebra,
  arithmetic, complex numbers, binary operations, {\it etc}.--- with 450+
  definitions and 1,000+ lemmas (Table~\ref{gen-data}). The Why3
  embedding comes with a series of semantic shortcuts designed to
  increase the overall level of proof automation, based on high-level
  composition rules and circuit subclasses with simple HOPS semantics
  (Section~\ref{sec:implem}).

\smallskip
\noindent
\textit{\textbf{Case studies.}} We present in Section \ref{sec:xp}  verified parametric  implementations
of the quantum part of \cite{shor1994algorithms}'s factoring algorithm
(Order Finding --Shor-OF),
Quantum Phase Estimation (QPE)~\cite{kitaev1995quantum,cleve1998quantum}\footnote{QPE is a major quantum building block,  at the heart of, e.g.,
   HHL~\cite{harrow2009quantum} logarithmic linear system solving
   algorithm or quantum simulation~\cite{georgescu2014quantum}.  }, Grover algorithm ~\cite{DBLP:conf/stoc/Grover96} (search) and 
Quantum Fourier Transform (QFT) -- all of them being much more complex than previous formally verified circuit-building implementations.        
Comparison with alternative approaches demonstrate a clear gain in proof effort (on Grover, factor 5 vs.~ the restricted proof on a high-level algorithm model~\cite{10.1007/978-3-030-25543-5_12}), 
while our method achieves a high level of proof automation (95\% on Shor-OF).   
Moreover, we are also able for Shor-OF to prove the polynomial complexity of the circuits produced by our implementation.  
%


\subsection{Discussion}
The scope of this paper is limited to proving
properties of circuit-building quantum programs. We do
not claim to support right now the interactions between classical data
and quantum data (referred to as ``classical control'' in the
literature), nor the probabilistic side-effect resulting from the
measurement. Still, we are already able to target realistic
implementations of famous quantum algorithms, and thanks to
equational theories for complex and real number we can \emph{reason} on
the probabilistic outcome of a measurement.

Also, we do not claim any novelty in the proofs for Shor-OF, QPE or Grover by themselves, but rather 
the first parametric
correctness proofs of the circuits produced by  programs implementing Shor-OF, QPE or Grover. 
%
%

%

That said, {\it we present the first non trivial, automated, parametric
  proofs of  significant circuit-building quantum  programs, where prior works were
  limited to toy examples, 
establishing  a new baseline for quantum verification of realistic programs. 
}  
%



\section{Background: Quantum Algorithms and Programs}

While in classical computing, 
the state of a bit is either $0$ or $1$, 
in quantum computing \cite{nielsen2002quantum} the state of a \emph{quantum
  bit} (or \emph{qubit}) is described by {\it amplitudes}
over  the two elementary values $0$ and $1$  (denoted in the Dirac notation with
$\ket{0}{}$ and $\ket{1}{}$),  i.e.~linear combinations   
\(\alpha_0 \ket{0}{} + \alpha_1\ket{1}{}\) where $\alpha_0$
and $\alpha_1$ are any {\it complex values} satisfying 
$\vert \alpha_0\vert^2 + \vert \alpha_1\vert^2 = 1$. In a sense,
amplitudes are generalization of probabilities.   

\smallskip

More generally, the state of a {\it qubit register} of $n$ qubits (``qubit-vector'')  is any {\it superposition} of
the $2^n$ elementary bit-vectors (``basis element'', where a bit-vector $k \in \{0 .. 2^n-1\}$ is denoted $\ket{k}{n}$), that is any
$\ket{u}{n}= \sum_{k = 0}^{2^n-1} \alpha_k\ket{k}{n}$ such that
$\sum_{k = 0}^{2^n-1} \vert \alpha_k \vert^2 = 1 $. 
For example, in the case
of two qubits, the basis is 
$\ket{00}{}$, $\ket{01}{}$, $\ket{10}{}$ and $\ket{11}{}$ (also abbreviated $\ket02$, $\ket12$, $\ket22$ and $\ket32$).
Such a (quantum state) vector  $\ket{k}{n}$  is
called a \emph{ket} of length  $n$ (and  dimension $2^n$).    

\smallskip

Technically speaking, we say that the quantum state of a register of $n$ qubits is represented by a normalized vector 
in a   Hilbert space of finite dimension $2^n$ (a.k.a.~finite-dimensional Hilbert space), whose basis is generated 
by the Kronecker product (a.k.a.~tensor product, denoted $\otimes$)
over the elementary bit-vectors. For instance, for $n=2$: $\ket0{}\otimes\ket0{}$, $\ket0{}\otimes\ket1{}$,
$\ket1{}\otimes\ket0{}$ and $\ket1{}\otimes\ket1{}$ act  as definitions  for 
 $\ket{00}{}$, $\ket{01}{}$, $\ket{10}{}$ and
$\ket{11}{}$.

\subsection{Quantum Data Manipulation}
\label{quantum-data}
The core of a quantum algorithm consists in  manipulating  a
qubit register through two main classes of operations.
(1) \textit{Quantum gate.}  Local
operation on a fixed number of qubits, whose action consists in the
application of a {\it unitary map} to  the corresponding quantum state vector
i.e.~a linear and  bijective operation preserving  norm and  orthogonality.
The fact that unitary maps are bijective ensures  that every unitary gate admits an 
\emph{inverse}. 
Unitary maps over $n$ qubits are usually represented as $2^n \times 2^n$ \emph{matrices}.  
%
~~(2) \textit{Measurement}. The retrieval of classical information out of the
quantum memory. This operation is probabilistic and modifies the global
state of the system: measuring the $n$-qubit system
$\sum_{k = 0}^{2^n-1} \alpha_k\ket{k}{n}$ returns the bit-vector $k$ of
length $n$ with probability $\vert \alpha_k \vert^2$.

Quantum gates might be applied in {\it sequence} or in {\it parallel}: sequence
application corresponds to {\it map  composition}  (or, equivalently, 
matrix multiplication), while parallel  application  
corresponds to the {\it Kronecker product}, or tensor product, 
of the original maps --- or,
equivalently, the Kronecker product of their matrix representations.\footnote{Given two matrices $A$ and $B$ with respectively $r(A)$ and $r(B)$ rows and  $c(A)$ and $c(B)$ columns, their Kronecker product is the matrix $A\otimes B = \begin{pmatrix} a_{11}B & & \dots a_{c(A)} B\\  \rotatebox{90}{\dots}& \rotatebox{135}{\dots}& \rotatebox{90}{\dots}
\\ a_{r(A)1}B & & \dots a_{r(A)c(A)} B\
  \end{pmatrix}$. This operation is central in quantum information representation. It enjoys a number of useful algebraic properties such as associativity, bilinearity or the equality $(A\otimes B)\cdot(C\otimes D) = (A\cdot C)\otimes (B\cdot D)$ for any matrices $A,B,C,D$ of adequate sizes -- where $\cdot$ denotes matrix multiplication.}

\subsection{Quantum Circuits}

In a way similar  to classical Boolean functions, the application of
quantum gates can be written in a diagrammatic notation: {\it quantum
circuits}. Qubits are represented with horizontal wires and gates with
boxes.  Circuits are built {\it compositionally}, from a given set of {\it atomic
gates} and by a small set of {\it circuit combinators}, including: parallel 
and sequential compositions, circuit inversing,
controlling, iteration, etc.

\begin{wrapfigure}[13]{r}{0.54\textwidth}
  \begin{center}
    \scalebox{.6}{
      \begin{tikzpicture}[xscale =.25,yscale =.7,decoration={brace}][scale=2] 
        \input{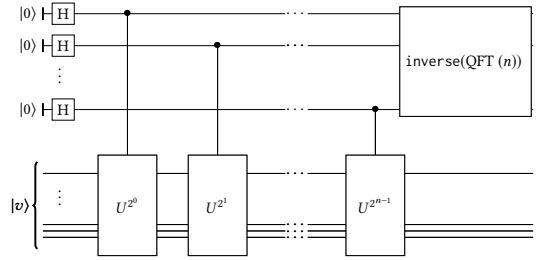}
      \end{tikzpicture}
    }
    \caption{The circuit for QPE}
    \label{phase-struct-simple}
  \end{center}
\end{wrapfigure}

As an example of a quantum circuit, we show in
Figure~\ref{phase-struct-simple} the bird-eye view of the
circuit of QPE, the (quantum) phase estimation algorithm, a standard primitive in many quantum algorithms. 
QPE is parametrized by $n$ (a number of wires) and
$U$ (a unitary --- the \emph{oracle}) and is built as follows.
First, a  register of $n$ qubits is initialized in state
$\ket{0}{}$, while another one is initialized in state $\ket{v}{n}$.
Then comes the circuit itself: a structured sequence of quantum
gates,  using the  unary Hadamard gate $H$ and the circuits
$U^{2^i}$ (realizing $U$ to the power $2^i$) 
and $\reverse{\qft(n)}$ -- the reversed Quantum Fourier Transform.
Both are defined as sub-circuits in a similar
way. 

For the purpose of the current discussion, one should simply note
two things: 
(1) the circuit is made of parallel compositions of
Hadamard gates and of sequential compositions of controlled $U^{2^i}$
(the controlled operation is depicted with vertical lines and 
symbol $\bullet$); ~~~(2) the circuit is \emph{parametrized} by $n$
and by $U$. This is very common: in general, a quantum algorithm
constructs a circuit whose size and shape depend on the parameters of 
the problem. It describes a \emph{family of quantum
  circuits}.

\subsection{Quantum algorithms}
\label{sec:q-algs}

Quantum algorithms intend to solve classical problems in a
probabilistic way, but with better performance than with classical algorithms. 
They generally  consist  in the generation of a quantum circuit
based on the problem parameters (e.g.~the size of the
instance), usually followed by iterating the following three
steps: (1) memory initialization, (2) run of the quantum circuit, (3) measure of
the memory to retrieve a classical piece of data.

The quantum circuit is seen as a predictive tool that
probabilistically gives some (classical) information from which one
can infer the targeted result. The fact that the probability is high
enough is a direct consequence of the mathematical properties of the
unitary map described by the quantum circuit. The essence of the
quantum algorithm ---and the reason for its efficiency--- consists in
describing an efficient circuit realizing this unitary map.

\smallskip
\textit{Obtaining guarantees on the families of circuits realized by
  quantum programs is therefore of uttermost importance.}

\smallskip 
We use the following  famous quantum algorithms throughout the article: 

\begin{itemize}
\item Phase Estimation algorithm
(QPE)~\cite{kitaev1995quantum,cleve1998quantum}: takes as parameter a
matrix $U$, an eigenvector $\ket{v}{}$ and answers the eigenvalue of
$U$ corresponding to $\ket{v}{}$. While the algorithm answers in a  deterministic way  
for some sub-cases, in general it only answers  up to some
probability.

\item  \citet{DBLP:conf/stoc/Grover96}'s
search algorithm: given a sparse
non-zero function $f : \{0\ldots2^n-1\}\to\{0,1\}$, Grover's algorithm
outputs one value $x$ such that $f(x)=1$ with a probability
high-enough to beat brute-force search.

\item \citet{shor1994algorithms}'s factoring algorithm: 
  We consider the  quantum part of the algorithm: the Order Finding
  algorithm. \emph{We shall refer to it as Shor-OF}.
  Concretely, given a non prime integer $N$ and a co-prime $x$, it outputs a
  period for x modulo $N$, in time \textsc{PolyLog}$(N)$ and probability of success $O(1)$. It
  is based on an application of QPE with a modular multiplication
  operator as oracle.

\end{itemize}


\subsection{Existing quantum programming languages: the circuit-building paradigm}

The standard model for quantum computation is the \emph{quantum
  co-processor model}: the quantum memory is stored in a dedicated
co-processor that the classical computer accesses through a dedicated
interface. This distinction between purely classical computation and
manipulation of quantum data is well-suited for most of the existing
quantum algorithms. Indeed, they usually consist in two  main parts:
a classical part, for dealing with classical information and
circuit-building, and a quantum part, where the circuit is sent to the
quantum co-processor.

Quantum programming languages follow this pattern: they usually come
up as domain-specific languages embedded in a classical computing
framework: The interaction with the quantum co-processor is handled
through an interface allowing for qubit initializations, measurements
and elementary quantum gates. Elementary
quantum gates are usually buffered into circuits before being sent to the
co-processor. Each language then has its own design-choices on how to
present the circuit to the programmer and which native
circuit-manipulation library to offer.

The bottom line is that, regardless of the language-design choices,
the core of a quantum programming language is to offer
circuit-building and circuit manipulation operators in order to build as
naturally as possible the circuits given in the descriptions of quantum
algorithms from the literature.

\subsection{Path-Sum Representation}
\label{sec:ps}

Historically the semantics for quantum circuits has been given in term
of unitary matrices~\cite{nielsen2002quantum} acting on Hilbert
spaces, that is, the canonical mathematical formalism for quantum
computation. If this semantics is well-adapted for representing simple
high-level circuit combinators such as the action of control or
inversion, it is cumbersome for specifying the semantics of general
circuits.


Path sums~\cite{dblp:phd/basesearch/amy19,amy2018towards} is a  recent symbolic representation that  has been
shown successful for proving {\it equivalence} of general
quantum circuits. Its
strength is to formalize the notation used in
e.g. Figure~\ref{ncpe}. The path-sum of a unitary matrix $U$ is then
written as
\( U : \ket{x}{} \mapsto \textit{PS}(x) \)
where $x$ is a list of booleans. $\textit{PS}(x)$ is defined with the
syntax of Figure~\ref{tab:syntax-ps}. The $P_k(x)$ are called
\emph{phase polynomials} while the $\ket{\phi_k(x)}{}$ are
\emph{basis-kets}.

This representation is {\it closed} under functional composition and Kronecker
product. For instance, if $V$ sends $y$ to
$PS'(y) = \frac{1}{{\sqrt2}^{n'}} \sum_{k=0}^{2^{n'}-1} \text{exp}
\left( \frac{ 2\cdot\pi\cdot{i}\cdot{P'_k(y)} }{ 2^{m'} } \right)
\ket{\phi'_k(y)}{}$, then $U\otimes V$ sends
$\ket{x}{}\otimes\ket{y}{}$ to
\begin{equation}\label{eq:ps}
  \frac{1}{{\sqrt2}^{n+n'}}
  \sum_{j=0}^{2^{n + n'}-1}
  e^{
    \frac{2\cdot\pi\cdot{i}(2^{m'}\cdot{P_{j/2^n}(x)} +
      2^m\cdot{P'_{j\%2^n}(y)})}{2^{m+m'}}
  }
  \ket{\phi_{j/2^n}(x)}{}\otimes\ket{\phi'_{j\%2^n}(y)}{}
\end{equation}
that is in the form shown in Figure~\ref{tab:syntax-ps}. 
However, if it has been shown  successful to  prove the equivalence of
large circuit instances~\cite{amy2018towards}, 
its main limitation
stands in the fact that  path-sum only address fixed-size
circuits.
Albeit a useful, compositional tool,  it
 cannot be used for proving properties of parametrized circuit-building 
quantum programs.

{

  \floatstyle{boxed}
  \restylefloat{figure}

  \begin{figure}
    \begin{alignat*}{100}
      \textit{PS}(x) &~~{:}{:}{=}&~&
      \frac{1}{{\sqrt2}^{n}}\sum_{k=0}^{2^{n}-1}
      e^{\frac{2\cdot\pi\cdot{i}\cdot{P_k(x)}}{2^m}}\ket{\phi_k(x)}{}
      \\
      P_k(x) &~~{:}{:}{=} && x_i \mid n \mid P_1(x)\cdot P_2(x) \mid
      P_1(x) + P_2(x)
      \\
      \ket{\phi_k(x)}{} &~~{:}{:}{=} && \ket{b_1(x)}{}\otimes\ldots\otimes\ket{b_n(x)}{}
      \\
      b_i(x) &~~{:}{:}{=} && x_i \mid \neg b(x) \mid b_1(x)\wedge b_2(x)
      \mid b_1(x)\oplus b_2(x) \mid \texttt{true}\mid \texttt{false}
    \end{alignat*}
    \caption{Syntax for regular path-sums~\cite{amy2018towards,dblp:phd/basesearch/amy19}}
    \label{tab:syntax-ps}
  \end{figure}
  
  \floatstyle{plain}
  \restylefloat{figure}
  
}


\smallskip
\textit{This paper proposes an extension of path-sum semantics to
  address the parametric verification of general quantum programs.}


\section{Motivating example}
\label{motiv}

\noindent
Let us consider the $n$-indexed family of circuits consisting of $n$
Hadamard gates, in sequence, as shown in Figure~\ref{motivating-example}.
Sequencing two Hadamard gates can easily be shown
equivalent to the identity operation. In other word, when fed with
$\ket{0}{}$, if $n$ is even the circuit outputs $\ket{0}{}$.

Albeit small, this circuit family together with its simple
specification exemplifies the typical framework we aim at in the
context of certification of circuit-building quantum programs:
\begin{wrapfigure}[6]{r}{0.4\textwidth}
  \scalebox{.8}{\begin{tabular}{|l|}
                  \hline\\
    A circuit $C_n$ defined as
      \\[1ex]
      ~\quad$
          \underbrace{\xymatrix@C=3ex{
          \ar@{-}[r]
      & *++[F]{H}\ar@{-}[r]
      & *++[F]{H}\ar@{-}[r]
      & \cdots\ar@{-}[r]
      & *++[F]{H}\ar@{-}[r]
      &
        }}_{n\text{ gates}}
        $
      \\
      Precondition:\hspace{1ex} $n \geq 0$ is even.
      \\
      Post-conditions:
      $\left\{\text{\begin{tabular}{l}
                      $C_n$ sends
                      $\ket{x}{}$ to $\ket{x}{}$
                      \\
                      $C_n$ consists on $n$ gates.
                    \end{tabular}}\right.$
                  \\[3ex]
                  \hline
    \end{tabular}}
  \caption{Motivating Example}
  \label{motivating-example}
\end{wrapfigure}
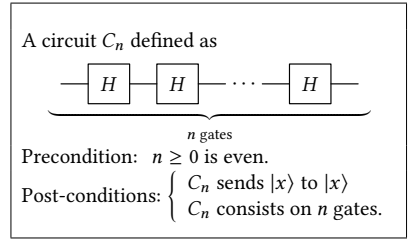
\begin{itemize}
\item The description of the circuit family is parametrized by a
  classical parameter (here, the non-negative integer $n$); 
 
\item The pre-condition imposes both constraints (here, the
  evenness of $n$) and soundness conditions (here, the
  non-negativeness of $n$) on the parameters;  

\item The post-condition can both refer to the semantics of the
  circuit result and to its form and shape (here, its size). 
\end{itemize}

\paragraph{Shortcomings of existing approaches}
In order to prove program specifications, the literature offers
several methods and tools, yet none of them fits our needs here.

(1) In the expressive Qwire
environment~\cite{paykin2017qwire,DBLP:journals/corr/abs-1803-00699}
embedded in Coq, the semantics of quantum programs is based on
matrices over complex numbers. If this toy motivating example would
be definable, the drawbacks of this approach is in general its lack of
scalability and automation. 
\,~~~(2) The quantum Hoare logic environment embedded in
Isabelle/HOL~\cite{10.1007/978-3-030-25543-5_12} offers a 
high-level  abstract view of quantum programs -- especially it does not offer circuits {\it per se}. It is thus not well-suited to detail the
fine-grained description of circuit-building implementations, in particular
the procedural description of a purely unitary circuit. Coding the toy
example in Figure~\ref{motivating-example} would be extremely ad-hoc and complicated 
as the framework does not natively offer elementary gates nor iteration over
classical integers.
\,~~~(3) The last possibility would be to use path-sums~\cite{amy2018towards}:
this framework is scalable and well-suited for sequence of gates, but
limited to \emph{fixed-size} circuits, while we target programs building families of circuits. 

It is to be noted that none of these approaches offer the possibility
to reason about the size of the produced circuit: either there is no such
notion of circuit~\cite{10.1007/978-3-030-25543-5_12}, or the circuit
is not an object one can reason on~\cite{amy2018towards,paykin2017qwire,DBLP:journals/corr/abs-1803-00699}.

\medskip
\noindent {\it By contrast, our own method allows to encode the Hadamard example in \qbrick\ with 
2 lines of codes and 11 lines of specification. Our deduction verification engine then produces 
18 proof obligations (ensuring the contract and circuit well-formedness) simple enough so that all of them   are automatically discharged 
by automated provers.  
}


\smallskip 

The circuit family presented in Figure~\ref{motivating-example} will
be used in the rest of the paper as a running, toy example for 
\qbricks. In particular, we show in Example~\ref{ex:motiv-dsl} how to
code it in our framework and how to express the specification in
Example~\ref{ex:motiv-hoare}.

\paragraph{Target for {\qbrick}} The goal of \qbrick\ is to offer a formal  verification framework for realistic circuit-building quantum programs, 
including parametric specification and proofs. We ultimately demonstrate this ability in Section \ref{sec:xp} on implementations of 
Grover, QPE and Shor-OF --     the experimental data are discussed in
Section~\ref{sec:xp}. 
%
%




\section{Our Proposal for Certified Quantum Programs}
\label{overview}

A common way to describe quantum algorithms in the literature is to
declare, or describe, either semi-formally or in natural language, a
sequence of unitary operations to be implemented, inter-crossed with a
sequence of formal assertions describing the evolution of the state of
the system along the performance of these functions. 
The example in Figure~\ref{ncpe} illustrates this case: it corresponds to
the exact description of the circuit for Shor's algorithm  at it is written
in~\cite[p.\,232]{nielsen2002quantum}.
The formal description of
state, left column in Figure~\ref{ncpe}, is interpretable as
specifications for the operations declared in the right column.  For
example, operation \emph{create superposition} is declared on the
right column, line 1. We interpret the formal expression of line 1,
left (framed in blue), as its precondition and the one of line 2, left
(framed in red), as its post-condition.



\smallskip 

{\it Our key observation is that standard quantum algorithm
  descriptions match perfectly with the process of {\it deductive
    program
    verification}~\cite{hoare1969axiomatic,DBLP:journals/sttt/Filliatre11},
  a well established formal method.
}

\subsection{Deductive verification for quantum programs}
\label{deduc} 

{\it Deductive program verification}~\cite{hoare1969axiomatic,
  DBLP:journals/sttt/Filliatre11, DBLP:journals/cacm/BarnettFLMSV11,
  Filliatre:2007:WPD:1770351.1770379} is probably the oldest formal
method technique, dating back to 1969~\cite{hoare1969axiomatic}.  In
this approach, programs are {\it annotated} with {\it logical
  assertions}, such as pre- and post-conditions for operations or loop
invariants, then so-called {\it proof obligations} are automatically
generated (e.g., by the weakest precondition algorithm) in such a way
that proving (a.k.a.~discharging) them ensures that the logical
assertions hold along any execution of the program.
These proof obligations are commonly proven by help of proof assistants
or automatic solvers.  

In more details, for any function $f$, annotations by precondition
$\mathit{pre}$ and postcondition $\mathit{post}$ are to be understood
as a contract for the implementation: the programmer commits to
ensuring that the output of $f$ satisfies $\mathit{post}$ for any
input satisfying $\mathit{pre}$, which, in essence, translates into
\[
\forall x. \mathit{pre}(x) \to \mathit{post}(f(x))\] 
Suppose that the function $f$ is defined, specified and verified. And
suppose one defines and specifies a further function $g$, using a call
$f(a)$ to $f$. This generates a new proof obligation
. To fulfill it, the already verified proof obligation
for $f$ 
is
assumed as an hypothesis.

\smallskip 

This compositional reasoning method is particularly well-suited to the compositional nature of circuits. 


\subsection{Rational for the design of {\qbrick}}
\label{sec:rational}

We adopt the methodology presented in Section~\ref{deduc} and adapt the key ingredients of deductive verification 
to the case of circuit-building quantum programs. Hence, 
\qbrick\  is equipped with:  a domain specific language
(DSL) for building circuit families, a dedicated logical specification language, 
a novel symbolic representation (HOPS) and a new Hoare-style logic
---called \emph{Hybrid Quantum Hoare Logic (HQHL)}.


\paragraph{Circuit representation}  
The language {\qbrickDSL} is only aimed at implementing quantum algorithms and
specifications provided in the form of Figure~\ref{ncpe}.  The circuits
used in quantum algorithms act in general on contiguous blocks of
memory registers and consist  of simple compositions and hierarchical
descriptions. Thus, unlike existing quantum programming languages such
as Quipper~\cite{green2013quipper} or Qwire~\cite{paykin2017qwire},
for \qbrick there is no need for complex wire manipulation.

Following this analysis, the low-level circuit-representation we
choose as a target {\qbrick} is akin to the one of qPCF~\cite{qpcf}: a
circuit is a simple compositional structure consisting of base gates
and circuit combinators such as sequential and parallel composition,
control, inversion, etc. These constructions are packaged withing a
domain-specific language (DSL) aimed at describing families of
circuits.

\paragraph{Symbolic representation of quantum circuits}
As mentioned in Section~\ref{sec:ps}, if path-sums offer a
compositional specification framework, they address the case of
fixed-size circuits. In particular, one cannot give specification to
general, parametrized programs describing families of circuits.

\qbrick proposes a solution to this limitation, by unifying what can
be done with the matrix and the path-sum semantics. Our proposal is a
\emph{higher-order path-sum semantics (HOPS)}. On one hand, we keep
the functional view on the action of circuits on quantum registers,
making it suitable for deductive verification. On the other hand, we
\emph{extend} its syntax to support parametric circuit construction,
high-level circuit combinators and reference to \qbrickDSL constructs.

\paragraph{Specification and verification} It turns out that HOPS offer a very convenient 
specification mechanism for quantum programs. They form the bedrock of 
the \qbrickDSL\ specification language -- together with a few other theories such as bit-vectors, integers or 
complex values. 

It also turns out that HOPS are convenient for verification purpose
too: we design a
proof-obligation generation procedure within HQHL dedicated to
circuit-building quantum languages, yielding  
proof obligations in the \qbrickDSL\ (first-order) logic. 

Finally, while part of the \qbrickDSL\ logic can be directly handled by automated solvers (such as SMT), 
we add dedicated equational theories and axioms for those parts of the logic (e.g., HOPS manipulation or complex values).

\paragraph{Implementation}  We implement \qbrick as a domain-specific
language embedded in the Why3 deductive verification
tool~\cite{bobot:hal-00790310, Filliatre:2007:WPD:1770351.1770379},
allowing to take advantage of its advanced programming, specification
and verification features for classical programs. We add all
quantum-related features on top of it. 

\smallskip 

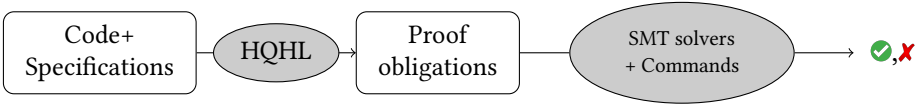
\begin{figure}
  \begin{center}
      \begin{tikzpicture}[xscale =.25,yscale =.7,decoration={brace}][scale=2] 

        \node(code)[draw,rectangle,rounded corners]at (-6,0){
\begin{tabular}{c}
Code+ \\
 Specifications\\
  \end{tabular}
        };
        \node(pos)[draw,rectangle,rounded corners]at (12,0){
\begin{tabular}{c}
        Proof\\ obligations\end{tabular}};
    \draw[->](code)to node[ellipse,draw,fill=gray!40]{\hqhl}(pos);
      \node(smt)[ellipse,draw,fill=gray!40] at (25,0){\footnotesize{\begin{tabular}{c}
            SMT solvers\\
             + Commands\\\end{tabular}}};
  \node(check) at (36,0){\Large{ \checkm,\xmark}};
      \draw[->](pos)to(smt)to(check);

      \end{tikzpicture}

    \caption{Overview of \qbrick certified implementation process}
    \label{process-overview}
  \end{center}
\end{figure}

An overview of \qbrick certified implementation process is depicted in
Figure~\ref{process-overview}: the user writes a \qbrickDSL programs 
together with \qbrickSPEC specifications. Our \hqhl\ engine
produces  proof obligations in \qbrickDSL\ logic.
Taking
advantage of the interface provided by Why3 embedding, the user can
either directly send these proof obligations to a series of
SMT-solvers for validation, together with
our circuits specifications  equational theories,
or facilitate solving by use of a series of interactive predicate
transformation commands (such as introduction of lemmas or hypotheses,
beta-reduction, etc).

\section{\qbrickDSL}
\label{sec:qdsl}


\qbrick is structured as a
domain-specific language (DSL), called \qbrickDSL, and a
domain-specific logical specification language called  \qbrickSPEC. 
The language is targeted for \emph{circuit description}: measurement
is out of the scope of the language, and all \qbrickDSL expressions
are terminating. We follow a very simple strategy for circuit
building: we use a regular inductive datatype for circuits, where the
data constructors are elementary gates, sequential and parallel
composition, and ancilla creation. In particular, unlike e.g. Quipper
or Qwire, a quantum circuit is not a function acting on qubits: it is
a simple, static object. Nonetheless, for the sake of implementing
quantum circuits from the literature this does not restrict
expressivity as they are usually precisely represented as sequences of
blocks.

Even if the language does not feature measurement, it is nonetheless
possible to \emph{reason} on probabilistic outputs of circuits, if we
were to measure the result of a circuit. Indeed, this can be expressed
in a regular theory of real and complex numbers.

\qbrickDSL is designed as a first-order, functional language: for the
purpose of circuit construction, this is enough and designing a
deductive system for it can then be done in a canonical way, following
e.g. Why3 strategy. Similarly, the specification language \qbrickSPEC
is a first-order predicate language, equipped with  various
equational theories. This makes proof-obligations more easily amenable
to automated solvers.

In the rest of this section, we present the language
\qbrickDSL. Section~\ref{sec:qspec} is then devoted to \qbrickSPEC
while Section~\ref{sec:deduction} to the deduction rules.

\subsection{Syntax of {\qbrickDSL}}
\label{syntax-qbdsl}
\begin{figure}
  \centering
  \includegraphics[scale=.7]{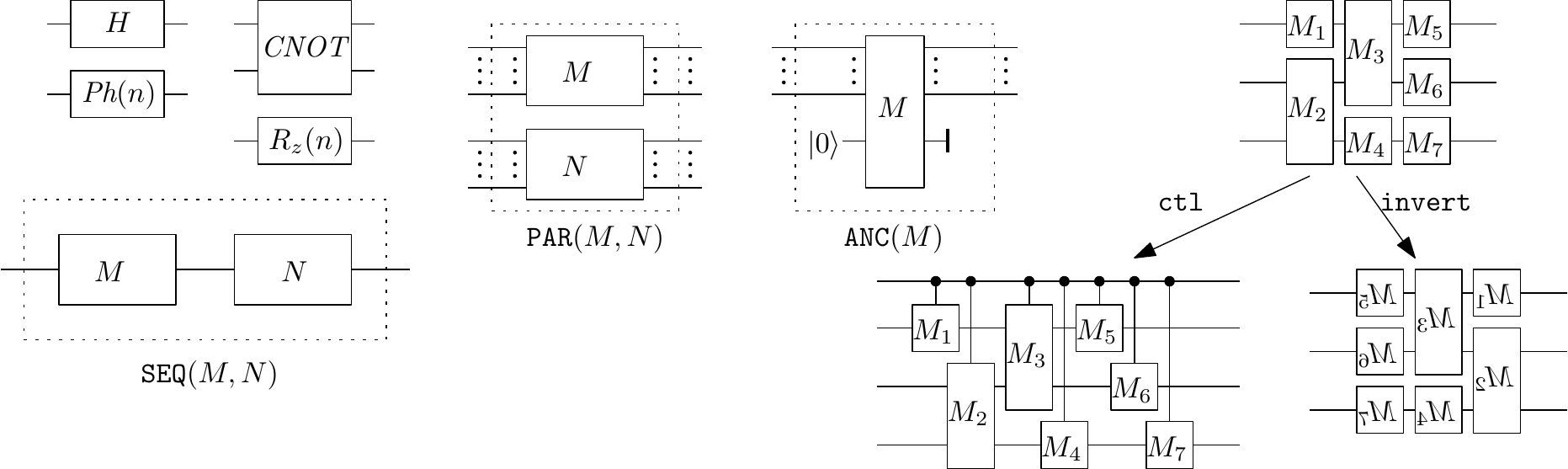}
  \caption{Circuit combinators}
  \label{fig:circblocks}
\end{figure}

The DSL \qbrickDSL is a small first-order functional, call-by-value
language with a special datatype $\texttt{circ}$ as the media to build
and manipulate circuits. The core of \qbrickDSL can be presented as a
simply-typed calculus, presented in Figure~\ref{tab:qbrickdsl}. The
basic data constructors for \texttt{circ} are \texttt{CNOT},
\texttt{SWAP}, \texttt{ID}, \texttt{H}, $\texttt{Ph}(e)$ and $\texttt{R}_z(e)$ (see
Table~\ref{matgates} in the Appendix for their semantics). The constructor for
high-level circuit operations are sequential composition \texttt{SEQ},
parallel composition \texttt{PAR} and ancilla creation/termination
\texttt{ANC} (see Figure~\ref{fig:circblocks} for details).

On top of \texttt{circ}, the type system of
\qbrickDSL features the type of integers \texttt{int} (with
constructors $\underline{n}$, one for each integer
$n$), booleans \texttt{bool} (with constructors \texttt{tt} and
\texttt{ff}), and n-ary products (with constructor
$\tuple{e_1,\ldots,e_n}$). This type system is not meant to be
exhaustive and it can be extended with usual constructs such as
floats, lists and other user-defined inductive datatypes --- its
embedding into WhyML makes it easy to use such types. The term
constructs are limited to function calls, \texttt{let}-style
composition, test with \texttt{if-then-else} and simple iteration:
$\texttt{iter}~f~n~a$ stands for $f(f(\cdots f(a)\cdots))$, with $n$
calls to $f$. We again stress out that this could easily be extended
--- we just do not need it for our purpose.

\begin{figure}
\scalebox{.9}{
  $
  \begin{array}{r@{\quad}lll}
    \text{Expression}
    & e
    & {:}{:}{=}
    & x \bor c \bor
      f(e_1,\ldots,e_n)
      \bor \letinterm{\tuple{x_1,\ldots,x_n}}{e}{e'}\bor \\
    &&& \iftermx{e_1}{e_2}{e_3}\bor
        \texttt{iter}~f~{e_1}~{e_2}
    \\[1ex]
    \text{Data Constructor}
    & c
    & {:}{:}{=}
    & \underline{n}\bor\ttrue\bor\ffalse\bor \tuple{e_1,\ldots, e_n}
      \bor \texttt{CNOT} \bor\texttt{SWAP} \bor \texttt{ID}\bor \texttt{H}\bor\texttt{Ph}(e)\bor\texttt{R}_z(e)
        \bor \\
    &&&
    \texttt{ANC}(e) \bor \texttt{SEQ}(e_1,e_2) \bor
        \texttt{PAR}(e_1,e_2)
    \\[1ex]
    \text{Function}
    & f
    & {:}{:}{=} 
    & f_{\text{d}} \bor f_{\text{c}}
    \\[1ex]
    \text{Declaration}
    & d
    & {:}{:}{=} 
    & \decl{f_{\text{d}}}{x_1,\ldots,x_n}{e}
    \\[1ex]
    \text{Type}
    & A
    & {:}{:}{=} 
    &   \bool \bor \inttype \bor \unittype \bor
      A_1 \times \cdots\times A_n \bor
      \circtype.
    \\[1ex]
      \text{Value}
  & v
  & {:}{:}{=}
  & x \bor
    \underline{n}\bor\ttrue\bor\ffalse\bor \tuple{v_1,\ldots, v_n}
      \bor \\
    &&&\texttt{CNOT} \bor \texttt{SWAP} \bor \texttt{ID}\bor \texttt{H}\bor\texttt{Ph}(\underline{n})\bor\texttt{R}_z(\underline{n})
        \bor
        \texttt{ANC}(v) \bor \texttt{SEQ}(v_1,v_2) \bor
        \texttt{PAR}(v_1,v_2)
    \\
    \text{Context}
    & C[-]
    & {:}{:}{=}
    & [-]\bor f(v_1,\ldots v_{i-1},C[-],e_{i+1},\ldots,e_n)
      \bor \\
    &&& \letinterm{\tuple{x_1,\ldots,x_n}}{C[-]}{e'}\bor
        \iftermx{C[-]}{e_2}{e_3}\bor
    \\
    &&& \texttt{iter}~f~{C[-]}~{e}\bor\texttt{iter}~f~{v}~{C[-]}
        \bor\tuple{v_1,\ldots v_{i-1},C[-],e_{i+1},\ldots,e_n}
      \bor \\
    &&&\texttt{CNOT} \bor \texttt{ID}\bor \texttt{H}\bor\texttt{Ph}(C[-])\bor\texttt{R}_z(C[-])
        \bor
        \texttt{ANC}(C[-]) \bor
    \\
    &&&\texttt{SEQ}(C[-],e) \bor \texttt{SEQ}(v,C[-]) \bor
        \texttt{PAR}(C[-],e)\bor\texttt{PAR}(v,C[-])
    \\
  \end{array}  
$}
\caption{Syntax for \qbrickDSL}
\label{tab:qbrickdsl}
\end{figure}

The language is essentially first-order: this is reflected
by the types $A$ of expressions. The type of a function is given by
the types of its arguments and the type of its output. By abuse of
notation, the type of a function with inputs of types $A_i$ and
output of type $B$ is written $A_1\times \cdots\times A_n\to B$.

A function $f$ is either a function $f_{\text{d}}$ defined with a
declaration $d$ or a constant function $f_{\text{c}}$.
The functions defined by declarations must not be mutually recursive:
this small, restricted language only features iteration.
Constant functions
consist in 
integer operators ($+$, $*$, $-$, \emph{etc}), boolean operators
($\wedge$, $\vee$, $\neg$, $\to$, \emph{etc}), Comparison operators
(${<}$, ${\leq}$, ${\geq}$, ${>}$ ,${=}$, ${\neq} :
\inttype\times\inttype \to \bool$)
and high-level circuit operators: $\texttt{ctl}, \texttt{invert} :
\circtype\to\circtype$ for controlling and inverting circuits,
and $\texttt{width},\noeg : \circtype \to \inttype$ for counting the
number of input and output wires, and the number of gates (not counting \texttt{ID}
nor \texttt{SWAP})
%
in the circuit $C$, $\texttt{width}(C)$ stands for the number of input
and output wires of $C$, $\texttt{invert}(C)$ stands for the invert of
$C$ and $\texttt{ctl}(C)$ the control of $C$. See
Figure~\ref{fig:circblocks} for the intuitive definition of circuit
combinators. 

Two very useful function declarations are the first and second
projections $\texttt{fst}:A\times B\to A$ and
$\texttt{snd}:A\times B\to B$. For instance, the former is defined with
$\texttt{let}~\texttt{fst}(x) =
(\texttt{let}~\tuple{y,z}=x~\texttt{in}~y)$.
%
    
%
%
%
\begin{figure}
  \begin{center}
    \hfill
    \begin{prooftree}
      \hfill
      \AxiomC{}
      \UnaryInfC{$\Gamma, x:A \vdash  x:A$}
      \DisplayProof
      \hfill
      \AxiomC{$\Gamma \vdash f:A_1\times\cdots\times A_n \to B$}
      \AxiomC{$\Gamma\vdash e_i:A_i $}
      \BinaryInfC{$\Gamma \vdash  f(e_1,\ldots,e_n): B$}
      \DisplayProof
      \hfill
      \AxiomC{$\Gamma \vdash e_i:A_i $}
      \UnaryInfC{$\Gamma \vdash  \tuple{e_1,\ldots,e_n}:
        A_1\times\cdots\times A_n$}
      \hfill
    \end{prooftree}
    
    \begin{prooftree}
      \AxiomC{$\Gamma \vdash e_1 :A_1\times\cdots\times A_n$}
      \AxiomC{$\Gamma, x_1: A_1,\ldots,x_n:A_n \vdash e_2 : B $}
      \BinaryInfC{$\Gamma \vdash  \letinterm{\tuple{x_1,\ldots,x_n}}{e_1}{e_2}: B$}
    \end{prooftree}

    \begin{tabular}{cc}
      \begin{minipage}{.45\textwidth}
    \begin{prooftree}
      \AxiomC{$\Gamma \vdash e_1 : \bool $}
      \AxiomC{$\Gamma \vdash e_2 : A $}
      \AxiomC{$\Gamma \vdash e_3 : A $}
      \TrinaryInfC{$\Gamma \vdash  \iftermx{e_1}{e_2}{e_3}: A $}
    \end{prooftree}
  \end{minipage}
      &
      \begin{minipage}{.45\textwidth}
    \begin{prooftree}
      \AxiomC{$\Gamma \vdash e_1 : \inttype $}
      \AxiomC{$\Gamma \vdash e_2 : A $}
      \AxiomC{$f : A\to{} A $}
      \TrinaryInfC{$\Gamma \vdash \texttt{iter}~f~{e_1}~{e_2} : A$}
    \end{prooftree}
  \end{minipage}
    \end{tabular}
  \end{center}
  \caption{Typing rules for \qbrickDSL}
  \label{tab:qbrickdsl-type}
\end{figure}

The typing rules are the usual ones (summarized for convenience in 
Table~\ref{tab:qbrickdsl-type}).

\subsection{Operational semantics}
As any other regular functional programming language, \qbrickDSL is
equipped with an operational semantics based on beta-reduction and
substitution. We define a notion of value and applicative context as
in Table~\ref{tab:qbrickdsl}.
We then define a rewriting strategy as the relation defined with
$C[e]\to C[e']$ whenever $e\to e'$ is one of the rule of
Table~\ref{tab:sem-op}. The table is split into the rules for the
language constructs and the rules defining the behavior of the
constant functions. We only give a subset of the latter rules. For
instance, the arithmetic operations are defined in a canonical manner,
and the boolean and comparison operators are defined in a similar
manner on values of type \texttt{int} and \texttt{bool}. The rules for
the constant functions acting on circuits are also for the most part
straightforward: the size of a sequence is the sum of the sizes of the
compounds for instance. The rules which we do not provide are the ones
for the control operation \texttt{ctl}: the intuition behind their
definition can be found in Figure~\ref{fig:circblocks}. For the
elementary gates, any definition can be used (see
e.g.~\cite{nielsen2002quantum}), as long as it can be written with the
chosen set of gates. One just has to then adjust the lemmas referring
to \texttt{ctl} in \qbrickSPEC. Similarly, the invert of elementary
gates are not given: we can choose the usual ones from the litterature
---and this definition is then parametrized by the choice of gates.

\begin{table}
  \scalebox{.9}{%
  \begin{tabular}{|c|}
    \hline
    \rowcolor{gray!25} 
    Language constructs
    \\
    Assuming that there is a declaration
    $f(x_1,\ldots,x_n)\triangleq e$.
    \\[1ex]
    $\begin{array}{r@{{}\to{}}l}
       f(v_1,\ldots,v_n)& e[x_1:=v_1,\ldots,x_n:=v_n]
       \\
       \texttt{let}~\tuple{x_1,\ldots,x_n}=\tuple{v_1,\ldots,v_n}~\texttt{in}~e
       &
         e[x_1:=v_1,\ldots,x_n:=v_n]
       \\
       \texttt{if~tt~then}~e_1~\texttt{else}~e_2
       &
       e_1
       \\
       \texttt{if~ff~then}~e_1~\texttt{else}~e_2
       &
         e_2
       \\
       \text{when $n\leq 0$:\quad}
       \texttt{iter}~f~\underline{n}~a & a
       \\
       \text{when $n> 0$:\quad}
       \texttt{iter}~f~\underline{n}~a & f(\texttt{iter}~f~\underline{n-1}~a)
     \end{array}$
    \\\hline
    \rowcolor{gray!25}
    Constant functions (subset of the rules)
    \\
    \begin{tabular}{@{}c@{}c@{}}
    $\begin{array}{r@{{}\to{}}l}
       \underline{n}+\underline{m}
       & \underline{n+m}
       \\
       \underline{n}-\underline{m}
       & \underline{n-m}
       \\
       \underline{n}*\underline{m}
       & \underline{n*m}
       \\
       \texttt{size}(\texttt{ID})
       & 0
       \\
       \texttt{size}(\texttt{SWAP})
       & 0
       \\
       \texttt{size}(g)
       & 1\text{\quad($g$ other gate)}
       \\
       \texttt{size}(\texttt{SEQ}(v_1,v_2))
       & \texttt{size}(v_1)+\texttt{size}(v_2)
       \\
       \texttt{size}(\texttt{PAR}(v_1,v_2))
       & \texttt{size}(v_1)+\texttt{size}(v_2)
       \\
       \texttt{size}(\texttt{ANC}(v))
       & \texttt{size}(v)
     \end{array}$
       &
         $\begin{array}{r@{{}\to{}}l}
       \texttt{width}(\texttt{CNOT})
       & 2
       \\
       \texttt{width}(\texttt{SWAP})
       & 2
       \\
         \texttt{width}(g)
       & 1\text{\quad($g$ other gate)}
       \\
       \texttt{width}(\texttt{SEQ}(v_1,v_2))
       & \texttt{width}(v_1)
       \\
       \texttt{width}(\texttt{PAR}(v_1,v_2))
       & \texttt{width}(v_1)+\texttt{width}(v_2)
       \\
       \texttt{width}(\texttt{ANC}(v))
       & \texttt{width}(v)-1
       \\
       \texttt{invert}(\texttt{SEQ}(v_1,v_2))
       & \texttt{SEQ}(\texttt{invert}(v_2),\texttt{invert}(v_1))
       \\
       \texttt{invert}(\texttt{PAR}(v_1,v_2))
       & \texttt{PAR}(\texttt{invert}(v_1),\texttt{invert}(v_2))
       \\
       \texttt{invert}(\texttt{ANC}(v))
       & \texttt{ANC}(\texttt{invert}(v))
     \end{array}$
    \end{tabular}
    \\
    \hline
  \end{tabular}}
  \caption{Operational semantics for \qbrickDSL}
  \label{tab:sem-op}
\end{table}

\subsection{Properties}
Note that the target low-level representation for an expression of
type $\texttt{circ}$ is a value made of the circuit data
constructors. This is derived from the safety properties of the
language:
\begin{property}[Safety properties and normalization]
  Provided that $\Gamma\vdash e:A$ and $e\to e'$, then
  $\Gamma\vdash e':A$.  Provided that $\vdash e:A$ is a closed
  expression, and provided that all the function in $e$ recursively
  admits (external) definitions, then either $e$ is a value or it
  reduces. Finally, the reduction strategy $(\to)$ is normalizing:
  there does not exist an infinite reduction sequence
  $e_1\to e_2\to\ldots$
  \qed
\end{property}

\begin{example}
  \label{ex:motiv-dsl}\rm
  The motivating example of Section~\ref{motiv} can be written in
  \qbrickDSL as
  \[
    \def\tt#1{\texttt{#1}}
    \begin{array}{l}
      \tt{let}~\tt{aux}(x)=\tt{SEQ}(x,\tt{H})
      \\
      \tt{let}~\tt{main}(n) = \tt{iter}~\tt{aux}~n~\tt{ID}
    \end{array}
  \]
  The function \texttt{aux} inputs a circuit and append a Hadamard gate
  at the end. The function \texttt{main} then input an integer parameter
  \texttt{n} and iterate the function \texttt{aux} to obtain \texttt{n}
  Hadamard in sequence.  In particular, one can show that for instance
  \[\def\ss{\texttt{SEQ}}\def\h{\texttt{H}}
    \texttt{main}~\underline{4}
    \to^*
    \ss(\ss(\ss(\ss(\texttt{ID},\h),\h),\h),\h),
  \]
  that is, a sequence of 4 Hadamard gates.
\end{example}

\subsection{Universality and usability of the chosen circuit
  constructs}
In \qbrickDSL, we use a restricted, small set of elementary circuit
building blocks. For instance, we have not included the NOT-gate
$(\begin{smallmatrix}0&1\\1&0\end{smallmatrix})$ . This is a design
choice: the chosen elementary gates are not meant to be convenient but
simple to specify yet forming a universal set of gates:
A \emph{universal} (resp. \emph{pseudo-universal}) set of elementary
gates is such that they can be composed thanks to sequence or
parallelism so as to perform (resp. approach arbitrarily close) any
quantum unitary matrix.

Other, maybe more convenient gates can then be defined as macros on top of
them. If one aims at using \qbrick inside a verification compilation
tool-chain, these macros can for instance be the gates of the targeted
architecture.

\subsection{Validity of circuits}
A circuit is represented as a rigid rectangular shape with a fixed
number of input and output wires. In particular, there is a notion of
validity: a \texttt{circ} object only makes sense provided two
constraints:
\begin{itemize}

\item in $\texttt{SEQ}(C_1,C_2)$, the two circuits $C_1$ and $C_2$
  should have the same number of wires (i.e. the same size). For
  instance, $\texttt{SEQ}(\texttt{CNOT},\texttt{H})$ is not valid: one
  cannot put in sequence a 2-qubit gates with a 1-qubit gate. This is
  a simple {\it syntactic} constraint;

\item in $\texttt{ANC}(C)$, the circuit $C$ should have
  $n+1$ wires. Moreover, if given as input a vector where the last
  qubit is in state $\ket{0}{}$, its output should also leave this
  qubit in state $\ket{0}{}$. This condition is on the other
  hand a \emph{semantic} constraint.
\end{itemize}

\noindent Note that even the syntactic constraints cannot be checked by a simple typing 
procedure, because of the higher-order reasoning involved here: the constraints must hold 
for any value of the parameters. 
All these syntactic and semantic constraints are thus expressed in
\qbrickSPEC, our domain-specific logical specification language, and
meant to be sent as proof obligations to a proof engine.

\begin{example}
  \label{ex:motiv-dsl-2}\rm
  Note how the circuit generated by \texttt{main} in
  Example~\ref{ex:motiv-dsl} is not necessarily a valid circuit
  (although in this case it is). This is one of the constraints that
  can be handled by \qbrickSPEC, as shown in
  Example~\ref{ex:motiv-hoare}.
\end{example}

\subsection{Denotational semantics}
As all expressions in \qbrickDSL are terminating, one can use regular
sets as denotational semantics for the language.
In order to be able to handle the partial definitions coming up in
Section~\ref{sec:qspec}, we include in the denotation of each type an
``error'' element $\bot$
We therefore define
the denotation of basic types as the set of their values: 
$\denot{\bool} = \{\ttrue,\ffalse,\bot\}$,
$\denot{\inttype} = \mathbb{Z}\cup\{\bot\}$
and
$\denot{\circtype} = \{ v~\bor~{}\vdash v : \circtype \}\cup\{\bot\}$.
Product types are defined as the set-product:
$\denot{A_1\times\cdots\times
  A_n}=(\denot{A_1}\times\cdots\times\denot{A_n})\cup\{\bot\}$ and
$\denot{\top}=\{\star,\bot\}$, the singleton set.
Finally, functions are defined as set-functions from the input set to
the output set. The denotation of the language constructs are the
usual one in a semantics based on sets ; for the constant functions,
the definitions are the canonical ones: arithmetic operations maps to
arithmetic operations for instance. In \qbrickDSL, everything is
well-defined and $\bot$ is only attainable from $\bot$ ---so for
instance, $\bot+x=\bot$.

Note that in the denotational semantics one can build non-valid
circuits. For instance, the circuit
$\texttt{SEQ}(\texttt{CNOT},\texttt{H})$ is a member of
$\denot{\circtype}$. This is to be expected as we aim at the
following property:

\begin{property}[Soundness]
  Provided that $\vdash e : A$, we have
  $\denot{e}\in\denot{A}\setminus\{\bot\}$. Moreover, provided that
  $e\to e'$ then we have $\denot{e}=\denot{e'}$.\qed
\end{property}

It is however possible to formalize the notion of syntactically valid
circuits as a subset of $\denot{\circtype}$.

\newcommand{\vwsem}{\mathcal{V}_{\text{syntax}}}

\begin{definition}\label{def:circ-valid}
  \def\tt#1{\texttt{#1}} We define the (syntactic) unary relation
  $\vwsem$ on $\denot{\circtype}$ as follows: Each one of the gate
  belongs to $\vwsem$; if $C_1$ and $C_2$ belongs to $\vwsem$ then so
  does $\tt{PAR}(C_1,C_2)$ and $\tt{ANC}(C_1)$; if moreover
  $\denot{\tt{width}}(C_1)=\denot{\tt{width}}(C_2)$ then
  $\tt{SEQ}(C_1,C_2)$ belongs to $\vwsem$.
\end{definition}

\section{\qbrickSPEC}
\label{sec:qspec}

In term of semantics of quantum circuits, the main novelty of \qbrick
is to be able to reason on \emph{open terms} seen as circuit
description, parametrized programs. To do so, we build upon the recent
proposal of
path-sums~\cite{amy2018towards,dblp:phd/basesearch/amy19}.
In \qbrickSPEC, we define the notion of \emph{higher-order path-sums
  (HOPS)}.
Original path-sums in \cite{dblp:phd/basesearch/amy19} are fixed
functionals, computed from concrete, fixed-size circuits. In HOPS, the
phase polynomials, ranges and basis kets are instead written as open
terms from \qbrickDSL, \emph{parametrized} with term variables. This
correspondence. This integration with \qbrickDSL brings two decisive
advantages compared to the original path-sums:
\begin{itemize}
  
\item {\bf Parametricity and compositionality}. Because of the sharing
  of term variables between \qbrickDSL terms and HOPS, \qbrickSPEC
  gives the ability to give specification to general programs
  describing families of circuits instead of fixed-size circuits.
  \qbrickSPEC opens path-sums to \emph{higher-order} specification and
  verification while retaining the vertical and horizontal
  compositional properties;

\item {\bf Versatility.} Thanks to the integration within a logical
  framework, \qbrickSPEC gives the ability to define ---and reason---
  upon logical macros asserting useful constraints related to
  probabilities, eigenvalues, etc (See
  Section~\ref{sec:spec-lib}.
  
\end{itemize}

\subsection{Syntax of \qbrickSPEC}
\label{syntax-qbrick-spec}
\qbrickSPEC consists of a set of dedicated relations and functions,
together with a language of algebraic expressions on top of a
first-order logic, together with logical
libraries to express constraints.
We define \qbrickSPEC as a first-order, predicate logic with the following syntax.
\[
  \begin{array}{r@{\qquad}lll}
    \text{Formula}
    & \phi, \psi
    & {:}{:}{=}
    & \phi \vee \psi \bor \phi \wedge \psi \bor \neg \phi \bor \phi \to \psi \bor R(\hat{e}_1,\ldots,\hat{e}_n)
    \\[1ex]
    \text{First-order expression}
    & \hat{e}
    & {:}{:}{=}
    & x \bor c(\hat{e}_1,\ldots,\hat{e}_n) \bor
      f(\hat{e}_1,\ldots,\hat{e}_n) \bor
      f_{\ell}(\hat{e}_1,\ldots,\hat{e}_n).
  \end{array}
\]
The first-order expressions $\hat{e}$ form a subset of \qbrickDSL:
they are restricted to variables and (formal) function calls to other
first-order expressions. Unlike regular, general expressions ---meant
to be computational vehicles--- these first-order expressions only aim
at being reasoned upon. The functions names are then expanded with
\emph{logic functions} $f_\ell$ with no computational content.  Among
these new functions, we introduce one function
$\texttt{iter}_{f} : \inttype\times A\to A$ for each function
$f:A\to A$, standing for the equational counterpart of the
iteration. The logic functions are defined equationally in the logic:
see Section~\ref{sec:deduction} for details.
The relation $R$ ranges over a list of constant relations over
first-order expressions. An important one is
\(
  \texttt{circ\_valid} : \circtype \to \bool,
\)
expressing the syntactic property $\vwsem$ of
Definition~\ref{def:circ-valid}.  In \qbrickSPEC, we identify
relations and functions of return type \texttt{bool}. Constant
relations are therefore simply constant functions of output type
\text{bool}.

The type system of \qbrickSPEC is extended with opaque types, equipped
with constant functions and relations to reason upon them. They come
with no computational content: the aim is purely to be able to
\emph{express and prove specification properties} of programs. This is
why we did not incorporate them in \qbrickDSL's type system.

The opaque types we consider in \qbrickSPEC are $\texttt{complex}$,
$\texttt{real}$, $\texttt{bitvector}$, $\hops$ and $\texttt{ket}$. The
operators and relations for these new types are given in
Table~\ref{type-ope}. Note that in the rest of the paper, by abuse of
notation we shall omit the casting operations $\texttt{i\_to\_r}$ and
$\texttt{r\_to\_c}$. We shall also use a declared exponentiation
function $[-]^{[-]}$ overloaded by abuse of notation with type
$\complextype\times \inttype\to \complextype$ and
$\realtype\times \inttype\to \realtype$.
The three types $\complextype$, $\realtype$ and $\bitvector$ are
standard; the types $\hops$ and $\kettype$ are novel and form the main
reasoning vehicle in \qbrickSPEC.

\subsection{The types $\hops$ and $\kettype$}
\label{sec:types-hops-ket}
In short, the type $\hops$ encodes our \emph{higher-order path sum}
(HOPS) representation for expressions of type {\circtype} in
\qbrickDSL, while $\kettype$ encodes the notion of ket-vector. As
these types are pure reasoning apparatus, we only need them in
\qbrickSPEC and they are defined uniquely though an equational theory.

As shown in Figure~\ref{tab:syntax-ps}, a regular path-sum is a static
abstract object consisting of 4 pieces: a global \emph{range} (the
value $n$), what we shall be calling its \emph{width}, i.e. the size
of the bit-vector $x$, and the \emph{phase polynomial}
$\frac{P_k(x)}{2^m}$ and the \emph{ket} vector $\phi_k(x)$, functions
of both the input $x$ and the index $k$.

To reflect this structure, the type {\hops} is equipped with 4
constant functions acting on an HOPS (see Table~\ref{type-ope}):
\texttt{hops\_width}, yielding its width, \texttt{hops\_range},
yielding its range, \texttt{hops\_angle}, yielding the real number
corresponding to $\frac{P_k(x)}{2^m}$, and \texttt{hops\_ket} yielding
the ket-vector $\phi_k(x)$. Note how the two last functions also
inputs two bit-vectors, corresponding to $k$ and $x$. If path-sums
compose nicely, a given linear map does not have a unique
representative path-sum (partly due to the fact that phase polynomials
are equal modulo $2\pi$). To capture this equivalence, we introduce
the constant relation \texttt{hops\_equiv}. In order to relate
circuits and HOPS, we introduce the constant function
$\texttt{circ\_to\_hops}$: it returns \emph{one possible} HOPS that
\emph{represents} the input circuit. The chosen HOPS is built in a
constructive manner on the structure of the circuit.
A useful relation is $\rcorrect{-}{-}$ relating a circuit and a HOPS:
it is defined as
$\rcorrect{h}{c} \triangleq
\texttt{hops\_equiv}(\texttt{circ\_to\_hops}(c),h)$.  Another useful
macro is the function
$\texttt{circ\_apply} : \circtype\times\kettype\to\kettype$, defined
as
$\texttt{circ\_apply}(C,k)\triangleq\texttt{hops\_apply}(\texttt{circ\_to\_hops}(C),k)$.

The type {\kettype} is a handle to easily manipulate ket-vectors, and
Table~\ref{type-ope} presents the constant functions meant to
manipulate it. {\bvtoket} builds a basis ket-vector out of the input
bit vector~; {\ketlength} returns the number of qubits in the ket ;
{\getket} returns the amplitude of the corresponding basis
ket-vector. The other operations are the usual operations on vectors:
addition, subtraction, tensors, scalar multiplication. Finally, the
function {\texttt{hops\_apply}} relates {\hops} and $\kettype$: it
identifies the input HOPS to a functional and applies it to the input
ket-vector.

\begin{table}
  \scalebox{.8}{%
  \begin{tabular}{|@{}c@{}|@{}c@{}|}
  \begin{tabular}{@{}c@{}}
      \hline
      $\complextype$ and $\realtype$\\
      \hline
      $\begin{array}{r@{~}l}
         i,\pi &: \complextype \\
         \texttt{i\_to\_r} &: \inttype\to\realtype\\
         \rtoc &: \realtype\to \complextype\\
         \real,\imag,\texttt{abs} &: \complextype \to \realtype\\
         e^{[]} &: \complextype\to \complextype\\
         -_c, +_c, *_c, /_c &: \complextype \times \complextype \to \complextype\\
         -_r, +_r, *_r, /_r &: \realtype \times \realtype \to \realtype\\
         \sqrt{-}&:\realtype\to\realtype
       \end{array}$\\
      \hline
      \bitvector\\
      \hline
      $\begin{array}{r@{~}l}
         \texttt{bv\_length}&: \bitvector\to\inttype \\
         \texttt{bv\_cst}&:  \inttype\times\texttt{bool} \to \bitvector \\
         \texttt{bv\_get} &: \bitvector\times\inttype \to\texttt{bool}\\
         \texttt{bv\_set} &:
                            \bitvector\times\inttype\times\texttt{bool}\to\bitvector
       \end{array}$\\
    \hline
  \end{tabular}
               &
                 \begin{tabular}{@{}c@{}}
                   \hline
      \hops\\
      \hline
      $\begin{array}{r@{~}l}
        \texttt{hops\_width} &: \hops \to \inttype \\
         \texttt{hops\_range} &: \hops \to \inttype \\
         \texttt{hops\_angle} &: \hops\times\bitvector\times\bitvector \to
                                \realtype\\
         \texttt{hops\_ket} &: \hops\times\bitvector\times\bitvector \to
                              \bitvector\\
         \texttt{hops\_equiv} &: \hops\times\hops \to \bool\\
         \texttt{circ\_to\_hops} &: \circtype \to\hops\\
         \texttt{hops\_apply} &: \hops\times\kettype \to \kettype
       \end{array}$\\
      \hline
      \kettype\\
      \hline
      $\begin{array}{r@{~}l}
      \ketlength &: \kettype \to \inttype\\
      \getket &: \kettype\times \bitvector \to \complextype\\
      \bvtoket &: \bitvector \to \kettype\\
      +_k,-_k,\otimes_k &: \kettype \times \kettype \to \kettype\\
                       *_k &: \complextype \times \kettype \to
                             \kettype
       \end{array}$\\
      \hline                       
  \end{tabular}
  \end{tabular}}
  \caption{Primary operators for \qbrickSPEC}
  \label{type-ope}
\end{table}

\subsection{Denotational semantics of the new types} 
The denotational semantics of $\realtype$ and $\complextype$ are
respectively the sets $\mathbb{R}\cup\{\bot\}$ and
$\mathbb{C}\cup\{\bot\}$, and the denotation of the operators are the
canonical ones. As for Section~\ref{sec:qdsl}, $\bot$ maps to $\bot$,
so for instance $\bot+_rx=\bot$.

The denotation of $\bitvector$ is defined as the set of all
bit-vectors, together with the ``error'' element $\bot$. The constant
functions are mapped to their natural candidate definition, using
$\bot$ as the default result when they should not be defined. So for
instance, $\denot{\texttt{bv\_cst}}(-1,\texttt{tt}) = \bot$.

An element of $\kettype$ is meant to be a ket-vector: we defined
$\denot{\kettype}$ as the set of \emph{all} possible ket-vectors
$\sum_{i=0}^{2^n}\alpha_n\ket{b_n}{m}$, for all possible
$m,n\in\mathbb{N}$, $\alpha_n\in\mathbb{C}$ and bit-vectors $b_n$ of
size $m$, together with the error element $\bot$.

Finally, $\hops$ is defined as the set of formal path-sums, as defined
in Table~\ref{tab:syntax-ps}, together with the error element $\bot$.
The denotation of the constant functions are defined as discussed in
Section~\ref{sec:types-hops-ket}. As an example,
$\denot{\texttt{hops\_range}}$ returns the range of the corresponding
HOPS. The map $\texttt{circ\_to\_hops}$ builds a valid HOPS out of the
input circuit, or $\bot$ if the circuit is not valid. The defined HOPS
follows the structure of the circuit. For instance,
$\texttt{circ\_to\_hops}(\texttt{SEQ}(C_1,C_2))$ is the sequential
composition of the two HOPS $\texttt{circ\_to\_hops}(C_1)$ and
$\texttt{circ\_to\_hops}(C_2)$ (as shown in Equation~\eqref{eq:ps}).

\subsection{Parametricity of HOPS}
A regular path-sum is not parametric: it represents \emph{one} fixed
functional. So why did we chose $\denot{\hops}$ to be a set of
path-sums? Let us consider an example.

\begin{example}
  Consider the motivating example of Section~\ref{motiv} and its
  instantiation in Example~\ref{ex:motiv-dsl}. The function
  $\texttt{main}$ describes a family of circuits indexed by an integer
  parameter $n$. Now, consider the typing judgment
  \(
    h:\hops,n:\inttype\vdash\rcorrect{h}{\texttt{main}(n)}:\bool.
  \)
  It can be regarded as a relation between HOPS $h$ and integers $n$,
  valid whenever $h$ represents $\texttt{main}(n)$. Technically, this
  relation is not quite the graph of a function (since several HOPS
  might match the circuit $\texttt{main}(n)$). Nonetheless, to each
  $n$ is associated a different set of HOPS $h$: in this sense, one
  can say that $h$ is ``higher-order'', as it is parametrized by
  $n$. Thus the ``HOPS'' terminology.
\end{example}

\subsection{Regular sequents and HQHL sequents in \qbrickSPEC}
\label{sec:logic-judg}
Formulas in \qbrickSPEC are typed objects ---and to a certain extent
one can identify them with first-order expressions of type
$\bool$. Due to this correspondence, we shall only say that logic
judgments in \qbrickSPEC are well-formed judgments of the form
$
  \Delta\vdash \phi
$
where the well-formedness means that $\Delta\vdash \phi:\bool$ is a
valid typing judgment in \qbrickDSL.
That being said, a well-formed judgment $\Delta\vdash\phi$ is valid
whenever it holds in the denotational semantics:
for every instantiation
$\sigma$ sending $x:A$ in $\Delta$ to $\denot{A}$,
the denotation
$\denot{\phi}_\sigma$ is valid.
In particular, the (free) variables of $\phi$ can be regarded as
universally quantified by the context $\Delta$.

Another useful kind of judgment is one allowing equational reasoning
on terms: if $\Delta\vdash e_1:A$ and $\Delta\vdash e_2:A$ are
valid $\qbrickDSL$ typing judgments, the judgment $\Delta\Vdash e_1=
e_2:A$ is valid whenever for every instantiation
$\sigma$ sending $x:B$ in $\Delta$ to $\denot{B}$,
the denotations of $e_1$ and $e_2$ with respect to $\sigma$ are equal
(as set-elements).

In order to be able to express program specifications with pre- and
post-conditions, we finally define a HQHL sequent of the
form $\Delta\Vdash\hoare{\phi}{e}{\psi}:A$ (we omit the type $A$ when
irrelevant or clear). The formula $\psi$ can make use of a reserved
free variable $\texttt{result}$ of type $A$. Such a sequent is then
well-formed provided that $\Delta\vdash\phi:\bool$,
$\Delta,\texttt{result}:A\vdash\psi:\bool$ and $\Delta\vdash e : A$
are valid typing judgments. Note how the reserved free variable
$\texttt{result}$ is being added to $\Delta$ for typing $\psi$. For
convenience, we also extend the notation to $\texttt{result}_i$ to
stand for the $i$th projection of \texttt{result} when $A$ is of the
form $A_1\times\cdots\times A_n$.

Then the point is that first, families of quantum circuits can be
described in \qbrickDSL and specifications in \qbrickSPEC, and second,
that one can come up with a set of meaningful axioms that are enough
to derive these specifications. This is the goal of
Section~\ref{sec:deduction}.

\begin{example}
  \label{ex:motiv-hoare}
  Consider the motivating example of Section~\ref{motiv} and its
  instantiation in Example~\ref{ex:motiv-dsl}. We can now give a
  specification to the function
  $\texttt{main}$, as follows:
  \[
    \begin{array}{l}
      n:\inttype,m:\inttype,x:\kettype\Vdash\\
      \quad\left\{n\geq 0\wedge\ketlength(x)=1\wedge n = 2*m\right\}~
      \texttt{main}(n)
      \left\{\texttt{circ\_apply}(C,x) = x\right\}.
    \end{array}
  \]
  The fact that \texttt{circ\_apply} is well-defined implies
  that $C$ is valid.
\end{example}

\section{The HQHL  Deduction System}
\label{sec:deduction}

Thanks to the logic judgments presented in
Section~\ref{sec:logic-judg}, it is possible to reason with
\qbrickSPEC formulas on expressions written in \qbrickDSL. If we
showed how to give a semantical meaning to the validity of a
judgment, we have not explained how to derive such a validity from
first principles. This is the purpose of the current section.

\begin{property}[Validity of the deduction rules]
  All of the deduction rules that we add in this section are sound
  with respect to the semantics.\qed
\end{property}

Note that we do not aim in this section at being exhaustive: we only
aim at giving an intuition as of how and why one can rely on an
automated deductive system to derive \qbrickSPEC judgments. In
particular, we show how to break down a program specification into a
set of proof obligations that can be sent to an SMT solver to be
automatically discharged.

\subsection{Deriving HQHL judgments}
In Example~\ref{ex:motiv-hoare}, we were able to state a HQHL
judgment for the motivating example of Section~\ref{motiv}. Using a bottom-up
strategy, it is possible to use the deduction rules of
Figure~\ref{tab:qbrickdsl-nog} to break down the judgments into
pieces reasoning on smaller terms. Along the way, there is the need
for introducing invariants and assertions.  As
usual, some of these assertions can be derived by
computing the weakest-preconditions: we do not necessarily have to
introduce every single one.

In the deduction rule $(\texttt{iter})$ of
Figure~\ref{tab:qbrickdsl-nog}, the first-order expression $\hat{e}_1$
is mechanically built from $e_1$ by inlining function calls, replacing
$P[\texttt{if}~e_1~\texttt{then}~e_2-\texttt{else}~e_3]$ with the
equivalent formulation
$(e_1\wedge P[e_2])\vee(\neg e_1\wedge P[e_3])$, changing
$\texttt{let}$'s with calls to projections, and replacing
each $\texttt{iter}\ f\ a\ i$ with its functional counterpart
${\texttt{iter}}_f(a,i)$.
The deduction rule $(\texttt{eq})$ states that whenever one can
show that two expressions are equal one can substitute one for the
other inside a HQHL judgment. Finally, we can derive from the
semantics the usual
substitution rules. For instance, provided that $\Gamma,x:A\vdash
\psi$ and $\Gamma\vdash e:A$ then $\Gamma\vdash \psi[x:=e]$.

The rules of Figure~\ref{tab:qbrickdsl-nog} make it possible to break
an expression down to a first-order expression from \qbrickSPEC. When
such a case is attained one can rely on the rule
\[
  \infer[(\texttt{f-o})]{
    \Gamma\Vdash\{\phi\}~\hat{e}~\{\psi\} : A
  }{
    \Gamma\vdash\phi\to\psi[\texttt{result}:=\hat{e}]
  }
\]
to generate a proof obligation as a regular sequent in \qbrickSPEC. Of
course, there is no guarantee that any such proof-obligation can
automatically be discharged. However, our work show that this strategy
can successfully be applied, as exemplified by the non-trivial case
studies we considered in Section~\ref{sec:xp}.

\begin{figure}
  \[
  \infer[(\texttt{iter})]{
    \Gamma\Vdash
    \{\phi\}~
    \texttt{iter}~f~e_1~e_2
    ~\{P[\hat{e}_1, \texttt{result}]\}
  }{
    \Gamma,x\Vdash \{\phi\wedge x\leq 0\}~e_2~\{P[x,\texttt{result}]\}
    &
    \Gamma,x,y\Vdash
    \{\phi\wedge P[x,y]\}
    ~f(y)~
    \{P[x+1,\texttt{result}]\}
  }
\]
~
\[
  \infer[(\texttt{let})]{
    \Gamma\Vdash
    \{P\}
    \letinterm{x_1,\ldots,x_n}{e_1}{e_2}
    \{R\}
  }{
    \Gamma\Vdash
    \{P\}
    e_1
    \{Q[x_i:=\texttt{result}_i\}
    &
    \Gamma,x_1,\ldots,x_n\Vdash
    \{Q\}
    e_2
    \{R\}
  }
\]
~
\[
  \infer[(\texttt{if})]{
    \Gamma\Vdash
    \{P\}
    \iftermx{e_1}{e_2[x:=e_1]}{e_3[x:=e_1]}
    \{R\}
  }{
    \Gamma\Vdash
    \{P\}
    e_1
    \{Q[x:=\texttt{result}\}
    &
    \Gamma,x\Vdash
    \{Q\wedge x\}
    e_2
    \{R\}
    &
    \Gamma,x\Vdash
    \{Q\wedge \neg x\}
    e_3
    \{R\}
  }
\]
~
\[
  \infer[(\texttt{tuple})]{
    \Gamma\Vdash
    \{P\}
    \langle e_1,\ldots,e_2\rangle
    \{R_1[\texttt{result}_1]\wedge\cdots\wedge R_n[\texttt{result}_n]\}
  }{
    \forall i ~\cdot~
    \Gamma\Vdash
    \{P\}e_i\{R_i[\texttt{result}]\}
  }
\]
~
\[
  \infer[(\texttt{decl})]{
    \Gamma\Vdash
    \{P\}
    f(e_1,\ldots,e_n)
    \{R\}
  }{
    f(x_1,\ldots,x_n) \triangleq e
    &
    \Gamma\Vdash
    \{P\}e[x_1:=e_1,\ldots,x_n:=e_n]\{R\}
  }
\]
~
\[
  \infer[(\texttt{weaken})]{
    \Gamma\Vdash
    \{P\}
    e
    \{Q\}:A
  }{
    \Gamma\vdash P\to P'
    &
    \Gamma\Vdash
    \{P'\}~e~\{Q'\}:A
    &
    \Gamma,\texttt{result}:A\vdash Q'\to Q
  }
\]
~
\[
  \infer[(\texttt{eq})]{
    \Gamma\Vdash
    \{P[e_2]\}~e[e_2]~\{Q[e_2]\}:A   
  }{
    \Gamma\Vdash e_1 = e_2 : A
    &
    \Gamma\Vdash
    \{P[e_1]\}~e[e_1]~\{Q[e_1]\}:A
  }
\]
  \caption{Deduction rules for \qbrick : HQHL rules for term constructs}
  \label{tab:qbrickdsl-nog}
\end{figure}

\subsection{Deduction rules for $\hops$}
\label{rcorrect-spec}
The main tools to relate circuits and HOPS are the constant function
$\texttt{circ\_to\_hops}$, its relational counterpart
$\rcorrect{-}{-}$, and the declared function
$\texttt{circ\_apply}$. They can be specified inductively on the
structure of the input circuit.
As an example of a lemma to be added for reasoning on
$\rcorrect{-}{-}$, consider the case of the Hadamard gate. The sequent
$
\Gamma \vdash  \rcorrect{h}{\texttt{H}}
$
can be derived from
\[
  \begin{array}{l}
    \Gamma,\vec{x},\vec{y} : \bitvector \vdash
    (  \bvlength(\vec{x}) =  1 \wedge 
    \bvlength(\vec{y}) =  1) \to   \\
    \qquad\qquad\qquad\hwidth (h) = 1 \wedge
    \hrange (h) = 1 \wedge\\
    \qquad\qquad\qquad\hangle(h)(\vec{x},\vec{y}) = (-1)^{\vec{x}_0 * \vec{y}_0 } \wedge
    \hket(h_3)(\vec{x},\vec{y}) = \vec{y}.
  \end{array}
\]
This simply states that the HOPS of the Hadamard gate is of the form
$
  \ket{x}{1}\mapsto \sum_{j=0}^1 (-1)^{\vec{x}_0 * \vec{y}_0}\ket{y}{1}
$.

For the circuit combinator $\texttt{PAR}$, one can state the rule
\begin{prooftree}
  \AxiomC{$\Gamma \vdash \rcorrect{h_1}{C_1}$}
  \AxiomC{$\Gamma \vdash \rcorrect{h_2}{C_2}$}
  \AxiomC{$\Gamma,\vec{x},\vec{y} : \bitvector \vdash \hopspar(h_1,h_2,h_3, \vec{x},\vec{y})$}
  \RightLabel{$\texttt{seq}$}
  \TrinaryInfC{$\Gamma \vdash  \rcorrect{h_3}{\texttt{PAR}(C_1,C_2)}$}
\end{prooftree}
where $\hopspar(h_1,h_2,h_3, \vec{x},\vec{y})$ is a formula
encapsulating the relationship between the phase polynomials, the kets,
the ranges and the widths of $h_1$, $h_2$ and $h_3$, as shown in
Equation~\eqref{eq:ps} in Section~\ref{sec:ps}.
The complete set of rules for $\rcorrect{-}{-}$ can be found in
Appendix~\ref{app:sec:rcorrect}

\subsection{Equational reasoning}
The SMT solvers we aim at using to discharge proof obligations
requires equational theories describing how to reason on the constant
functions that were introduced. Some of these equational theories are
standard and well-known in verification: for instance, 
bit-vectors and algebraic fields. Together with a few properties on
square-root, exponentiation, real and imaginary parts, the latter is
all we need for $\realtype$ and $\complextype$: in quantum
computation, the manipulations of real and complex numbers turns out
to be quite limited. In particular, we do not need anything related to
real or complex analysis. 

The main difficulty in the design of \qbrick has been to lay out
equational theories and lemmas for $\circtype$, $\hops$ and $\kettype$
that can efficiently help in automatically discharging proof
obligations.  Many of these equations and lemmas are quite
straightforward. For instance, we turn the rewriting rules of
Table~\ref{tab:sem-op} into equations, such as
$(x,y:\circtype)\Vdash\texttt{width}(\texttt{PAR}(x,y)) =
\texttt{width}(x)+\texttt{width}(y)$, or $n:\inttype\Vdash
\texttt{iter}_f(a,n+1) = f(\texttt{iter}_f(a,n))$. These equations
maps the (syntactic) computational behavior of expressions into the logic.
Other equations are expressing purely semantic properties. 
For instance, the equation
\[
\Gamma \Vdash \texttt{circ\_apply}(\texttt{SEQ}(C_1,C_2),k) =
\texttt{circ\_apply}(\texttt{circ\_apply}(C_1,\texttt{circ\_apply}(C_2,k)))
\]
is correct with respect to the semantics and is part of the equational
theory.

\paragraph{Circuit size}
An important equational theory that we need in \qbrickSPEC is one
for reasoning on the size of circuits. Being able to reason on
circuit size is one of the important feature for the certifying an
quantum algorithm implementation: one of its main point is for instance
the polynomiality of the number of gates over the size of the input.
We again stress out that this aspect of certification is not addressed
by existing approaches~\cite{dblp:journals/corr/abs-1904-06319,10.1007/978-3-030-25543-5_12,dblp:phd/basesearch/amy19}.

As we discussed above, the equational theory for circuits reflects the
rewrite rules. For instance, we have
$(x,y:\circtype)\Vdash\texttt{size}(\texttt{PAR}(x,y)) =
\texttt{size}(x)+\texttt{size}(y)$.  A complete set of rules can be
found in Appendix, in Figure~\ref{tab:qbrickdsl-nog-size}.

\begin{example}
  \label{ex:motiv-hoare-2}
  Consider the motivating example of Section~\ref{motiv} and its
  instantiation in Example~\ref{ex:motiv-dsl}. We can specify the size
  of the resulting circuit, as follows:
  \[
      n:\inttype\Vdash
      \left\{n\geq 0\right\}
      \texttt{main}(n)
      \left\{\texttt{size}(\texttt{result}) = n \right\}.
  \]
\end{example}

\subsection{Specialized libraries}
\label{sec:spec-lib}
For \qbrick we also derived more specialized, non-trivial libraries of
lemmas. In particular:

\smallskip
\noindent
\textit{Subclasses of circuits.} Shortcuts for reasoning on specific
  classes of circuits.  Among them are, in particular, the circuits
  with an \hops\ of range (1). We call them \emph{flat} circuits. Flat
  circuits enjoy a number of simplifying properties that are exposed
  in Section~\ref{app:sec:flat}. Section~\ref{app:sec:flat} also introduce a
  further simplified subclass of circuits, called \emph{diagonal}
  circuits.

\smallskip
\noindent
\textit{Eigenvectors and eigenvalues.} One of the common need in the
  specification of quantum programs (and in particular in the case of
  QPE) is the need for asserting that a particular ket
  $k$ is a eigenvector of the unitary map described by some circuit
  $c$, with eigenvalue the complex number $\rho$.
  In \qbrickSPEC we define the relation
  \(
    \texttt{eigen}(c,k,\rho)
  \)
  as $\texttt{circ\_apply}(c,k)= \rho *_k k$

\smallskip
\noindent
\textit{Probabilities.} A quantum program is usually a probabilistic
  program: it returns the desired result with a probability that
  depends on the problem parameters (number of iteration, structure of
  the problem, etc). If one aims at fully specifying such programs,
  this probability therefore needs to be expressible as
  post-condition.

  As recalled in Section~\ref{quantum-data},
  the probability of obtaining a
  result by a measurement is correlated with the amplitudes of the
  corresponding ket-basis vectors in the quantum state of the
  memory. In \qbrickSPEC we define a relation
  \(
  \texttt{proba\_partial\_measure} :
  \circtype\times\ket\times\bitvector \to \real
  \)
  meaning that when the input circuit is applied to the input ket,
  if we were to measure the result the probability of obtaining the
  given vector would be the result of the function. We define it as
  \(
  \texttt{proba\_partial\_measure}(c,k,x) =
  (\texttt{abs}(\getket(\texttt{circ\_apply}(c,k),x)))^2.
  \)
  In the specifications of Grover, QPE and Shor-OF we define specialized
  functions based on \texttt{proba\_partial\_measure} specifying the
  constraints specific to each algorithm.
  
\smallskip
\noindent
\textit{Algebraic operations on operators.}  In the course of the proof
  of quantum specification, it is sometimes useful to be able to
  manipulate algebraic expressions containing building blocks such as
  \emph{rotations} or \emph{projectors}. In \qbrickSPEC, with the use of
  HOPS it is possible to define such objects as macros, and then prove
  algebraic equalities between them, that we then add as auxiliary
  lemmas to be used by the automated provers. The use of HOPS gives us
  the possibility to specify ---and prove--- equalities parametrized
  by problem instances. This is for instance used extensively in the
  proof of the Grover specification.

\section{Implementation}
\label{sec:implem}

The \qbrick\ framework described so far is implemented as a DSL embedded inside the Why3 
deductive verification
platform~\cite{bobot:hal-00790310,filliatre2013why3},
written in the WhyML programming language.\footnote{\it  Implementation statistics and description of mathematical libraries can be found  in Appendix, Section \ref{app:sec:implem}.}
This allows us to benefit from several strengths of Why3, such as
efficient code extraction toward Ocaml, generation of proof obligations (to implement the \hqhl\ mechanism) 
and access to several proof means. 
Indeed, a  dedicated interface enables direct access to proof obligations and 
 either send them  to a set  of automatic SMT-solvers (CBC4, Alt-Ergo,
Z3, etc.), or enter a number of interactive proof transformation
commands (calls for lemmas or hypotheses, term substitutions, etc.) or
even send them towards  proof assistants (Coq, Isabelle/HOL) ---we do not use this option in our case-studies. 
The development itself  counts 17,000+ lines
of code, including  400+ definitions and 1700+ lemmas, all proved
within Why3. 
Most of the development concerns the  (verified) mathematical libraries (14,000+ loc for a total of 17,000+
loc ---see Table~\ref{gen-data} in Appendix, Section
\ref{app:sec:implem}).
They    cover the mathematical structures at stake in quantum computing (complex, Kronecker product, bit-vectors, etc.), together with a
{\it formally verified collection} of mathematical results concerning
them.  
They count  14,000+ lines
of code: 300+ definitions, 1600+ lemmas and 32 axiomatic definitions.

\section{Case studies and experimental evaluation}
\label{sec:xp}
Thanks to our implementation of \qbrick, we could develop and prove 
parametric implementations of several emblematic quantum algorithms,
namely 
Grover search, QFT, QPE and Shor-OF. We also implemented Deutsch-Jozsa  for the sake of comparison with prior work. 

\subsection{Case studies}
\label{sec:xp:cases}
Before discussing the experimental evaluation in Section~\ref{sec:xp:stats}, let us first introduce our implementations of
QPE, Shor-OF and Grover  algorithms\footnote{{\it More details can be found in Appendix, Section \ref{app:sec:xp:cases}.}}.

\paragraph{Quantum Phase Estimation (QPE)} Developed by
\citet{kitaev1995quantum,cleve1998quantum}, this
procedure inputs a unitary operator $U$ and an eigenvector
$\ket{v}{}$ of $U$ and finds the eigenvalue $e^{2\pi i \Phi_v}$
associated with $\ket{v}{}$. It
is a central piece in many emblematic algorithms, such as quantum
simulation~\cite{georgescu2014quantum} or HHL
algorithm~\cite{harrow2009quantum} -- resolution of linear systems of
equations in time \textsc{PolyLog}. 
We implemented two different versions of QPE:

\begin{itemize}
  \item In the first case (\emph{core case}) we assume that
$\Phi_v$ admits a binary writing with $n$ bits.  Then there is
$\varphi_v\in \tofset{2^n}$ such that
$\Phi_v = \frac{\varphi_v}{2^n}$, and the eigenvalue associated with
$\ket{v}{}$ is $e^{2\pi i \Phi_v}$ (also written $\omega_n^{\varphi_v}$).
The goal is to seek this value
$\varphi_v$ and the algorithm deterministically outputs this value.
\item In the second version  (\emph{general case}), no assumption is made over 
  $\Phi_v$  which can take any real value such that   $0\leq\Phi <1$.
  The goal is to seek the value $k\in\tofset{2^n}$ that minimizes the distance $\Phi - \frac{k}{2^n}$ (modulo $1$). The output of the algorithm is non deterministic. The proved specification is that it outputs $k$ with probability at least $\frac{4}{\pi^2}$.
  \end{itemize}

\paragraph{\citet{shor1994algorithms}'s algorithm} Certainly the most emblematic of all quantum algorithms,
%
it implements integer factoring by polynomial reduction to Order
Finding.
%
%
The quantum circuit ---the quantum part of the algorithm, the one we
are certifying--- consists in an application of QPE to the unitary operator
$U : \ket{y}{} \to \ket{x\cdot y\,\mod\, N}{}$. 
We developed a certified \emph{concrete} implementation following  the implementation proposed in~\cite{beauregard2002circuit} ---a reference in term of complexity. 
Besides proving the functional requirements of the order-finding problem (including achieving it with probability  $O(1)$), we also prove some appreciable \textbf{\textit{complexity results}}: our implementation requires applying $O(\textbf{Log}_2(N))^4$ quantum gates and requires $4\textbf{Log}_2(N)+2$ qubits.\footnote{A further refinement is possible by, as indicated in~\cite{beauregard2002circuit}, using an hybrid version of the Quantum Fourier Transform, but it  would require adding effective measure operation and classical control to \qbrick.} 

\paragraph{Grover search algorithm} Developed by \citet{DBLP:conf/stoc/Grover96}, in its
  first version, it enables to find a distinguished element in an
  unstructured data base with a quadratic speedup with regards to the
  best known equivalent classical approaches. It was then generalized
  by \citet{boyer1998tight} to the similar problem with a parameterized number $k$
of distinguished elements.
We implemented this most general case, without any restriction over $k$.

\subsection{Experimental evaluation}
\label{sec:xp:stats}

 In addition to QPE, Shor-OF and Grover algorithm, we also consider 
implementations of the Quantum Fourier Transform (QFT) and the Deutsch-Josza
algorithm. 

\smallskip 

{\it We have been successful on all the cases, providing the first
  verified parametric implementation of  QFT, QPE, Shor-OF and Grover}. 

\smallskip 

Different metrics about our formal developments are reported in Table~\ref{data-cs}\footnote{Experiments were run on Linux, on a PC equipped with an Intel(R)
Core(TM) i7-7820HQ 2.90GHz and 15 GB RAM. We used Why3 version 1.2.0
with solvers Alt-Ergo-2.2.0, CVC 3-2.4.1, CVC4-1.0, Z3-4.4.1.}:  lines of decorated code, lemmas, proof obligations (PO), automatically proven PO (within
time limit 5 seconds), remaining POs and the number of interactive commands we
entered to discharge them -- this last metric is detailed in  Table~\ref{app:com-stats} in Appendix, Section \ref{app:sec:xp:stats}.
%
%
%


Note that metrics for each implementation strictly concern the code that is {\it proper} to it (eg., QPE contains calls to QFT but QPE line in Table~\ref{data-cs} does not include the QFT implementation. Similarly, Shor-OF calls both QPE and QFT) -- the whole Shor-OF development is reported in the ``Shor-OF full'' row and row ``Total'' sums it with lines DJ and Grover.

\begin{table}[ht]
  \begin{center}
    \begin{tabular}{|l|c|c|c|c|c|c|c|c|c|}
      \cline{2-10}
      \rowcolor{gray!50}
      \multicolumn{1}{c|}{\cellcolor{white}
      }     &\#LoC & \#Extr. &\#Def.&\#Lem. & \#POs & \multicolumn{3}{c|}{Automation} &  \#Cmd\\
      \cline{7-9}
      \rowcolor{gray!50}
      \multicolumn{1}{c|}{\cellcolor{white}
      }     & &  && &  &\# Aut.& \% Aut.& Non aut. &  \\
      
      \hline
      \rowcolor{gray!25}
      DJ &53&11&2&1&72&61&>84\%&11&39\\
      Grover &416&42&9&20&698&668&>95\%&30&167\\
      \rowcolor{gray!25}
      QFT &65&18&3&0&62&53&>85\%&9&37\\
      QPE &319&33&6&15&423&396&>93\%&27&155\\
      \rowcolor{gray!25}
      Shor-OF &809&132&34&1&1522&1456&>95\%&66&264\\

      Shor-OF (full)    &  1193  & 183   & 43  & 16  & 2007  &  1905  & 95\%  & 102   & 456  \\ 

      \hline 
      \hline 

      \rowcolor{gray!50}
      Total    &  1662  & 236   & 54  & 37  & 2777  &  2634  & >94\%  & 143   & 662  \\ 

      \hline    \end{tabular}
    
    \#LoC.: lines of decorated code  --- \# Extr.: lines of extracted code (OCaml) \linebreak
    \#Aut.: automatically proven POs  --- \#Cmd: interactive commands

    \smallskip 
    \caption{Implementation \& verification for case studies with \qbrick} 
    \label{data-cs}
  \end{center}
\end{table}

\paragraph{Result}   {\it 
\qbricks did allow us to implement and verify in a parametric 
manner the Shor-OF, QPE and Grover algorithms, at a rather smooth cost and with high proof automation (94\% in average, 95\% for full Shor-OF). 
}

%
%
%


\smallskip 

Interestingly, it should be noted that while Grover relies on mechanisms and arguments  
significantly different from those of QPE or Shor-OF, we were able to implement, specify and prove it 
without adding anything new to \qbrick\ -- demonstrating the genericity of our approach. 

\smallskip 

A last interesting observation concerns the time we spent on the different implementations of these case studies, given in chronological order: 
1.5 person.year to build \qbrick\ and be able to implement, specify  and prove QFT and a restricted case of  QPE --  1 person.day for DJ -- 2 person.weeks for each of QPE (full case) and Grover -- 
and finally 10 person.days to add Shor-OF on top of QPE.   
 %
%
%
Considering how significantly the initial development effort factorizes over further developments,
we have good reasons to believe that \qbricks\ provides a generic and convenient environment for specifying, developing and proving quantum programs in 
a reasonably fast and easy way.

\subsection{Comparison with prior works}

Table~\ref{app:com} in Appendix (Section \ref{app:sec:xp:stats}) provides a summary of the comparison with these preexisting formal
verification efforts for quantum programs.

\paragraph{Regular path-sums} \citet{amy2018towards,dblp:phd/basesearch/amy19} uses path sums  for the verification of several circuits 
of complexity similar to that of QFT (QFT, Hidden shift, Toffoli generalized, etc). Yet, these experiments consider  fixed circuits (up to 100 qubits) and the technique cannot be applied to parametric families of circuits or circuit-building languages. 
 


\paragraph{QHL} \citet{10.1007/978-3-030-25543-5_12} report about  the parametric verification of 
Grover search algorithm, on a {\it restricted case} 
\footnote{Given a predicate value with $k$ true value in $\tofset{2^n}$, Grover algorithm outputs one of these true values with good probability. The case in~\cite[p.\,232]{10.1007/978-3-030-25543-5_12}  concerns cases where  $k = 2^j$ for a given integer $j$.}
and in the {\it high-level} algorithm description  formalism of QHL -- especially QHL has no notion of circuit. 
Interestingly, the generic verification of our low-level implementation  is  5$\times$ smaller than
theirs: 3184 total lines vs  583 total lines (416 lines + 167 commands).    


\paragraph{Qwire} Finally, ~\citet{dblp:journals/corr/abs-1904-06319} have presented in their 2019 preprint
a fully verified parametric (circuit-building) implementation of the Deutsch-Josza algorithm in Coq.  This implementation is given two independent correction proofs.
As already pointed out, DJ is  more a standard teaching example (with no practical application) than a representative quantum algorithm -- roughly speaking, 10$\times$ smaller than Grover and 20$\times$ smaller than (full) Shor-OF.   
Moreover, our development requires significantly less efforts than
theirs: less code (53 vs  74 lines) and 5.5$\times$ less proof commands (39 vs 222) for their
textbook-style version -- the one closest to our own
version\footnote{We still have a strong gain over their optimized
  encoding 
   in terms of proof commands (39
  vs 112), while roughly equivalent for code (53 vs 59 lines).}.


\subsection{Summary}

It appears that \qbricks provides a powerful framework for the development of formally verified parametric implementations of realistic quantum algorithms, setting up a new 
baseline in terms of quantum program formal proofs  and achieving  high scores in terms of proof automation.

\section{Related works}
\label{sec:related}

\paragraph{Formal verification of quantum circuits}
In the last couple of years, several efforts have been led toward introducing
formal methods in quantum programming.
Prior efforts regarding quantum circuit verification
\cite{10.1007/978-3-540-70545-1_51,%
  10.1007/978-3-030-25543-5_12,%
  Ying:2014:MLP:2648783.2629680,%
  DBLP:journals/corr/BoenderKN15,%
  paykin2017qwire,%
  DBLP:journals/corr/abs-1803-00699,%
  randthesis,%
  dblp:phd/basesearch/amy19,%
  amy2018towards} have been described throughout the paper, especially
in Sections~\ref{intro}, \ref{motiv} and \ref{sec:xp}. 

\smallskip 

{\it We build on top of these seminal works and propose the first
  deductive verification framework  for circuit-building quantum programs -- the current consensus in quantum programming languages,  featuring
  clear separation between code \& proof, parametric  
  specifications \& proofs and a high degree of proof automation.} 

\smallskip 

Especially, our technique is more automated than those based on interactive proving \cite{dblp:journals/corr/abs-1904-06319,DBLP:journals/corr/abs-1803-00699} thanks to the new \hqhl\ Hoare-style deduction system,  
borrows and extends the path sum representation \cite{amy2018towards}  to the parametric case, 
and do consider a circuit-building language rather than a high-level algorithm description language \cite{10.1007/978-3-030-25543-5_12}.



\paragraph{Quantum Languages  and Deductive Verification}
\citet{10.1007/978-3-030-25543-5_12} introduce Quantum Hoare Logic for high-level description of quantum algorithms. 
QHL and our own \hqhl\ are very different, as the underlying formalisms are very different. Combining the two approaches on a combined formalism is an exciting 
research direction.


\paragraph{Optimized compilation of circuits}
Formal methods and other program analysis techniques are also used in
quantum compilation, in order to build highly optimized 
circuits~\cite{parentrs17,%
  bhattacharjeesd19,%
  haaswijk2019,%
  DBLP:conf/cav/AmyRS17,%
  DBLP:journals/corr/abs-1803-01022,%
  DBLP:journals/corr/abs-1901-10118,
  hietala19:verif_optim_quant_circuit}.
This is a crucial current research area. Indeed, the quantum hardware
available in the near future is expected to be highly constrained in
terms of qubits, connectivity and quality: the so-called NISQ
era~\cite{nisq,tannuq19,paler17}.

Especially, the ZX-calculus~\cite{coecke2017picturing} 
represents quantum circuits 
by diagrams, amenable to automatic simplification through dedicated rewriting rules. This
framework leads to a graphical proof
assistant~\cite{kissinger2015quantomatic} geared at certifying the semantic
equivalence between circuit diagrams, with application to circuit
equivalence checking and certified circuit compilation and optimization
\cite{fagan2018,beaudrap2019,kissinger2019}. Yet, ZX-calculus is restricted to fixed circuits 
and cannot handle parametrized families of circuits or circuit-building languages.  

\paragraph{Other quantum applications of formal methods} 
Huang \textit{et
  al.} \cite{DBLP:journals/corr/abs-1811-05447,huang:2019:sav:3307650.3322213}
proposes a ``runtime-monitoring like'' verification method for quantum
circuits, with an annotation language restricted to structural
properties of interest (e.g., superposition or entanglement). 
Verification of these
assertions is led by statistical testing instead of formal proofs. 
The recent Silq language~\cite{silq} also represents an advance in the way
toward  automation in quantum programming. It
automatizes uncomputation operations, enabling the programmer to abstract from low level implementation details. 
Another line of research is concerned with the
development of specialized type systems for quantum programming
languages. In particular, frameworks based on linear
logic~\cite{selinger05lambda,ross2015algebraic,lago2010quantum} and
dependent types~\cite{qpcf,paykin2017qwire} have been developed  to
tackle the non-duplicability of qubits and the constraints on the
structure of circuits. 
Finally, formal methods are also at stake for the verification of protocols
using quantum information, such as cryptographic 
protocols~\cite{nagarajan2002formal,%
  gheorghiu2019verification,%
  broadbent2015verify,%
  mahadev2018classical,%
  davidson2012formal}.


\section{Conclusion}
\label{conclusion}

We have presented \qbrick, the first  
formal verification environment for   circuit-building quantum programs --  featuring clear separation
between code and proof, parametric  specifications and proofs, high
degree of proof automation and allowing to encode quantum programs in
a natural way. 
We build on best practices of formal verification for the classical
case and tailor them to the quantum case. Especially, while relying on the general framework of deductive verification, 
we  design
a  new domain-specific
  circuit-building language \qbrickDSL\  for quantum programs 
together with a new logical specification language 
  \qbrickSPEC\ 
and the novel  \hqhllong. 
Finally, we introduce   and intensively build upon  
{\it higher-order path sums} (HOPS), a symbolic representation  convenient  for both specification (parametricity) and automation (closure).  

In order to demonstrate the interest of the full framework, we develop  the first 
verified parametric implementations of 
three famous non-trivial quantum algorithms (namely Shor-OF, QPE and Grover search),  
proving by fact that applying formal verification to realistic (circuit-building) quantum
programs is possible and should be further developed.

\newpage
\bibliography{main}
\newpage

\appendix


\section{Matrix semantics for gates}
The following Table gives the matrix semantics for gates that are used in this article, bringing complements to 
Section~\ref{syntax-qbdsl}.
\begin{table}[!h]
  \begin{minipage}{1.0\linewidth}
    \centering
    \begin{tabular}{|c|c|c|c|c|c|}
      \hline 
      $\texttt{CNOT}$
      &$\texttt{SWAP}$ 
      &$\texttt{ID}$ 
      &$\texttt{H}$ 
      &$\texttt{Ph}(n)$ 
      &$\texttt{R}_z(n)$
      \\
      \hline
      $\begin{pmatrix} 1&0&0&0 \\ 0&1&0&0\\ 0&0&0&1\\
        0&0&1&0  \end{pmatrix}$
      &
      $\begin{pmatrix} 1&0&0&0 \\ 0&0&1&0\\ 0&1&0&0\\
        0&0&0&1  \end{pmatrix}$
      &
      $\begin{pmatrix}
        1 &0 \\0 &1
      \end{pmatrix}$
      &
      $\frac{1}{\sqrt{2}} 
      \begin{pmatrix}
        1 &1 \\ 1 &-1
      \end{pmatrix}$
      &$\begin{pmatrix} e^{\frac{2i\pi}{2^n}} &  0 \\ 0 &
        e^{\frac{2i\pi}{2^n}}  \end{pmatrix}$ 
      &$\begin{pmatrix} e^{\frac{-2i\pi}{2^n}} & 0 \\ 0 &
        e^{\frac{2i\pi}{2^n}} \end{pmatrix}$
      \\\hline
    \end{tabular}
  \end{minipage}
  \caption{Matrix semantics for gates}
  \label{matgates}
\end{table}

\section{Complete deduction rules for relation $\rcorrect{\_}{\_}$} 
\label{app:sec:bv-ket}

In this section we first introduce some additional definitions and notations about
bit-vectors and kets, completing Section \ref{syntax-qbrick-spec}. Here and in the following we commit the abuse to treat boolean values \texttt{tt} and \texttt{ff} as, respectively, $0$ and $1$, omitting the call for the required cast function.

\subsection{Additional notations and definitions for bit-vectors and kets}
\paragraph{Bit vectors}
Type \bitvector\
is an opaque type with two operators
\[\begin{array}{rcl}
    \bvlength&:& \bitvector \to{} \inttype\\
    \getbv&:& \bitvector \to{} \inttype\to{} \inttype\\
  \end{array}\]

and invariants
\[
  \begin{array}{rcl}
    \Gamma, \vec{x}: \bitvector &\vdash&\ahoare{\top}{\bvlength(\vec{x})}{\result\geq 0}\\
    \Gamma, \vec{x}: \bitvector, i:\inttype&\vdash&\ahoare{i\in \tofset{\bvlength(\vec{x})}}{\getbv(\vec{x},i)}{\result \in \tofset{2}}\end{array}\]

We usually abbreviate $\getbv(\vec{x},i)$ by    $\vec{x}_i$. 
Type \bitvector\  is given with constructors \texttt{bv\_set} and  \texttt{bv\_cst}
Let $f$ be an integer function with values in $\{0,1\}$ and let $n$ be a positive integer.
Then we define $\makebv(f,n)$ as the sequence $(f(0))(f(1)) \dots(f(n-1))$. Formally, let us first introduce predicate \bin:

\[
  \Gamma, f: \inttype\to{}\inttype, i:\inttype\vdash_\defin\ahoare{i\in\tofset{s}}{\bin_s(f)}{\result = f(i)\in \tofset{2}}\]
then:
\begin{multline*}\Gamma, f \inttype\to{}\inttype, i:\inttype\vdash\\
  \ahoare{i\in\tofset{s},\bin_s(f)}{\makebv(f,s)}{\bvlength(\result) = s \wedge \result_i = f(i)}
\end{multline*}
We also define concatenation of bit vectors $\vec{x}$ and $\vec{y}$. It satisfies:
\[
  \begin{array}{rcl}
    \multicolumn{3}{l}{\Gamma,\vec{x},\vec{y},i:\inttype : \bitvector\vdash}\\
    \{\top\}&\concat(\vec{x},\vec{y})&
                                       \left\{\begin{array}{l}\bvlength(\result) = \\ \quad\bvlength(\vec{x}) +\bvlength(\vec{y})\end{array}\right\}\\
    \{i\in\tofset{\bvlength(\vec{x})}\}&\concat(\vec{x},\vec{y})&
                                                                  \{\result_i = \vec{x}_i\}\\
    \left\{\begin{array}{l}i\in \rrbracket\bvlength(\vec{x}),\\ \quad \bvlength(\vec{x})+\bvlength(\vec{y})\rrbracket \end{array}\right\}&\concat(\vec{x},\vec{y})&
                                                                                                         \{\result_i = \vec{y}_{i-\bvlength(\vec{x})}\}
  \end{array}
\]

Let $\vec{x}$ be a bit vector and let $i,j$  integers $\llbracket 0 ,\bvlength(\vec{x})$ such that $i<j$, then we introduce sections of $\vec{x}$  in, respectively, $\llbracket 0, i\llbracket$,  $\llbracket i+1, \bvlength{\vec{x}}\llbracket$ and  $\llbracket i+1, j\llbracket$. They satisfy the following rules:

\begin{multline*}
  \Gamma,\vec{x}:\bitvector, i,j,k: \inttype\vdash
  \\
  \begin{array}{rcl}
    \Big\{\begin{array}{l}j\in \tofset{\bvlength(\vec{x})}\\ k\in \tofset{\bvlength(\vec{x} - i)} \end{array}\Big\}& \vec{x}_{\upharpoonleft i}&\Big\{\begin{array}{l}\bvlength(\result) = \vec{x} - i\\
                                                                                                                                                        \result_k = \vec{x}_{k+i} \end{array}\Big\},\\
    \Big\{\begin{array}{l}j\in \tofset{\bvlength(\vec{x})}\\ k\in \tofset{j} \end{array}\Big\}& \vec{x}_{j\upharpoonright }&\Big\{\begin{array}{l}\bvlength(\result) = j,\\
                                                                                                                                    \result_k = \vec{x}_k \end{array}\Big\}\\
    \Bigg\{\begin{array}{l}i\in \tofset{\bvlength(\vec{x})}\\j\in \tofsett{i}{\bvlength(\vec{x})}\\ k\in \tofset{\bvlength(j - i)} \end{array}\Bigg\}& \vec{x}_{\upharpoonleft i ; j \upharpoonright}&\Big\{\begin{array}{l}\bvlength(\result) = i - j\\
                                                                                                                                                                                                              \result_k = \vec{x}_{k+j} \end{array}\Big\}
  \end{array}
\end{multline*}

At last we introduce two functions \bvtoint and \inttobv, from bit vectors to integers and back, enabling the interpretation of a bit vector as a binary integer :

\[\begin{array}{lll}
    \Gamma, x:\bitvector &\vdash& \\\multicolumn{3}{c}{ \ahoare{\top}{\bvtoint(\vec{x})}{\result = \sum_{k=0}^{\bvlength(\vec{x})-1} \vec{x}_k * 2^{\bvlength(\vec{x})-k-1}}}\\
    \Gamma, x:\bitvector, k,n,i : \inttype &\vdash&\\\multicolumn{3}{c}{ \ahoareBigr{0\leq n,k\in\tofset{2^n},i\in \tofset{n}}{\inttobv(k,n)}{\begin{array}{l}\bvlength(\result) = n\\\result_i =  \div\ (\mod\ k\ 2^{n-i+1})\ 2 \end{array}}}

  \end{array}
\]

\subsection{Kets}
Type \kettype\ is an opaque type, similar as bit-vectors with complex values.
Kets have a number of cells equal to $2^n$, for a positive integer $n$ called its logarithmic length (also called, for simplicity, its length).  They are defined without norm restriction. Usually, we designate kets using the standard Dirac notation $\ket{v}{n}$, with $n$ being the logarithmic length of $\ket{v}{}$. We drop this $n$ index when unnecessary.

Dirac notation is also used as notation for functions casting integers and bit vectors to kets : let $k,n$ be integers such that $k\in \llbracket 0,2^n\llbracket$ and let $\vec{x}$ be a bit vector. Then:
\[
  \begin{array}{rcrcl}
    \multicolumn{3}{l}{    \Gamma, k,n,i:\inttype \vdash}\\    
    &&\{0\leq n, k\in \tofset{2^n}\}&\ket{k}{n}&\{\ketlength(\result) = n\}\\
    &&\{0\leq n, k\in \tofset{2^n}\}&\ket{k}{n}&\{\getket(\result,k) = 1\}\\
    &&\{0\leq n, k\in \tofset{2^n} i\in \tofset{2^n} i<>k\}&\ket{k}{n}&\{\getket(\result,i) = 0\}\\
    
  \end{array}
\]

Functions $+, -,  \otimes$ are introduced and used the usual way over kets :
{\small
\[
  \begin{array}{rcl}
    \multicolumn{3}{l}{    \Gamma,\ket{v}{},\ket{w}{} :\kettype, i :\inttype \vdash}\\    
    \{\top\} &\ket{v}{}\otimes\ket{w}{}&\Big\{\begin{array}{l}\ketlength(\result) = \\\qquad\ketlength(\ket{v}{}) + \ketlength(\ket{w}{})\end{array}\Big\}\\
    \Big\{\begin{array}{l}i\in \llbracket 0, \ketlength(\ket{v}{})\ +\\\quad \ketlength(\ket{w}{})\rrbracket\end{array} \Big\} &\ket{v}{}\otimes\ket{w}{}&\Big\{\begin{array}{l}\getket(\result,i) =\\\qquad  \getket(\ket{v}{}, \div\ i\ 2^{n_2}) * \getket(\ket{w}{}, \mod\ i\ 2^{n_2}\end{array}\Big\}\\
             &&\\
    \{\top\} &\ket{v}{} + \ket{w}{}&\{\ketlength(\result) = \ketlength(\ket{v}{})\}\\
    \Bigg\{\begin{array}{l}i\in \tofset{\ketlength(\ket{v}{})} \\ \ketlength(\ket{v}{}) = \\\qquad\ketlength(\ket{w}{}) \end{array} \Bigg\} &\ket{v}{} + \ket{w}{}&\{\begin{array}{l}\getket(\result,i) = \getket(\ket{v}{}, i) + \getket(\ket{w}{}, i)\end{array}\}\\
             &&\\
    \{\top\} &\ket{v}{} - \ket{w}{}&\{\ketlength(\result) = \ketlength(\ket{v}{})\}\\
    \Bigg\{\begin{array}{l}i\in \tofset{\ketlength(\ket{v}{})} \\ \ketlength(\ket{v}{}) = \\\qquad\ketlength(\ket{w}{}) \end{array} \Bigg\} &\ket{v}{} - \ket{w}{}&\{\begin{array}{l}\getket(\result,i) = \getket(\ket{v}{}, i) - \getket(\ket{w}{}, i)\end{array}\}
  \end{array}
\]
}

\section{Deductive rules for relation $\rcorrect{\_}{\_}$ }
\label{app:sec:rcorrect}

In this section, we complete the introduction of deductive rules for
relation $\rcorrect{\_}{\_}$ from Section~\ref{rcorrect-spec}.
We first give rules for gates, then rules for composed circuits.

\subsection{Gates \hops\ characterization predicates}
\label{gates-char}
For any gate in the language, we first introduce a specific predicate characterizing an \hops\ for this gate. Predicate \hopshad,
for Hadamard gate, was given in Section~\ref{rcorrect-spec}, below are similar predicates for the additional gates;

\begin{description}

  \item[\texttt{ID}]\begin{multline*}
  \hopsid(h,\vec{x},\vec{y}) :=   \bvlength(\vec{x}) =  1 \wedge 
  \bvlength(\vec{y}) =  0) \to   \\       \hwidth (h) = 1 \wedge
  \hrange (h) = 0 \wedge\\
  \hangle(h)(\vec{x},\vec{y}) = 1 \wedge
  \hket(h)(\vec{x},\vec{y}) = \vec{x} 
\end{multline*}

  \item[\texttt{SWAP}]\begin{multline*}
  \hopsswap(h,\vec{x},\vec{y}) :=   \bvlength(\vec{x}) =  2 \wedge 
  \bvlength(\vec{y}) =  0) \to   \\       \hwidth (h) = 2 \wedge
  \hrange (h) = 0 \wedge\\
  \hangle(h)(\vec{x},\vec{y}) = 1 \wedge
  \hket(h)(\vec{x},\vec{y}) = \texttt{bv\_set}(\texttt{bv\_get}(\vec{x},0 ), 0,\texttt{bv\_get}(\vec{x},1 ))) 
\end{multline*}

  \item[\texttt{Ph}]\begin{multline*}
  \hopsphase(h,n,\vec{x},\vec{y}) := 0\leq n \wedge\  \bvlength(\vec{x}) =  1 \wedge 
  \bvlength(\vec{y}) =  0) \to   \\       \hwidth (h) = 1 \wedge
  \hrange (h) = 0 \wedge\\
  \hangle(h)(\vec{x},\vec{y}) = e^{\frac{2\pi i}{2^n}} \wedge
  \hket(h)(\vec{x},\vec{y}) = \vec{x} 
\end{multline*}

  \item[\texttt{Rz}]\begin{multline*}
  \hopsrz(h,n,\vec{x},\vec{y}) := 0\leq n \wedge\   \bvlength(\vec{x}) =  1 \wedge 
  \bvlength(\vec{y}) =  0) \to   \\       \hwidth (h) = 1 \wedge
  \hrange (h) = 0 \wedge\\
  \hangle(h)(\vec{x},\vec{y}) = e^{(-1)^{1-\vec{x}_0}* \frac{2\pi i}{2^n}} \wedge
  \hket(h)(\vec{x},\vec{y}) = \vec{x} 
\end{multline*}

  \item[\texttt{CNOT}]\begin{multline*}
  \hopscnot(h,\vec{x},\vec{y}) := (  \bvlength(\vec{x}) =  2 \wedge 
  \bvlength(\vec{y}) =  0) \to   \\       \hwidth (h) = 2 \wedge
  \hrange (h) = 0 \wedge\\
  \hangle(h)(\vec{x},\vec{y}) = 1 \wedge
  \hket(h)(\vec{x},\vec{y}) = \vec{x}_0 \cdot ((1- \vec{x}_0) \vec{x}_1 +\vec{x}_0 (1-\vec{x}_1)) 
\end{multline*}

\end{description}
\subsection{Combining circuits}  
We also introduce binary and ternary relations between \hops, so as to characterize the conditions for relation 
$\rcorrect{\_}{\_}$ in combined circuits. This induces one \hops\ relation per circuit combinator:

  \begin{description}

  \item[Sequence composition]
    \[\begin{array}{c}
        \multicolumn{1}{l}{\hopssequence(h_1,h_2,h_3,\vec{x},\vec{y}):=}\\
        \left(\begin{array}{c}\bvlength(\vec{x}) =  \hwidth (h_1) ,\\ \bvlength(\vec{y}) =  \hrange (h_1) +  \hrange (h_2) \wedge\\   \hwidth (h_1) = \hwidth (h_2)  \wedge\\
              \end{array} \right)\\
        \to\\
        \left(
        \begin{array}{c}
          \hwidth (h_3) = \hwidth (h_1)  \wedge\\
          \hrange (h_3) = \hrange (h_1) + \hrange(h_2) \wedge \\
          \big(\hangle(h_3)(\vec{x},\vec{y}) =
          \hangle(h_1)(\vec{x},\vec{y}_{ \texttt{hops\_range}(h_2)\upharpoonright}) +_a\\
          \hangle(h_2) (\hket(h_1)(\vec{x},\vec{y}_{ \hrange(h_2)\upharpoonright}),\vec{y}_{\upharpoonleft \texttt{hops\_range}(h_2)})\big) \wedge\\ 
          \big(\hket(h_3)(\vec{x},\vec{y}) = 
          \hket(h_2)( \hket(h_1)(\vec{x},\vec{y}_{ \texttt{hops\_range}(h_2)\upharpoonright}),\vec{y}_{\upharpoonleft \texttt{hops\_range}(h_2)})\big)
        \end{array}
        \right)
      \end{array}\]
    
  \item[Parallel composition] 
    
    \[
      \begin{array}{c}
        \multicolumn{1}{l}{\hopspar(h_1,h_2,h_3,\vec{x},\vec{y}) := }\\
        \left(
        \begin{array}{c}
          \bvlength(\vec{x}) =  \hwidth (h_1)+  \hwidth (h_2) \wedge \\
          \bvlength(\vec{y}) =  \hrange (h_1) + \hrange (h_2)
        \end{array}
        \right)\\
        \to\\
        \left(
        \begin{array}{c}
          \hwidth (h_3) = \hwidth (h_1) + \hwidth(h_2) \wedge\\
          \hrange (h_3) = \hrange (h_1) + \hrange(h_2) \wedge\\
          \big(\hangle(h_3)(\vec{x},\vec{y}) =
          \texttt{hops\_angle}(h_1)(\vec{x}_{ \texttt{hops\_width}(h_2)\upharpoonright},\vec{y}_{ \texttt{hops\_range}(h_2)\upharpoonright})\ +_a\\
          \hangle(h_2) (\vec{x}_{\upharpoonleft \texttt{hops\_width}(h_2)},\vec{y}_{\upharpoonleft \texttt{hops\_range}(h_2)} )\big) \wedge\\ 
          \big(\hket(h_3)(\vec{x},\vec{y}) = 
          \hket(h_1)(\vec{x}_{ \hwidth(h_2)\upharpoonright},\vec{y}_{ \texttt{hops\_range}(h_2)\upharpoonright}) \otimes\\
          \hket (h_2) (\vec{x}_{\upharpoonleft \texttt{hops\_width}(h_2)},\vec{y}_{\upharpoonleft \texttt{hops\_range}(h_2)} ) 
        \end{array}
        \right)
      \end{array}\]

  \item[Control composition]
    \[\begin{array}{c}
        \multicolumn{1}{l}{\hopsctl (h_1,h_2,c,\vec{x},\vec{y}) :=}\\
        \left(
        \begin{array}{c}
          \bvlength(\vec{x}) =  \hwidth(h_1) +1  \wedge \bvlength(\vec{y}) =  \hrange (h_1)
        \end{array}
        \right)
        \\
        \to\\
        \left(
        \begin{array}{c}
          \hwidth (h_2) = \hwidth(h_1) +1   \wedge  \hrange (h_2) = \hrange (h_1)\  \wedge \\
          \big(\hangle(h_2)(\vec{x},\vec{y}) =
          \vec{x}_0 * \hangle(h_1)(\vec{x}_{\upharpoonleft 1 } ,\vec{y})+ (1-\vec{x}_0)\big)\  \wedge\\
          \big(\hket(h_2)(\vec{x},\vec{y}) = \textbf{ if } 
          \vec{x}_0 = 1 \textbf{ then } \texttt{concat} (\texttt{bv\_cst} (1, 1), \hket(h_1)(\vec{x}_{\upharpoonleft 1 } ,\vec{y})) \textbf{ else } \vec{x}
        \end{array}
        \right)
      \end{array}\]
    
  \item[Ancilla composition]
    \[\begin{array}{c}
        \multicolumn{1}{l}{\hopsanc(h_1,h_2,\vec{x},\vec{y}) :=}\\
        \left(
        \begin{array}{c}
          \{\hwidth (h_2) = \hwidth (h_1)-1\wedge\ \\
               \bvlength(\vec{x}) =  \hwidth (h_2) \wedge\ \bvlength(\vec{y}) =  \hrange (h_2)\wedge\\ 
          \vec{x}_{\hwidth (h_1-1)} = 0) \to  (\hket(h_1)(\vec{x},\vec{y}))_{\hwidth (h_1-1)} = 0
        \end{array}
        \right)
        \\
        \to\\
        \left(
        \begin{array}{c}
          \hwidth (h_2) = \hwidth(h_1) -1 \wedge    \hrange (h_2) = \hrange (h_1)\wedge\\
          
          \hangle(h_2)(\vec{x},\vec{y}) = \hket (h_1)(\concat(\vec{x},\texttt{bv\_cst} (1, 0)),\vec{y}) \wedge\\
          \concat(\hket(h_2)(\vec{x},\vec{y}), \texttt{bv\_cst} (1, 0)) = \hangle (h_1)(\concat(\vec{x},\texttt{bv\_cst} (1, 0)),\vec{y})
        \end{array}   \right) \end{array}\]

\end{description}

\subsection{Deduction rules}
As introduced in Section~\ref{rcorrect-spec}, the characterization
predicates introduced so far serve as hypotheses for compositional
deduction rules for relation $\rcorrect{}{}$. These rules are given in Table~\ref{tab:qbrickdsl-circuits}.

\floatstyle{boxed}
\restylefloat{figure}

\begin{figure}

      \begin{center}
    
    \begin{prooftree}
      \AxiomC{$\Gamma,\vec{x},\vec{y} : \bitvector \vdash \hopsid(h,\vec{x},\vec{y})$}
      \RightLabel{$\texttt{id}$}
      \UnaryInfC{$\Gamma \vdash  \rcorrect{h}{\texttt{ID}}$}
    \end{prooftree}
    
    \begin{prooftree}
      \AxiomC{$\Gamma,\vec{x},\vec{y} : \bitvector \vdash \hopsswap(h,\vec{x},\vec{y})$}
      \RightLabel{$\texttt{swap}$}
      \UnaryInfC{$\Gamma \vdash  \rcorrect{h}{\texttt{SWAP}}$}
    \end{prooftree}
    
    \begin{prooftree}
      \AxiomC{$\Gamma,\vec{x},\vec{y} : \bitvector \vdash \hopshad(h,\vec{x},\vec{y})$}
      \RightLabel{$\texttt{had}$}
      \UnaryInfC{$\Gamma \vdash  \rcorrect{h}{\texttt{H}}$}
    \end{prooftree}
    
    \begin{prooftree}\AxiomC{$\Gamma,\vec{x},\vec{y} : \bitvector \vdash \hopsphase(p,h,\vec{x},\vec{y})$}
      \RightLabel{$\texttt{phase}$}
      \UnaryInfC{$\Gamma \vdash  \rcorrect{h}{\texttt{Ph}(p)}$}
    \end{prooftree}
    
    \begin{prooftree}
      \AxiomC{$\Gamma,\vec{x},\vec{y} : \bitvector \vdash \hopsrz(p,h,\vec{x},\vec{y})$}
      \RightLabel{$\texttt{rz}$}
      \UnaryInfC{$\Gamma \vdash  \rcorrect{h}{\texttt{R}_z(p)}$}
    \end{prooftree}
    
    \begin{prooftree}\AxiomC{$\Gamma,\vec{x},\vec{y} : \bitvector \vdash \hopscnot(h,\vec{x},\vec{y})$}
      \RightLabel{$\texttt{cnot}$}
      \UnaryInfC{$\Gamma \vdash  \rcorrect{h}{\texttt{CNOT}}$}
    \end{prooftree}

    \begin{prooftree}
      \AxiomC{$\Gamma \vdash \rcorrect{h_1}{C_1}$}
      \AxiomC{$\Gamma \vdash \rcorrect{h_2}{C_2}$}
      \AxiomC{$\Gamma,\vec{x},\vec{y} : \bitvector \vdash \hopssequence(h_1,h_2,h_3, \vec{x},\vec{y})$}
      \RightLabel{$\texttt{seq}$}
      \TrinaryInfC{$\Gamma \vdash  \rcorrect{h_3}{\texttt{SEQ}(C_1,C_2)}$}
    \end{prooftree}

    \begin{prooftree}
      \AxiomC{$\Gamma \vdash \rcorrect{h_1}{C_1}$}
      \AxiomC{$\Gamma \vdash \rcorrect{h_2}{C_2}$}
      \AxiomC{$\Gamma,\vec{x},\vec{y} : \bitvector \vdash \hopspar(h_1,h_2,h_3, \vec{x},\vec{y})$}
      \RightLabel{$\texttt{par}$}
      \TrinaryInfC{$\Gamma \vdash \rcorrect{h_3}{\texttt{PAR}(C_1,C_2)}$}
    \end{prooftree}
    \begin{prooftree}
      \AxiomC{$\Gamma\vdash \rcorrect{h_1}{C}$}
      \AxiomC{$\Gamma,(\vec{x},\vec{y} : \bitvector) \vdash \hopsctl(h_1,h_2, \vec{x},\vec{y})$}
      \RightLabel{$\texttt{ctl}$}
      \BinaryInfC{$\Gamma \rcorrect{h_2}{\texttt{CTL}(C)}$}
    \end{prooftree}

    \begin{prooftree}
      \AxiomC{$\Gamma\vdash \rcorrect{h_1}{C}$}
      \AxiomC{$\Gamma,(\vec{x},\vec{y} : \bitvector)\vdash \hopsanc(h_1,h_2,k, \vec{x},\vec{y})$}
      \RightLabel{$\texttt{ancilla}$}
      \BinaryInfC{$\Gamma \vdash \rcorrect{h_2}{\texttt{ANC}(C)}$}
    \end{prooftree}
    
    \begin{prooftree}
      \AxiomC{$\Gamma\vdash \rcorrect{h_1}{C}$}
      \AxiomC{$\Gamma\vdash \hopsequiv(h_1,h_2)$}
      \RightLabel{$\texttt{hosp\_subst}$}
      \BinaryInfC{$\Gamma \vdash \rcorrect{h_2}{C}$}
    \end{prooftree}

  \end{center}
  
  \caption{Deduction rules for relation $\rcorrect{\_}{\_}$}
  
  \label{tab:qbrickdsl-circuits}
\end{figure}

  \subsection{Circuit size deduction rules}
We end up this section by providing the compositional deduction rules
for circuit sizes,  in
Figure~\ref{tab:qbrickdsl-nog-size} : \texttt{SWAP} and \texttt{ID} gates
have size $0$, the other gates have size $1$, composing two circuits
either in sequence or in parallel adds up their sizes, controlling a
circuit raises its size by factor at most constant \ctlconst and
discarding an ancilla qubit does not affect the size of a circuit.

\floatstyle{boxed}
\restylefloat{figure}

\begin{figure}[!h]
  
  \begin{center}
    \begin{prooftree}
    \hfill
      \AxiomC{$C\in \{\texttt{ID,SWAP}\}$}
      \RightLabel{$\texttt{Id-SWAP-size}$}
      \UnaryInfC{$\size{C} = 0$}
\DisplayProof
    \hfill
      \AxiomC{$C\in \{\texttt{H}, \texttt{Ph}(n), \texttt{R}_z(n)\}$}
      \RightLabel{$\texttt{gate-size}$}
      \UnaryInfC{$\noeg(C) = 1$}
\hfill
    \end{prooftree}
\hfill
    
    \begin{prooftree}
      \AxiomC{$\Gamma \vdash\noeg(C_1) = n_1$}
      \AxiomC{$\Gamma \vdash\noeg(C_2) = n_2$}
      \AxiomC{$\Gamma \vdash\size{C_1} = \size{C_2} $}
      \RightLabel{$\texttt{seq-size}$}
      \TrinaryInfC{$\Gamma \vdash\noeg((\texttt{SEQ}(C_1,C_2))) = n_1 +  n_2$}
    \end{prooftree}
    \begin{prooftree}
      \AxiomC{$\Gamma \vdash\noeg(C_1) = n_1$}
      \AxiomC{$\Gamma \vdash\noeg(C_2) = n_2$}
      \RightLabel{$\texttt{par-size}$}
      \BinaryInfC{$\Gamma \vdash\noeg((\texttt{PAR}(C_1,C_2))) = n_1 +  n_2$}
    \end{prooftree}
    
    \begin{prooftree}
      \AxiomC{$\Gamma \vdash$\begin{tabular}{l}$ 0\leq k,  k + \size{C} < n ,$\\ $0\leq c <\qsize(\gc), k\leq c \to i+ \size{G} < c$\end{tabular}}
      \AxiomC{$\Gamma \vdash \noeg(C) = n$}
      \RightLabel{$\texttt{ctl-size}$}
      \BinaryInfC{$\Gamma \vdash \noeg(\texttt{CTL}(C,c,k,n)) \leq \ctlconst * n$}
    \end{prooftree}

    \begin{prooftree}
      \AxiomC{$\Gamma\vdash \rcorrect{H_1}{C}$}
      \AxiomC{$\Gamma\vdash \hopsanc(H_1,H_2,k)$}
      \RightLabel{$\texttt{anc-size}$}
      \BinaryInfC{$\Gamma \vdash \noeg(\texttt{ANC}(C,k) = \noeg(C)$}
    \end{prooftree}
  \end{center}

  \caption{Deduction rules for \qbrick : size (number of  gates)}
  \label{tab:qbrickdsl-nog-size}
\end{figure}

\section{Reasoning about HOPS: special circuit subclasses}
\label{app:sec:flat}

In this section, we complement Section \ref{sec:spec-lib}  by presenting 
reasoning rules dedicated to specific cases of circuits, namely flat and diagonal circuits.

\subsection{Flat circuits}
Observe from definitions in Section~\ref{gates-char} that all gates but Hadamard have a correct \hops\ with \hrange\ equal to $0$. We call an \hops\ with range $0$ an \hop, and we call the circuit property of admitting a correct \hop\ being \emph{flat}.

One easily observes that flatness is stable through \texttt{sequence}, \texttt{parallel}, \texttt{control} and \texttt{ancilla} composition.
In  an \hop\ $h$,  function \applyps\ is a pseudo-sum of one term, indexed by the only bit-vector $\vec{y}$ of length $0$ (written ($\vec{\cdot}^0$)). Then we defined functions $\flathket$ and $\flathangle$ over \hop. They ensure
\begin{multline*}\Gamma, x:\bitvector, h : \hops \vdash \\
  \begin{array}{lcr}
    \left\{\begin{array}{l}\bvlength(\vec{x})
             = \hwidth(h) \wedge \\\bvlength(\vec{y}) = 0\ \wedge\\\ \hrange(h =0)\end{array}\right\}
    &\flathangle(h,\vec{x})
    &\{\result = \hangle(h)(\vec{x},\vec{y})\},
    \\&&\\
    \left\{\begin{array}{l}\bvlength(\vec{x}) = \hwidth(h)\ \wedge\\\bvlength(\vec{y}) = 0\wedge \\ \hrange(h =0)\end{array}\right\}
    &\flathket(h,\vec{x})
    &\{\result = \hket(h)(\vec{x},\vec{y})\}\\
  \end{array}
\end{multline*}

and we have:

{\small
  \begin{multline*}\Gamma, x:\bitvector, h:\hops \vdash\\
    \ahoarelr{
      \begin{array}{l}
        \bvlength(\vec{x}) = \hwidth(h), \\
        \hrange(h) = 0
      \end{array}}
    {\applyps(h,\ket{\vec{x}}{})}
    {\begin{array}{l}\result = \\\quad\flathangle (h,\vec{x})\ket{\flathket(h,\vec{x})}{}
     \end{array}}
 \end{multline*}
}

Furthermore, for any given flat circuit $C$, functions \flathangle\ and \flathket\ are actually
independent of the choice of an $\hop$ correct for $C$. Formally, we
have the following theorem :

\begin{multline*}\Gamma, h_1,h_2 : \hops, C: \circuit,\vec{x} : \bitvector \vdash
  \\
  \ahoarelr{
    \begin{array}{l}
      \hrange(h_1) = 0,\\
      \hrange(h_2) = 0,\\
      \rcorrect{C}{h_1}, \rcorrect{C}{h_2} 
    \end{array}
  }{h_1}{
    \begin{array}{l}
      \flathket(\result, \vec{x}) = \flathket(h_2, \vec{y}) ,\\
      \flathangle(\result, \vec{x}) = \flathangle(h_2, \vec{y}) ,\\
    \end{array}
  }
\end{multline*}

Thus, we can directly use functions $\flatangle : \circuit \to
\bitvector \to \complex$ and $\flatket : \circuit \to \bitvector \to
\kettype$ for flat circuits:

\begin{multline*}\Gamma, C : \circuit, \vec{x}, h:\hops : \bitvector\vdash\\
  \begin{array}{lcr}
    \left\{\begin{array}{l}
             \bvlength(\vec{x}) = \size{C},\\
             \hrange(h) = 0,\\
             \rcorrect{C}{h}
           \end{array}
    \right\}&\flatangle(C,\vec{x})&\{\result = \flathangle (h, \vec{x})\},\\
            &&\\
    \left\{\begin{array}{l}
             \bvlength(\vec{x}) = \size{C},\\
             \hrange(h) = 0,\\
             \rcorrect{C}{h}
           \end{array}
    \right\}&\flatket(C,\vec{x})&\{\result = \flathket
                                  (h, \vec{x})\}
  \end{array}
\end{multline*}

Hence, for the semantics of flat circuits, one can reason directly with functions \flatangle\ and \flatket, along the rules given in Figures~\ref{flatgates} (flat gates), \ref{flatcircuits} (flat combined circuits) and \ref{hop} (interpreting \flatket\ and \flatangle\ characterizations in terms of \hops).

\floatstyle{boxed}
\restylefloat{figure}

{\footnotesize

\begin{figure}[!h]
  \begin{flushleft}
    $ $
  \end{flushleft}
  \begin{center}
    \begin{prooftree}
      \AxiomC{}
      \RightLabel{$\texttt{id-flat}$}
      \UnaryInfC{$
        \ahoarelrc{\Gamma, \vec{x} : \bitvector \vdash}{\begin{array}{c}\bvlength(\vec{x})=1\end{array}}{\texttt{ID}}
        {\begin{array}{c}\flat(\result),\\\size{\result} = 1,\\\flatket (\result,\vec{x})  = \vec{x},\\
           \flatangle (\result,\vec{x})  = 1
         \end{array}}$}
   \end{prooftree}
   
    \begin{prooftree}
      \AxiomC{}
      \RightLabel{$\texttt{swap-flat}$}
      \UnaryInfC{$
        \ahoarelrc{\Gamma, \vec{x} : \bitvector \vdash}{\begin{array}{c}\bvlength(\vec{x})=1\end{array}}{\texttt{SWAP}}
        {\begin{array}{c}\flat(\result),\\\size{\result} = 1,\\\multicolumn{1}{l}{\flatket (\result,\vec{x})=}   \\\qquad\texttt{bv\_set}(\texttt{bv\_get}(\vec{x},0 ), 0,\texttt{bv\_get}(\vec{x},1 ))) ,\\
           \flatangle (\result,\vec{x})  = 1
         \end{array}}$}
   \end{prooftree}
   
    \begin{prooftree}
      \AxiomC{}
      \RightLabel{$\texttt{phase-flat}$}
      \UnaryInfC{$
        \ahoarelrc{\Gamma, \vec{x} : \bitvector, n : \inttype  \vdash}{\begin{array}{c}\bvlength(\vec{x})=1 \\ 0\leq n\end{array}}{\texttt{Ph}(n)}
        {\begin{array}{c}\flat(\result),\\\size{\result} = 1,\\\flatket (\result,\vec{x})  = \vec{x},\\
           \flatangle (\result,\vec{x})  = e^{2\pi i \frac{1}{2^n}}
         \end{array}}$}
   \end{prooftree}
   
   \begin{prooftree}
     \AxiomC{}
     \RightLabel{$\texttt{rz-flat}$}
     \UnaryInfC{$
       \ahoarelrc{\Gamma, \vec{x} : \bitvector , n: \inttype  \vdash}{\begin{array}{c}\bvlength(\vec{x})=1 \\0\leq n\end{array}}{\texttt{R}_z(n)}
       {\begin{array}{c}\flat(\result),\\\size{\result} = 1,\\\flatket (\result,\vec{x})  = \vec{x},\\
          \flatangle (\result,\vec{x})  = e^{(-1)^{1-\vec{x}_0}2\pi i \frac{1}{2^p}}
        \end{array}}$}
  \end{prooftree}
  
  \begin{prooftree}
    \AxiomC{}
    \RightLabel{$\texttt{cnot-flat}$}
    \UnaryInfC{$
      \ahoarelrc{\Gamma, \vec{x} : \bitvector  \vdash}{\begin{array}{c}\bvlength(\vec{x})=1 \end{array}}{\texttt{CNOT}}
      {\begin{array}{c}\flat(\result),\\\size{\result} = 1,\\\flatket (\result,\vec{x}), \\\quad = \vec{x}_0\cdot((1-\vec{x}_0)\vec{x}_1 +   \vec{x}_0(1-\vec{x}_1)),\\
         \flatangle (\result,\vec{x})  = 1
       \end{array}}$}
 \end{prooftree}
\end{center}

\caption{Deduction rules for \qbrick : Flat gates}

\label{flatgates}
\end{figure}
}
{\footnotesize
\begin{figure}[!h]
  \begin{center}
    
    \begin{prooftree}
      \AxiomC{$\Gamma \vdash \flat(C_1)$}
      \AxiomC{$\Gamma \vdash \flat(C_2)$}
      \RightLabel{$\texttt{seq-flat}$}
      \BinaryInfC{
        $
        \begin{array}{c}
          \multicolumn{1}{l}{
          \Gamma, x:\bitvector,  \vdash}\\\left\{
          \begin{array}{c}
            \size{C_1} = \size{C_2},\bvlength(x) = \size{C_1} 
          \end{array}\right\}\\
          \texttt{SEQ}(C_1,C_2)\\
          \left\{
          \begin{array}{c}
            \flat(\result)\ \wedge\ \size{\result} = \size{C_1}\wedge\  \\
            \flatangle(\result,\vec{x}) = \flatangle (C_1,\vec{x})\ + 
            \flatangle (C_2,\flatket(C_1,x)) \wedge \\
            \flatket(\result,\vec{x}) =\flatket (C_2,\flatket(C_1,x))
          \end{array}
          \right\}
        \end{array}
        $
      }
    \end{prooftree}

    \begin{prooftree}
      \AxiomC{$\Gamma \vdash \flat(C_1)$}
      \AxiomC{$\Gamma \vdash \flat(C_2)$}
      
      \RightLabel{$\texttt{par-flat}$}
      \BinaryInfC{
        $
        \begin{array}{c}
          \multicolumn{1}{l}{
          \Gamma, x:\bitvector \vdash}\\\{\bvlength(x) = \size{C_1} + \size{C_2}\}\\
          \texttt{PAR}(C_1,C_2)\\
          \left\{
          \begin{array}{c}
            \flat(\result)\wedge\ \size{\result} = \size{C_1} + \size{C_2} \wedge \\
            \flatangle(\result,\vec{x}) = \flatangle (C_1,\vec{x}_{\size{C_1}\upharpoonright})\ * 
            \flatangle (C_2,\vec{x}_{\upharpoonleft\size{C_1}})\wedge \\
            \flatket(\result,\vec{x}) =\texttt{concat}(\flatket (C_1,\vec{x}_{\size{C_1}\upharpoonright})(\flatket (C_2,\vec{x}_{\upharpoonleft\size{C_1}})                    \end{array}
          \right\}\end{array}$}
    \end{prooftree}
    
    \begin{prooftree}
      \AxiomC{$\Gamma\vdash \flat(C)$}
      \RightLabel{$\texttt{ctl-flat}$}
      \UnaryInfC{
        $
        \begin{array}{c}
          \multicolumn{1}{l}{
          \Gamma, x:\bitvector\vdash}\\
          \left\{\begin{array}{l}\bvlength(x) = \size{C}+1\end{array}\right\}\\
          \texttt{CTL}(C)\\
          \left\{\begin{array}{c}
                   \flat(\result)\wedge\ \size{\result} = \size{C}+1 \wedge \\
                   \flatangle(\result,\vec{x}) = \vec{x}_0 * \flatangle (C,\vec{x}_{\upharpoonleft 1})\ +\ (1 - x_0)\ \wedge\\
                   \multicolumn{1}{l}{\flatket(\result,\vec{x}) =} \\\textbf{if }\vec{x}_0 = 1 \textbf{ then }\texttt{concat} (\texttt{bv\_cst} (1, 1), \flatket(C,\vec{x}_{\upharpoonleft 1 } )) \textbf{ else } \vec{x}
                 \end{array}
          \right\}
        \end{array}$
      }
    \end{prooftree}
  
    \begin{prooftree}
      \AxiomC{$
        \ahoarelrc{
          \Gamma, (x :\bitvector)\vdash}{
          \begin{array}{l}
            \bvlength(x) = \size{C},\vec{x}_i = 0\\
        \end{array}}{C}{
\begin{array}{c}
\flat(\result)\ \wedge \\
  \flatket(\result, \vec{x} )_i = 0
\end{array}
        }$}
      \RightLabel{$\texttt{anc-flat}$}
      \UnaryInfC{
        $
        \begin{array}{c}
          \multicolumn{1}{l}{
          \Gamma, (\vec{x}:\bitvector)  \vdash}\\
          \left\{\begin{array}{l}
            \bvlength(x) = \size{C}-1\\
          \end{array}\right\}\\
          \texttt{ANC}(C)\\
          \left\{\begin{array}{c}
                   \flat(\result) \wedge\ \size{\result} = n -1 \wedge \\
                   \flatangle(\result,\vec{x}) =  \flatangle (C, (\concat(\vec{x},\text{bv\_cst}(1,0))))\\
                   \texttt{concat}(\flatket(\result,\vec{x}))(\text{bv\_cst}(1,0))  = \\ \qquad  \big(\flatket(C,\concat(\vec{x},\text{bv\_cst}(1,0)))\big))
                 \end{array}
          \right\}
        \end{array}$
      }
    \end{prooftree}

  \end{center}
  
  \caption{Deduction rules for \qbrick : Flat circuits}
  
  \label{flatcircuits}
\end{figure}
}
{\footnotesize
\begin{figure}[!h]
  \begin{center}
    
    \begin{prooftree}
      \AxiomC{$\Gamma \vdash \rcorrect{h}{C}$}
      \AxiomC{$\Gamma\vdash
        \ahoarelr{\top}{h}{
          \hrange(\result) = 0}$}
      \RightLabel{$\texttt{flat-by-hops}$}
      \BinaryInfC{
        $\Gamma\vdash
        \ahoarelr{
          \begin{array}{c}
            \bvlength(x) = \size{C}\\
          \end{array}}{C}{
          \begin{array}{c}
            \flat(\result)\\
            \flatangle(\result,\vec{x})=         \hangle(h)(\vec{x},\vec{\cdot}^0) \\
            \flatket(\result,\vec{x}) = \hket(h)(\vec{x},\vec{\cdot}^0)\\
          \end{array}}$
      }
    \end{prooftree}
    
    \begin{prooftree}
      \AxiomC{$\Gamma\vdash\flat(C)$}
      \AxiomC{$
        \Gamma  \vdash  
        \begin{array}{c}
          \left\{\begin{array}{l}\bvlength(\vec{x}) = \size{C}\\
                 \end{array}
          \right\} 
          \\h\\
          \left\{
          \begin{array}{c}
            \hrange(\result) = 0,\\ \hwidth(\result) = \size{C}\\
            \hangle(\result)(\vec{x},\vec{\cdot}^0) = \flatangle(C,\vec{x})\\
            \hket(\result)(\vec{x},\vec{\cdot}^0)= \flatket(C,\vec{x}) \\
          \end{array}\right\}
        \end{array}$
      }
      \RightLabel{$\texttt{flat-to-hops}$}
      
      \BinaryInfC{$\Gamma\vdash\rcorrect{h}{C}$}
      
    \end{prooftree}
    \color{black}
  \end{center}
  \caption{Deduction rules for \qbrick : \hop}
  \label{hop}
\end{figure}
}
\subsection{Diagonal circuits}

A further refinement of the semantics is possible for flat circuits
whose function $\flatket$ behaves as the identity. These circuits are called \emph{diagonal}
circuits. Since this property is again preserved
through \texttt{sequence}, \texttt{parallel}, \texttt{control} and
\texttt{ancilla} composition, we can reason about the semantics of diagonal circuits without care for function \flatket. 
~~

\section{Implementation: more details}
\label{app:sec:implem}

We provide here more details about the implementation, complementing Section \ref{sec:implem}. 

\medskip

\subsection{Framework overview}

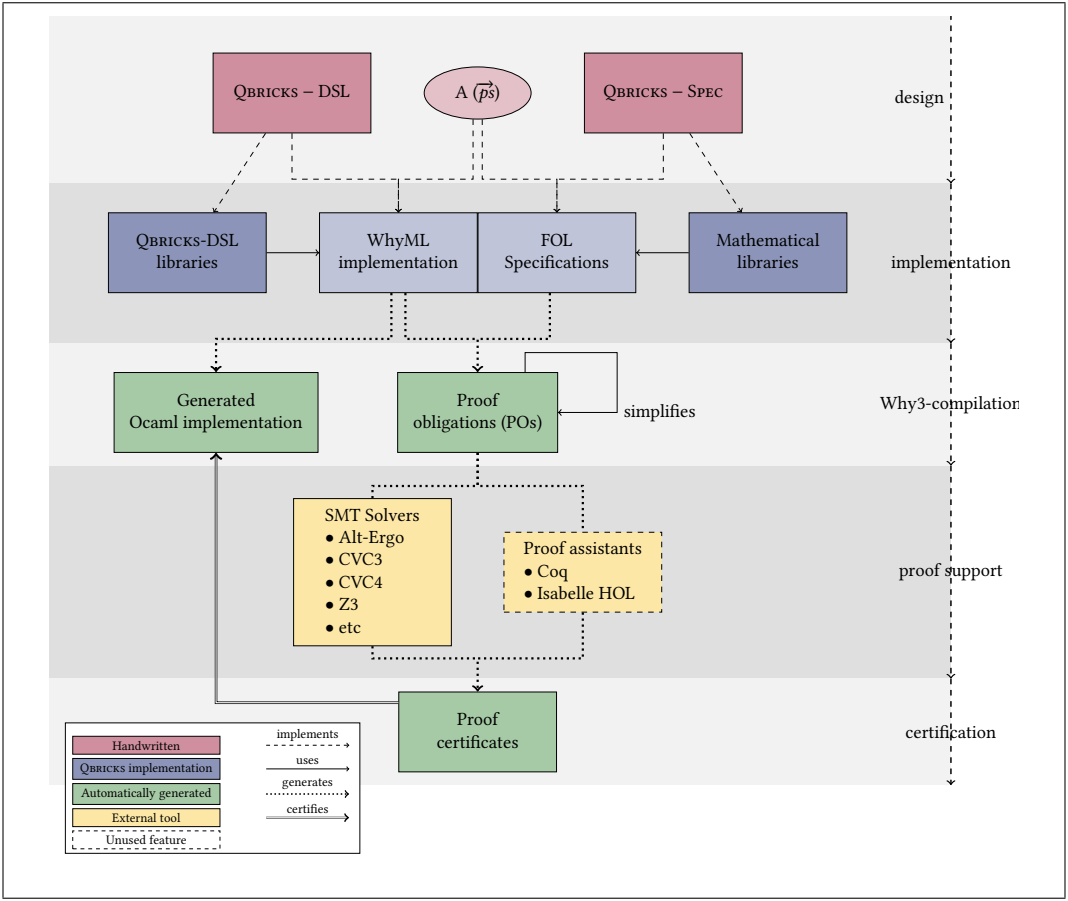
\begin{figure}
  \begin{center}
    \begin{tikzpicture}
      \clip(-5.7,1)rectangle(7.2,-10.4);
      \scalebox{.7}{
        
        \fill[color=gray!10]
        (-9,1.5)--(9,1.5)--(9,-1.7)--(-9,-1.7);
        \fill[color=gray!25]
        (-9,-1.7)--(9,-1.7)--(9,-4.7)--(-9,-4.7);
        \fill[color=gray!10]
        (-9,-4.7)--(9,-4.7)--(9,-7)--(-9,-7);
        \fill[color=gray!25]
        (-9,-7)--(9,-7)--(9,-11)--(-9,-11);
        \fill[color=gray!10]
        (-9,-11)--(9,-11)--(9,-13)--(-9,-13);
        
        \draw[thick,dashed,|->](9,1.5) -- node[left]{design}(9,-1.7);
        \draw[thick,dashed,->](9,-1.7)to node{implementation}(9,-4.7);
        \draw[thick,dashed,->](9,-4.7)to node{Why3-compilation}(9,-7);
        \draw[thick,dashed,->](9,-7)to node{proof support}(9,-11);
        \draw[thick,dashed,->](9,-11)to node{certification}(9,-13);
        
        \begin{scope}
          \node(algo)[draw,ellipse,fill= re!25]at(0,0){\begin{tabular}{c}A ($\overrightarrow{\textit{ps}}$) \end{tabular}};
          \node(dsl)[draw,rectangle, minimum width = 3 cm, minimum height = 1.5 cm,left = 10mm of algo,fill= re!45]{$\qbrickDSL$};  
          \node(spec)[draw,rectangle, minimum width = 3 cm, minimum height = 1.5 cm,right = 10mm of algo,fill= re!45]{$\qbrickSPEC$};  
        \end{scope}
        
        \begin{scope}[yshift =-3cm]
          \node(ref)at(0,0){};
          \node(implem)[draw,rectangle, minimum width = 3 cm, minimum height = 1.5 cm,left = 0mm of ref.center,fill= blu!25] {\begin{tabular}{c}WhyML\\ implementation\end{tabular}};  
          \node(fol)[draw,rectangle, minimum width = 3 cm, minimum height = 1.5 cm,right = 0mm of implem,fill= blu!25]{\begin{tabular}{c} FOL\\Specifications\end{tabular}};  
          \node(matlib)[draw,rectangle, minimum width = 3 cm, minimum height = 1.5 cm,right = 10mm of fol,fill= blu!45]{\begin{tabular}{c} Mathematical \\ libraries\end{tabular}};
          \node(qlib)[draw,rectangle, minimum width = 3 cm, minimum height = 1.5 cm,left = 10mm of implem,fill= blu!45]{\begin{tabular}{c} \qbrickDSL \\ libraries\end{tabular}};
        \end{scope}
        
        \begin{scope}[yshift =-6cm]
          \node(po)[draw,rectangle, minimum width = 3 cm, minimum height = 1.5 cm,fill= gree!35] at (0,0){\begin{tabular}{c}Proof\\ obligations (POs)\end{tabular}};  
          \node(ec)[draw,rectangle, minimum width = 3 cm, minimum height = 1.5 cm, left = 15 mm of po,fill= gree!35]{\begin{tabular}{c}Generated\\ Ocaml implementation \end{tabular}};  
        \end{scope}
        
        \begin{scope}[yshift =-9cm]
          \node(solver)[draw,rectangle, minimum width = 3 cm, minimum height = 1.5 cm,fill= yell!35] at (-2,0){\begin{tabular}{l}SMT Solvers \\$\bullet$ Alt-Ergo  \\$\bullet$ CVC3 \\$\bullet$  CVC4 \\$\bullet$ Z3 \\$\bullet$ etc\end{tabular}};  
          \node(pa)[draw,rectangle,dashed, minimum width = 3 cm, minimum height = 1.5 cm,fill= yell!35] at (2,0){\begin{tabular}{l}Proof assistants\\ $\bullet$ Coq\\ $\bullet$ Isabelle HOL\end{tabular}};  
        \end{scope}
        
        \begin{scope}[yshift =-12cm]
          \node(pc)[draw,rectangle, minimum width = 3 cm, minimum height = 1.5 cm,fill= gree!35] at (0,0){\begin{tabular}{c}Proof\\ certificates \end{tabular}};  
        \end{scope}
        
        \node(whynorth)[above=5mm of implem]{};
        \node(folnorth)[above=5mm of fol]{};
        \node(ponorth)[above=5mm of po]{};
        \node(pcnorth)[above=5mm of pc]{};
        \node(ecnorth)[above=5mm of ec]{};
        \node(posouth)[below=5mm of po]{};

        \draw[->,dashed](dsl)to(qlib);
        \draw[->,dashed](spec)to(matlib);
        \draw[->,dashed](dsl)|-(whynorth.center)to(implem);
        \draw[->,dashed](algo.-100)|-(whynorth.center)to(implem);
        \draw[->,dashed](spec)|-(folnorth.center)to(fol);
        \draw[->,dashed](algo.-80)|-(folnorth.center)to(fol);
        \draw[->](matlib)to(fol);
        \draw[->](qlib)to(implem);
        \draw[very thick,dotted](implem.-80)|-(ponorth.center);
        \draw[->,very thick,dotted](fol.-100)|-(ponorth.center)to(po);
        \draw[very thick,dotted](po) to (posouth.center)-|(solver);
        \draw[very thick,dotted](po) to (posouth.center)-|(pa);
        \draw[very thick,dotted](solver)|-(pcnorth.center);
        \draw[->,very thick,dotted](pa)|-(pcnorth.center)to(pc);
        \draw[->,very thick,dotted](implem.-100)|-(ecnorth.center)to(ec);
        
        \draw[double,<-](ec)|-(pc.160);
        \node(pone)[above right  = 10 mm and 10 mm of po.east]{};
        \draw[->](po.40)|-(pone.center)|-node[right]{simplifies}(po.east);
        
        \begin{scope}[yshift=-17.5cm,xshift=-9cm]
          \scalebox{.7}{
            \node[draw,rectangle,fill=white,minimum width=8cm,minimum height=3.5cm]at(1.8,-1.15){};
            \node(r)[draw,rectangle,fill=re!45,minimum width=4cm]at (0,0){Handwritten};
            \node(b)[draw,rectangle,fill=blu!45,below = 1mm of r,minimum width=4cm]{\qbrick implementation};
            \node(g)[draw,rectangle,fill=gree!35,below = 1mm of b,minimum width=4cm]{Automatically generated};
            \node(y)[draw,rectangle,fill=yell!35,below = 1mm of g,minimum width=4cm]{External tool};
            \node(d)[draw,dashed,below = 1mm of y,minimum width=4cm]{Unused feature};
            \node(i1)[right = 10 mm of r.east]{};
            \node(i2)[right = 35 mm of r.east]{};
            \node(u1)[right = 10 mm of b.east]{};
            \node(u2)[right = 35 mm of b.east]{};
            \node(g1)[right = 10 mm of g.east]{};
            \node(g2)[right = 35 mm of g.east]{};
            \node(c1)[right = 10 mm of y.east]{};
            \node(c2)[right = 35 mm of y.east]{};
            \draw[dashed,->](i1) to node[above]{implements} (i2);
            \draw[very thick,dotted,->](g1) to node[above]{generates} (g2);
            \draw[->](u1) to node[above]{uses} (u2);
            \draw[double,->](c1) to node[above]{certifies} (c2);
          }
        \end{scope}  
        
      }    \end{tikzpicture}
  \end{center}
  \caption{Implementation of \qbrick\ within Why3}
  \label{implem-scheme}
\end{figure}

The overall \qbrick verification chain is  depicted  in
Figure~\ref{implem-scheme}.

\paragraph{Implementation} The first step concerns the design of an
algorithm $A$ with parameters $\overrightarrow{\textit{p}}$ 
thanks to \qbrickDSL 
and \qbrickSPEC.  \qbrick 
is embedded in the Why3 programming language (called WhyML).  We provide for it a set of mathematical libraries, enabling reasoning about
\qbrickSPEC formulas (see Section~\ref{matlib} below for further
details about these).  Note that specifications are inserted within
the implementation, as decorations for functions. The specification
material may also contain, still within the implementation,
intermediary lemmas and ghost material (either ghost definition for
functions objects that are  used exclusively for the specifications, or
ghost elements inserted as parameters for non-ghost functions).

\paragraph{Code extraction, PO generation}  Next, we can perform two different Why3 compilation operations:
\begin{itemize}
\item Code extraction, which basically consists in deleting all
  the ghost material from the implementation. It results in a simple
  Ocaml implementation of $A$ with parameters
  $\overrightarrow{\textit{p}}$.
\item Compilation of the specification code, generating:
  \begin{itemize}
  \item first
    order logic proof obligations (POs). 
  \item first
    order logic axioms, formalizing  definitions from the algorithm implementation. They contribute to the logical context to evaluate PO satisfaction, together with, for any function $f$, specifications
    for functions above $f$ or from imported files.
  \end{itemize}  
\end{itemize}  

\paragraph{Solving} Then \qbrick takes advantage of various means to prove  the
generated POs modulo our equational theories. A dedicated interface enables to directly access them
and either  send them  to a set  of automatic SMT-solvers (CVC4, Alt-Ergo,
Z3, etc.), or enter a number of interactive proof transformation commands (calls for lemmas or hypotheses, term substitutions, etc.) or even to  proof assistants (Coq, Isabelle/HOL) -- we do not use this option in our case-studies.

\medskip 

Statistics about \qbricks implementation are given in
Table~\ref{gen-data}. 
The development itself  counts 17,000+ lines
of code, including  400+ definitions and 1700+ lemmas -- all proved within Why3 but 32 axioms in the mathematical libraries (see next section). 
Notice the volume of (verified) mathematical libraries in
the overall development (14,695 loc for a total of 17,313 loc). These libraries are described in the next section.

\subsection{Mathematical libraries}
\label{matlib}

 When we began developing \qbricks,
there was no native automated support for data-types commonly
manipulated in quantum, such as complex numbers, angles, matrices,
kets, vectors, etc. Hence, aside the implementation of \qbrickDSL and
\qbrickSPEC, \qbricks
development also required the development of a WhyML library for these
data-types {\it in the quantum context}.

The goal, for this specific part, was to provide the
mathematical structures at stake in quantum computing, together with a
{\it formally verified collection} of mathematical results concerning
them. Then these developments are led as mathematical theories,
elaborating proofs from a restricted number of axiomatic definitions.

As shown Table~\ref{gen-data}, this library contains 14,000+ lines
of code: 300+ definitions, 1600+ lemmas and 32 axioms.
Most of the axioms are actually axiomatic definitions giving structure for  data types $\complex$ and angle. These are listed in Table~\ref{axioms}:
\begin{itemize}
\item Complex numbers are given a field structure. We introduce a
  casting function \rtoc\ from the Why3 standard library type
  \textsc{real} to \complex. It preserves the field structure. Then
  constant $i$ is introduced and any complex $x$ is uniquely
  identified as the addition of a real value  $\real(x)$ and a
  pure imaginary value $i*\imag(x)$. Real numbers ordering are imported
  for pure real complexes (with $\imag = 0$) and $\pi$ is introduced
  as a positive constant. At last we axiomatically define integer
  exponentiation of complexes and the standard complex exponentiation
  of constant $e$.
\item Angles are given a group structure. A casting function \ctoa\ 
  from \complex to angles is introduced. Given value $x$, it builds
  the angle of measure $2i\pi*x$ radians. Conversely, function $\atoc$,
  given an angle $\theta$, computes the complex value corresponding to
  $\theta$ on the unity circle. Mind that angles are uniquely
  determined by this complex value. At last, we introduce a single
  division over 2 operation for angles.
\end{itemize}

\begin{table}[h]
  \begin{center}
    \scriptsize
    \begin{tabular}{|l|c|c|c|c|}
      \cline{2-5}
      \rowcolor{gray!50}
      \multicolumn{1}{c|}{\cellcolor{white}
      }      &Lines of code & Lemmas & Modules & Definitions\\
      \hline
      \Xhline{1pt}
      \rowcolor{gray!25}
      \textbf{Mathematics libraries} &14695&1614&77&328\\
      \Xhline{1pt}
      Sets&532&59&4&14\\
      \rowcolor{gray!25}
      Algebra&2091&190&10&37\\
      Arithmetics&538&77&4&7\\
      \rowcolor{gray!25}
      Binary arithmetics&1778&189&8&42\\
      Complex numbers&2226&344&15&57\\
      \rowcolor{gray!25}
      Quantum data&3335&310&12&68\\
      Exponentiation &843&100&4&4\\
      \rowcolor{gray!25}
      Iterators &861&72&6&30\\
      Functions &259&33&3&8\\
      \rowcolor{gray!25}
      Kronecker product &420&41&2&8\\
      Unity circle&1812&199&9&53\\
      \hline
      \hline
      \rowcolor{gray!25}
      \textbf{Qbricks core}&1357&50&5&35\\
      \Xhline{1pt}
      \textbf{Semantics reasoning}&744&55&3&35\\
      \Xhline{1pt}
      \rowcolor{gray!25}
      \textbf{Generic functions}&517&12&5&34\\
      \Xhline{1pt}
      \multicolumn{5}{c}{\tiny{}}\\
      \Xhline{1pt}
      \rowcolor{gray!50}
      \textsc{Total}&17313&1731&78&410\\
      \Xhline{1pt}
    \end{tabular}
    \caption{Metrics about \qbricks implementation}
    \label{gen-data}
  \end{center}
\end{table}  

\begin{table}[!h]
  {\small
    \begin{tabular}{lc|c}
      \Xhline{2\arrayrulewidth}
      \rowcolor{gray!20}
      \multicolumn{3}{c}{\textbf{Complex numbers}}\\
      \Xhline{2\arrayrulewidth}
      \multirow{1}{*}{\textbf{Field:}}& \multicolumn{2}{c}{  
                                        Type $c(-,+,*,0,1)$, is  a field$^{*}$}\\
      \hline
      \multirow{3}{*}{\textbf{\rtoc.} : $r\to c$}&  $\rtoc (0.0) = 0 $&    $\forall x, y:r. \rtoc (x*y) = x*y$
      \\&  $\rtoc (1.0) = 1$ &    $\forall x, y:r. \rtoc (x+y) = x+y$\\
                                      & $\forall x:r.\rtoc (- x) = - \rtoc(x)$ & $\forall x, y:r. y \neq 0.0 \to\rtoc (x/y) = x/y$\\
      \hline
      \textbf{ $i$ : t}&
                         \multicolumn{2}{c}{\(i*i=-1\)}\\
      \hline
      \multirow{4}{*}{\textbf{$\real,\imag$}: $c\to c$} 
                                      &\multicolumn{2}{c}{$\forall x,y: c.\ \real(x+y) = \real(x)+\real(y)$} \\
                                      &\multicolumn{2}{c}{$\forall x,y: c.\ \imag(x+y) = \imag(x)+\imag(y)$} \\
                                      &\multicolumn{2}{c}{$\forall x: c.\ x = \real(x)+i *\imag(x)$}\\
                                      &\multicolumn{2}{c}{$\forall x: c.\ \forall y,z: c.\ x = y+i* z \to y = \real(x)\wedge z=\imag(x) $}\\
      \hline
      \textbf{ $\pi$: t}&
                          \multicolumn{2}{c}{\(0<\pi\)}\\
      \hline
      \multirow{2}{*}{\textbf{$[]^{[]}$}}: $c\times \Int \to c$
                                      &
                                        $\forall x:c. x^0 = 1$& $\forall x:c.\forall i,j: \Int.x^{(i+j)} = x^i*x^j $\\
                                      & $\forall x:c. x^1 = x$& 
                                                                $\forall x:c.x\neq 0 \to 0^x=0 $ \\
      \hline
      \multirow{3}{*}{\textbf{$e^{[]}$}: $c \to c$}
                                      &
                                        
                                        $\forall x:c. e^x\neq 0$ &
                                                                   $ e^0 = 1$  \\
                                      & 
                                        $e^{i\pi/4} = \sqrt{2}(1+i)$ &$e^1 = e$ \\
                                      & \multicolumn{2}{c}{$\forall x,y:c.e^{(x+y)} = e^x*e^y $} \\
      
\hline
      \Xhline{2\arrayrulewidth}
      \rowcolor{gray!20}
      \multicolumn{3}{c}{\textbf{Angles}}\\
      \Xhline{2\arrayrulewidth}
      
                                      & \multicolumn{2}{c}{Type $a(+_{a},-_{a},0_{a})$, is  a group $^{*}$}\\
      
      \hline
      
      \multirow{2}{*}{\textbf{\ctoa} : $c\to a$}
                                      & \multicolumn{2}{c}{$\forall x,y:c. \imag(x) = 0 \to \imag(y) = 0 \to \ctoa(x+y) = \ctoa(x)+_a \ctoa(y)$ }\\
                                      & \multicolumn{2}{c}{$\forall x:c\_r. \imag(x) = 0 \to \ctoa(-x) = -_a\ctoa(x)$}\\
      \hline
      \multirow{2}{*}{\textbf{\atoc} : $a\to c$}&$\atoc(0_a) = 1$ &
                                                                    $\forall \theta,\theta':a. \theta = \theta' \leftrightarrow \atoc(\theta) = \atoc(\theta') $\\
                                      & \multicolumn{2}{c}{ $\forall x:c. \imag(x) = 0 \to  \atoc(\ctoa(x)) = e^{2i\pi*x}$}\\
      \hline
      $[]\slash_a 2$ : $a\to a$& \multicolumn{2}{c}{$\forall \theta:a. \theta\slash_a 2 +_a \theta\slash_a 2 = \theta$}
      \\
      \hline
    \end{tabular}
    
    $(*)$: theory cloned from the \texttt{algebra} Why3 standard library 
  }
  \caption{Axiomatic definitions introduced in \qbrickSPEC libraries}
  \label{axioms}
\end{table}

\qbrickSPEC libraries use two additional axioms : the matrix data type
is a record type with two positive integer arguments (for the number
of rows and the number of columns of the matrix) and a value function
of type $(\Int\times\Int)\to \complex$. An axiom identifies matrices
if they share the same number of rows $r$ and columns $c$, and their
value coincide for positive integers bounded by $r$ and
$c.$. Kets are introduced as matrices respecting certain conditions (to have $2^n$ rows, for a given positive integer $n$, and $1$ column). Similarly,
the \bitvector\ opaque data-type is introduced with parameter a positive integer length
and value  from \Int to $\{0,1\}$. And we identify, by use of a last
axiom, bit vectors that share the same length and whose value coincide
up to this bound.

\section{Experimental evaluation: more details}
\label{app:sec:xp}

This section complements Section \ref{sec:xp}, with both more detailed commented descriptions of case-studies (Section \ref{app:sec:xp:cases}) and additional statistics (Section \ref{app:sec:xp:stats}).

\subsection{Case studies: description and comments}
\label{app:sec:xp:cases}

This section complements Section \ref{sec:xp:cases}. 
We introduce and comment the specifications for our implementations of QPE, Shor-OF, Deutsch-Josza and Grover search algorithms.

\paragraph{Quantum Phase Estimation}
All the eigenvalues of a unitary operator $U$ are equal to
$e^{2\pi i \Phi}$ for a given real $\Phi$ such that $0\leq\Phi <1$.
Phase estimation (QPE)~\cite{kitaev1995quantum,cleve1998quantum} is a
procedure that, given a unitary operator $U$ and an eigenvector
$\ket{v}{}$ of $U$, finds the eigenvalue $e^{2\pi i \Phi_v}$
associated with $\ket{v}{}$.  It
is a central piece in many emblematic algorithms, such as quantum
simulation~\cite{georgescu2014quantum} or HHL
algorithm~\cite{harrow2009quantum} -- resolution of linear systems of
equations in time \textsc{PolyLog}.

\begin{figure}[th]
  \begin{center}
    \scalebox{.8}{
      \begin{tikzpicture}[xscale =.25,yscale =.35,decoration={brace}][scale=2] 
        \input{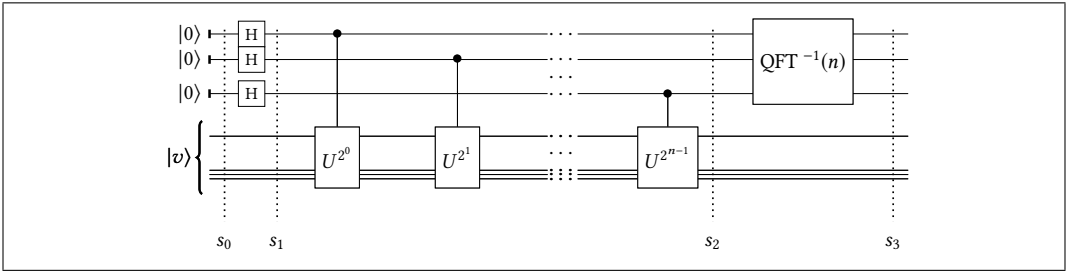}
      \end{tikzpicture}
    }
    \caption{The circuit for QPE}
    \label{phase-struct-simple2}
  \end{center}
\end{figure}

An overall view on the circuit for QPE is given on
Figure~\ref{phase-struct-simple2} and the successive intermediary register states are given in Table~\ref{qpe-spstates}.
It uses two registers, one of size $n$ and initialized to
$\ket{0}{n}$ and one of size $s$ and initially in state $\ket{v}{s}$:
First, we superpose the first register of the entry state, by applying an Hadamard gate $H$ to each of its qubits. It results in state $s_1$.

\begin{figure}[th]
  \begin{center}
    \[
    \begin{array}{lcr}
     s_0  = \ket{0}{n}\otimes \ket{v}{s}
& \qquad 
 s_1 = \left(\frac{1}{\sqrt{2^n}}\sum_{k=
  0}^{2^n-1}\ket{k}{n}\right)\otimes\ket{v}{s}
& \qquad
s_2 = \left(\frac{1}{\sqrt{2^n}}\sum_{j = 0}^{2^n-1} e^{\frac{2\pi i*\bitrev{j} \phi_v}{2^n}}
\ket{j}{n}\right)\otimes\ket{v}{s}
\\ \multicolumn{3}{c}{ s_3 = \bigg(\frac{1}{2^n}\sum_{j = 0}^{2^n-1}\sum_{l = 0}^{2^n-1} e^{\frac{2\pi i*\bitrev{j} (l-\varphi_v)}{2^n}} \ket{i}{n}\bigg)\otimes\ket{v}{}}

\end{array}
\]
    \caption{Successive states of the register along QPE circuit}
    \label{qpe-spstates}

  \end{center}
  \end{figure}

Then we perform, on the second register,
a sequence of $U$ circuits elevated to the successive powers of $2$ and
controlled by qubits from the first register. It results   in
 state $s_2 $.
At last, we apply the reversed quantum Fourier transformation, $\qft^{-1}(n)$, to the first register and get state $s_3$.

\begin{figure}[th]
  \begin{center}
\[
\begin{array}{c}
\multicolumn{1}{l}{\Gamma,  (f: \hops) , (C: \circtype) ,( \ket{v}{} : \kettype),  (k,n: \inttype),(\ghost\ \theta:\realtype), (j : \ghost\ \inttype) \vdash }\\
  \left(
  \begin{array}{c}
\rcorrect{f}{C}\ \wedge\  \width(C) = n\ \wedge\ 0< k\ \wedge \ \eigen(f,\ket{v}{},e^{2\pi i*\theta})\\

  \end{array}
\right)\\
\texttt{QPE}(C,k,n)\\
\left(\begin{array}{c}
  \probmeaspartp (\result, {k}\ket{v}{}, \texttt{error}<\frac{1}{2^{k+1}}) \geq \frac{4}{\pi^2}\ \wedge\ \\ 
  \theta = \frac{j}{2^k}\to\probmeaspart (\result, \ket{v}{},\ket{j}{k}) = 1\
   \\ 
   \end{array}
\right)
\end{array}
\]
    \caption{Specification for our  implementation of Quantum Phase estimation}
    \label{qpe-spec}
  \end{center}
\end{figure}

The specifications for this algorithm are given in
Figure~\ref{qpe-spec}. It inputs an oracle HOPS $f$ together
with a circuit  $C$ implementing it, an eigenvector $\ket{v}{}$ of $f$
together with the associated ghost eigenvalue $e^{2\pi i*\theta}$ and two
register length parameters $k$ and $n$.  The first postcondition  then certifies that,
in the general case, the probability to find an $x$ such that
$x-\theta < \frac{1}{2^{n+1}}$ after applying circuit
$\texttt{QPE}(C,k,n)$ on entry $\ket{0}{k}\ket{v}{}$ and measuring
qubits $0$ to $n$ is at least $\frac{4}{\pi^2}$. The second postcondition  concerns the
particular case when $\theta$ is a multiple of $\frac{1}{2^n}$. Then
this probability rises to $1$. 

\paragraph{Shor-OF} \cite{shor1994algorithms}'s Order Finding is the central procedure in Shor's integer prime factor
decomposition algorithm. It consists in a
particular instantiation of QPE.
This algorithm is certainly the most famous of all quantum computing existing algorithm,  raising worldwide interest for the domain.
Given a non prime
integer $N$, it outputs a factor of it in time \textsc{PolyLog}$(N)$. Its most spectacular application is that it enables to break the RSA encryption protocol in polynomial time.

As a result from  Bézout's identity,  for any integer
$a<N$ co-prime with $N$, if $r$ is the modular order of $a$ in $N$
(that is the least integer such that $\mod\ a^r N = 1$) and if $r$ is even (which happens with probability $\bigcirc(1)$), then
either $\gcd(a^{r/2}-1)$ or $\gcd(a^{r/2}+1)$
divides $N$.
Furthermore, if $U$ is an unitary implementing the multiplication modulo $N$ then for any $k \leq r, e^{\frac{-2i\pi*sk}{r}}$ is an eigenvalue of $U$.
The quantum part of Shor-OF algorithm then consists in:
\begin{itemize}
\item implementing, for each $j<k$, the circuit $U^{2^j}$,
  \item applying quantum phase estimation, with oracle these $U^{2^j}$, on input  ket $\ket{1}{n}$ (which is  the uniform superposition of $U$ eigenstates).
  \end{itemize}

Note the importance of the implementation of  $U^{2^j}$ circuits. In particular, a naive implementation of them as  iterative sequences
of $2^j$ instances of $U$ would introduce an exponential complexity
factor and loose any advantage of using a quantum computer. Hence, the functional correction of an implementation is not enough. In the case of Shor-OF for example, a relevant implementation must also certify the \textsc{PolyLog} complexity requirement.

We implemented the version of the oracle $U^j$ described in the reference implementation~\cite{beauregard2002circuit}.
The overall specification for our implementation  is given in Figure~\ref{phase-struct-simple2}.

\begin{figure}[th]
  \begin{center}

      \[
\begin{array}{c}
\multicolumn{1}{l}{\Gamma (a,b,n: \inttype), (j: \ghost\ \inttype) \vdash}\\
\left(\begin{array}{c}
\coprime (a,b)\wedge\ 1 \leq b < 2^n  \wedge 1\leq j < b \ \wedge
\mod\ a^j\ b = 1\\

  \end{array}
\right)\\
\texttt{Shor-circ}(a,b,n)\\
 \left(\begin{array}{c}
  \probmeaspartp \big(\ket{1}{n}, \texttt{error}\leq_{\frac{1}{2^{2n+1}}} \big) \geq \frac{4}{\pi^2}\  \wedge\\
   \nogates(\result ) = \polyshor(n)\  \wedge\\  \ancillas(\result) = n+2 \wedge\ 
   \width(\result )= 3 *n
 \end{array}
\right)
\end{array}
\]
    \caption{Specification for our implementation of Shor-OF algorithm}
    \label{shor-spec}
  \end{center}
\end{figure}
\begin{enumerate}
\item As for QPE, the probability that the measurement will give an
  output close to an eigenvalue is more than $\frac{4}{\pi^2}$. In
  this case, recovering the actual value $e^{\frac{-2i\pi*sk}{r}}$
  requires another classical manipulation, the \emph{continuous
    fraction expansion},
\item the second postcondition concerns the size of the circuit,
  proved by use of rules from Figure~\ref{tab:qbrickdsl-nog-size}. Note
  that the specification is not given as a complexity class (such as
  $\bigcirc(n^4)$), but gives the actual polynomial $\polyshor(n)$. It
  is equal to $(n+1)^4(28 \ctlconst^2 + 12\ctlconst) + (n+1)^2(10
  \ctlconst + 1)+ 4n+ 8\ctlconst+3 $. Furthermore, the implementation
  of oracles in~\cite{beauregard2002circuit} uses several instances of the QFT. For them, it leaves
  open the possibility to use
   approximations for QFT with $k_{\textit{max}}$ gates. Hence the
  author provides a size bound  that is in $\bigcirc(n^2)$ and linear in
  $k_{\textit{max}}$. In our implementation we used the standard exact
  implementation of QFT, with number of gates bounded by $(n^2)$.
\item The third and fourth postcondition specify the width of the
  circuit and the number of additional ancilla qubits it
  requires. For the final application of QFT$^{-1}$ (that appearing  in Figure~\ref{phase-struct-simple2}), \cite{beauregard2002circuit} uses a
  semi-classical version of QFT$^{-1}$, which makes use of a single qubit instead of $2*n$.
Since \qbrick does not contain classical data operations, here also we implemented the standard 
version of  QFT$^{-1}$, so that the whole circuit has width $3*n$ and uses an additional $n+2$ ancilla qubits.

\end{enumerate}

\paragraph{Deutsch-Josza algorithm}
Deutsch-Josza algorithm is a toy algorithm, mainly used in
introductory courses as an illustration of quantum state
superposition. Given a positive integer $n$, given a function $f$ from \inttype\ to \bool\ that is guaranteed to be either constant or balanced on $\tofset{2^n}$ and given a circuit $C$ \emph{implementing} $f$\footnote{ which formally means  that for any bit vectors $x$ of length $n$  and $y$ of length $1$,
\[\iogate(C, \ket{x}{}\otimes \ket{y}{} = \ket{x}{}\otimes \ket{y\oplus f(\bvtoint(x))}{}\]
}, Deutsch-Josza algorithm decides, in time linear in $n$, whether $f$ is constant or balanced on  $\tofset{2^n}$. Its complete specifications is given as:

\[
\begin{array}{c}
\multicolumn{1}{l}{\Gamma, (C: \circtype) , (f:  \inttype \to \bool), (n : \inttype) \vdash }\\
  \left(
\begin{array}{c}
\implements(C,f)\ \wedge  1< n\ \\ \wedge
 (\neg\constant(f,n) \to \balanced(f,n))\\

  \end{array}
\right)\\
\texttt{Deutsch-Jozsa}(C,n)\\
 \left(\begin{array}{c}
   \multicolumn{1}{l}{ \constant(f,n)\to \probmeaspart}\\ \qquad  (\result, \texttt{concat}(\texttt{bv\_cst }(n,0))  (\texttt{bv\_cst }(1,1))) = 1\  \wedge\ \\ 
   \multicolumn{1}{l}{ \balanced(f,n)\to \probmeaspart}\\ \qquad  (\result, \texttt{concat}(\texttt{bv\_cst }(n,0))  (\texttt{bv\_cst }(1,1)),  \texttt{bv\_cst }(n,0)) = 0\  \wedge\ \\ 
   \nogates(\result )= 2 * n + \nogates(C)+2\ \wedge\   \\ 
   \width(\result )= n +1\ \wedge \ancillas(\result = \ancillas(C))\\ 
 \end{array}
\right)
\end{array}
\]
Hence, the two first postconditions ensures that the partial
measurement outputs $\ket{0}{}$ exactly when the function $f$ is
constant, the third postcondition gives a size bound for the circuit (that is
linear in both $n$ and the size of the oracle implementation), and last
line states that the width of the circuit is $n+1$ and it does not use
any extra ancilla from those (maybe) required for $C$.

\paragraph{Grover algorithm}
Grover algorithm enables to search distinguished elements in a
non-structured data base. It notably provides a quadratic acceleration
in the resolution of NP-hard problems.  Given positive integers $n$
and $k$ such that $k\in\tofsett{1}{2^n}$, given a function $f$ from
\inttype\ to \bool\ that is guaranteed to be satisfied by exactly $k$
elements in $\tofset{2^n}$ and given a circuit $C$ \emph{implementing}
$f$ and a positive index $i$, the algorithm consists in an iterative
sequence of $i$ times a sequence combination of a specific operator
(called \emph{Grover diffusion operator}) and oracle $U$. It is such
that the probability to obtain, after measurement, a
result satisfying $f$, is equal to
$\sin^2\left(\arcsin\left(\frac{k}{2^n}\right)\dot (1 +
2i)\right)$. This specification is formalized below together with  the size of the circuit (that is linear in both $n$ and the number of iterations $i$) and  the width of the circuit (equal to $n$).


\[
\begin{array}{c}
\multicolumn{1}{l}{\Gamma, (C: \circtype) , (f:  \inttype \to \bool),(n,i,k : \inttype) \vdash }\\
  \left(
\begin{array}{c}
\implements(C,f)\ \wedge  1< n\ \wedge 1 \leq k < 2^n -1\ \wedge 1 \leq i  \\ \wedge\ 
  \texttt{Card}(\{j\mid 0\leq j < 2^n \wedge  f(j) = \texttt{true}\}) = k \\
  \end{array}
\right)\\
\texttt{Grover}(C,k,n)\\
\left(\begin{array}{c}
 \probmeaspart_f  (\result,   \texttt{bv\_cst }(n,0),f) = \sin^2\left(\arcsin\left(\frac{k}{2^n}\right)\dot (1 + 2i)\right)\ \wedge \\
   \nogates(\result )= i*(\nogates(C)* \bigcirc(n))\\ \wedge\ 
   \width(\result )= n\wedge \ancillas(\result) =1\\ 
 \end{array}
\right)
\end{array}
\]

In this specification, we used expression $ \probmeaspart_f  (\result,   \texttt{bv\_cst }(n,0),f)$ as a syntactic sugar for \[\sum_{\tiny \begin{array}{c} j\in \tofset{2^n}\\ f(j)=\texttt{tt}\end{array}}
\probmeaspart  (\result,   \texttt{bv\_cst }(n,0), \inttobv(n,i))\]

  
  
      



\subsection{Case studies: metrics}
\label{app:sec:xp:stats}

This section complements Section \ref{sec:xp:stats}. 

\paragraph{Details on interactive proof commands} Table~\ref{app:com-stats} reports a classification of  
the interactive (proof) commands used in our case-studies to discharge the remaining PO. 
Interactive commands
divide mainly into:  
\begin{itemize}
\item calls for deduction rules (Rules), 
\item calls for hypotheses from the context (Hyp.), 
\item $\beta-$reductions, 
\item splitting commands
(splitting conjunction in either goal or context and elimination of
case distinctions), 
\item and substitutions (Subst). 
\end{itemize}

\noindent Column Misc. contains introduction of constant values for  existential quantification, 
case distinctions and intermediate assertions.

\begin{table}[htbp]
  \begin{center}
    \begin{tabular}{|l|c|c|c|c|c|c|c|}
      \cline{2-8}
      \rowcolor{gray!50}
      \multicolumn{1}{c|}{\cellcolor{white}
      }      &\#Rules & \#Hyp. &\#$\beta$-red.&\#Split. & \#Subst. & Misc.
      &\textbf{Total}\\
      \hline
      \rowcolor{gray!25}
      DJ &17&9&10&2&1&0&\textbf{39}\\
      Grover &91&26&23&12&15&0&\textbf{167}\\
      \rowcolor{gray!25}
      QFT &15&14&5&3&0&0&\textbf{37}\\
      QPE &66&36&17&15&19&2&\textbf{155}\\
      \rowcolor{gray!25}
      Shor-OF  & 93 &62 &25 &57 &19 &8 &\textbf{264} \\
      Shor-OF (full)    &   174 &112 &47 &75 &38 &10 &\textbf{456} \\

      \hline 
      \hline 

      \rowcolor{gray!50}
      Total    &  282 &147 &80 &89 &54 &10 &\textbf{662}  \\ 
      \hline    \end{tabular}
    
    \#Rules: calls for deduction rules ---  \#Hyp.: calls for hypotheses
    
    \smallskip 
    \caption{Verification for case studies: repartition of interactive commands} 
    \label{app:com-stats}
  \end{center}
\end{table}  

As a complement to these figures, it should be mentioned that the
specifications for these algorithms all contain some pure mathematical
proofs which are not directly linked to circuit composition. For example,  198
out of the 321 lines of decorated code declared for QPE in
Table~\ref{data-cs} concerns rewriting lemmas for the measurement
outcomes. Also, our implementation for Grover follows the standard
algorithm explanation. Doing so, it introduces,  in addition to circuit building and
\hops\ specifications, vectorial space embedding and geometry material which, again, actually represents the major part of the proof effort.

\paragraph{Prior work} Table \ref{app:com} gives a summary of anterior efforts for the verification of  quantum programs.

{\footnotesize
\begin{table}[th]
  \begin{center}
    \begin{tabular}{|l|c|c|c|c|c|c|}
      \cline{3-7}
      \rowcolor{gray!50}
      \multicolumn{2}{c|}{\cellcolor{white}
      } &Circuit building&Parameters&Restrictions&Size&Automation\\
\hline
    \cline{3-7}\multirow{5}{*}{\qbrick}&DJ&\checkm&$U,n$&None&53LoC + 39Cmd&Partial\\
&Grover&$\checkm$&$U,k,n$&None&416 LoC+ 167 Cmd&Partial\\
    
      &QFT&\checkm&$n$&None&65 LoC + 37 Cmd&Partial\\
      &QPE&\checkm&$U,n,\ket{v}{}$&None&319 LoC + 155 Cmd&Partial\\
      &Shor-OF&\checkm&$n,a,b$&None&809 LoC + 264 Cmd&Partial\\
    \hline
    \hline
    \multirow{2}{*}{Path-sums}  &QFT&\checkm&Not parameterized&Only instances &Unknown&Full\\    
    &Others&\xmark&\xmark&\xmark&\xmark&\xmark\\
      
\hline
      \multirow{2}{*}{Qwire}&DJ&\checkm&$U,n$&None&\begin{tabular}{c}74 LoC + 222Cmd $^{\dagger}$\\59 LoC + 112Cmd $^{\dagger}$\end{tabular}&None\\
      &Others&\xmark&\xmark&\xmark&\xmark&\xmark\\
        \hline  \multirow{2}{*}{QHL}
      &Grover& None & $U,k,n$& $\exists j.  k = 2^j$&$3000+ LoC$ $^{\ddagger}$&Partial\\
      &Others&\xmark&\xmark&\xmark&\xmark&\xmark\\
\hline
    \end{tabular}
    \#LoC.: lines of decorated code  ---  \#Cmd: interactive commands \linebreak
    \# $U$ : circuit oracle -- \# $n$ : width of the input register
\# $\ket{v}{}$ eigen vector of $U$  -- \# $a,b,k$ : further integer parameters      
    
    \smallskip 
    \caption{Comparison with quantum formal verification from the literature} 
    \label{app:com}

  \end{center}

$^{\dagger}$: The authors provide two proofs: the textbook-style proof and a new clever proof. The proof in \qbrick\ corresponds to  textbook style.

$^{\ddagger}$: QHL does not provide any information on proof commands or tactics, but the lines of code include intermediate lemmas.  

\end{table}  
}

\end{document}
